\newif\ifdraft
\draftfalse
\newif\iffull
\fulltrue

\documentclass{llncs}
\usepackage{times}
\usepackage{makeidx}
\usepackage{hyperref}
\usepackage[users=CYM, \ifdraft \else hide \fi]{uchanges}
\ChangesSetUserColor{C}{blue}
\ChangesSetUserColor{M}{red}
\ChangesSetUserColor{Y}{purple}

\usepackage{listings}
\usepackage{infer}
\usepackage{mathtools}
\usepackage{amsfonts}
\usepackage{ amssymb }
\usepackage{color}
\usepackage{xargs}
\definecolor{lightgray}{rgb}{.9,.9,.9}
\definecolor{darkgray}{rgb}{.4,.4,.4}
\definecolor{purple}{rgb}{0.65, 0.12, 0.82}
\definecolor{darkgreen}{rgb}{0, 0.5, 0}
\definecolor{turquoise}{rgb}{0, 0.5, 0.5}
\usepackage{float}
\usepackage{subcaption}
\usepackage{multirow}
\usepackage{booktabs}

\lstdefinelanguage{JavaScript}{
  keywords={typeof, new, true, false, catch, function, return, null, catch, switch, var, if, in, while, do, else, case, break},
  keywordstyle=\color{blue}\bfseries,
  ndkeywords={class, export, boolean, throw, implements, import, this, contract},
  ndkeywordstyle=\color{turquoise}\bfseries,
  keywords={[3]bool, address},
  keywordstyle= [3] \color{darkgreen}\bfseries,
  identifierstyle=\color{black},
  sensitive=false,
  comment=[l]{//},
  morecomment=[s]{/*}{*/},
  commentstyle=\color{purple}\ttfamily,
  stringstyle=\color{red}\ttfamily,
  morestring=[b]',
  morestring=[b]"
}

\newcommand{\myparagraph}[1]{\medskip\noindent\textbf{#1}}

\newcommand{\wei}{\textit{wei}}
\newcommand{\define}{:=}
\newcommand{\nil}{\epsilon}
\newcommand{\cons}[2]{{#1}::{#2}}
\newcommand{\BB}{\mathbb{B}}
\newcommand{\NN}{\mathbb{N}}
\newcommand{\arrayof}[1]{[{#1}]}
\newcommand{\sequenceof}[1]{\mathcal{L}(#1)}
\newcommand{\setof}[1]{\mathcal{P}(#1)}
\newcommand{\lam}[2]{\lambda {#1}. \, {#2}}
\newcommand{\none}{\bot}

\newcommand{\access}[2]{{#1}. {#2}} 
\newcommand{\size}[1]{|{#1}|}

\newcommand\pto{\mathrel{\ooalign{\hfil$\mapstochar$\hfil\cr$\to$\cr}}}

\newcommand{\bytearray}{\arrayof{\BB^8}}
\newcommand{\byte}{\BB^8}
\newcommand{\integer}[1]{\NN_{#1}}
\newcommand{\logevents}{\textit{Ev}_{\textit{log}}}
\newcommand{\memorytype}{\BB^{256} \to \BB^8}
\newcommand{\storagetype}{\BB^{256} \to \BB^{256}}
\newcommand{\word}{\BB^{256}}
\newcommand{\addresses}{\mathcal{A}}

\newcommand{\accounts}{\mathbb{A}}
\newcommand{\blockheaders}{\mathcal{H}}
\newcommand{\transactions}{\mathcal{T}}
\newcommand{\teffects}{N}
\newcommand{\transenvs}{\mathcal{T}_{\textit{env}}}
\newcommand{\exstates}{\mathcal{S}}
\newcommand{\contracts}{\mathcal{C}}
\newcommand{\exenvs}{I}
\newcommand{\mstates}{M}
\newcommand{\actions}{\textit{Act}}

\newcommand{\stackof}[1]{\mathcal{L}(#1)}
\newcommand{\callstacks}{\mathbb{S}}

\newcommand{\annotatedcallstacks}{\callstacks_n}
\newcommand{\contractsbot}{\contracts_\bot}
\newcommand{\untrustedaddresses}{\mathcal{A}_C}
\newcommand{\gstates}{\Sigma}

\newcommand{\project}[2]{{#1}\downarrow_{#2}}
\newcommand{\filtercallscreates}[1]{\textsf{calls}_{#1}}

\newcommand{\sstep}[3]{{#1} \vDash {#2} \, \rightarrow {#3}} 
\newcommand{\ssteps}[3]{{#1} \vDash {#2} \, \rightarrow^* {#3}} 
\newcommand{\ssteptrace}[4]{{#1} \vDash {#2} \, \xrightarrow[]{#4} {#3}} 
\newcommand{\sstepstrace}[4]{{#1} \vDash {#2} \, \xrightarrow[]{#4}^* {#3}} 

\newcommand{\sstepslocalupdatetrace}[5]{{#1} \vdash {#2} \xrightarrow[{#4}]{#5}^* {#3}}

\newcommand{\callstack}{S} 
\newcommand{\transenv}{\Gamma} 
\newcommand{\originator}{\textsf{o}} 
\newcommand{\gasprize}{\textit{prize}}
\newcommand{\blockheader}{H}
\newcommand{\parent}{\textit{parent}}
\newcommand{\beneficiary}{\textit{beneficiary}}
\newcommand{\difficulty}{\textit{difficulty}}
\newcommand{\blocknumber}{\textit{number}}
\newcommand{\gaslimit}{\textit{gaslimit}}
\newcommand{\timestamp}{\textit{timestamp}}
\newcommand{\mstate}{\mu}
\newcommand{\exenv}{\iota}
\newcommand{\gstate}{\sigma}
\newcommand{\regstate}[3]{(#1, #2, #3)}
\newcommand{\regstatefull}[4]{(#1, #2, #3, #4)}
\newcommand{\excstate}{\textit{EXC}}

\newcommand{\haltstatefull}[4]{\textit{HALT}(#1, #2, #3, #4)}

\newcommand{\lgas}{\textit{gas}}
\newcommand{\callstackplain}{S_{\textit{plain}}}
\newcommand{\smstate}[5]{(#1, #2, #3, #4, #5)}
\newcommand{\sexenv}[5]{(#1, #2, \allowbreak #3, #4, #5)}
\newcommand{\accountstate}[4]{(#1, #2, #3, #4)}
\newcommand{\transenvinit}{\transenv_{\textit{init}}}

\newcommand{\gas}{\textsf{gas}}
\newcommand{\data}{\textsf{d}}
\newcommand{\pc}{\textsf{pc}}
\newcommand{\stack}{\textsf{s}}
\newcommand{\actw}{\textsf{i}}
\newcommand{\memo}{\textsf{m}}
\newcommand{\activeaccount}{\textsf{actor}}
\newcommand{\inputdata}{\textsf{input}}
\newcommand{\sender}{\textsf{sender}}
\newcommand{\tvalue}{\textsf{value}}
\newcommand{\activecode}{\textsf{code}}

\newcommand{\balance}{\textsf{b}}

\newcommand{\accountcode}{\textsf{code}}

\newcommand{\inittrans}[3]{\textit{initialize}\,({#1}, {#2}, {#3})}
\newcommand{\finalizetrans}[4]{\textit{finalize} \, ({#1, {#2}, {#3}})}
\newcommand{\transaction}{T}
\newcommand{\transstep}[4]{{#3} \xrightarrow{#1, #2} {#4}}
\newcommand{\transeffects}{\eta}
\newcommand{\exconf}[2]{(#1, #2)}
\newcommand{\emptytranseffects}{\epsilon_\transeffects}
\newcommand{\concatstack}[2]{{#1}++{#2} }

\newcommand{\recipient}{\textit{to}}

\newcommand{\contract}[2]{(#1, #2)}
\newcommand{\contractaddress}{a}
\newcommand{\contractcode}{\textit{code}}
\newcommand{\getcontractaddress}[1]{\textit{address}\,({#1})}
\newcommand{\getcontractcode}[1]{\textit{code} \, ({#1})}
\newcommand{\annotate}[2]{{#1}_{#2}}
\newcommand{\exstate}{s}

\newcommand{\transenvcomponents}{\mathcal{C}_{\transenv}}
\newcommand{\transenvcomponent}{c_{\transenv}}
\newcommand{\finalstate}[1]{\textit{final} \, ({#1})}

\newcommand{\equalupto}[1]{=_{/{#1}}}

\newcommand{\ISZERO}{\textsf{ISZERO}}
\newcommand{\SUICIDE}{\textsf{SELFDESTRUCT}}
\newcommand{\PUSH}[1]{\textsf{PUSH}{#1}}
\newcommand{\ADD}{\textsf{ADD}}
\newcommand{\CALLCODE}{\textsf{CALLCODE}}
\newcommand{\DELEGATECALL}{\textsf{DELEGATECALL}}

\newcommand{\pluseq}{\mathrel{+}=}
\newcommand{\minuseq}{\mathrel{-}=}
\newcommand{\cond}[3]{{#1} \, ? \,  {#2}\, : \,  {#3}}
\newcommand{\mini}[2]{\textit{min} \, (#1, #2)}
\newcommand{\update}[3]{{#1}[{#2} \rightarrow {#3}]}
\newcommand{\getinterval}[3]{#1\, [#2, #3]}
\newcommand{\updateinterval}[4]{{#1}[[{#2, #3}] \rightarrow #4]}
\newcommand{\arraypos}[2]{{#1} \, [#2]}
\newcommand{\getaccount}[2]{#1 (#2)}
\newcommand{\updategstate}[3]{{#1} \big \langle {#2} \rightarrow {#3} \big \rangle}
\newcommand{\optionof}[1]{ {#1} \cup \{ \bot \}}
\newcommand{\concat}[2]{{#1} \cdot {#2}}
\newcommand{\extract}[3]{{#1}[#2, #3]}
\newcommand{\inc}[3]{{#1}[{#2}\pluseq{#3}]}
\newcommand{\dec}[3]{{#1}[{#2}\minuseq{#3}]}

\newcommand{\account}[4]{(#1, #2, #3, #4)}
\newcommand{\accountnonce}{\textsf{nonce}}
\newcommand{\accountbalance}{\textsf{balance}}
\newcommand{\accountstor}{\textsf{stor}}
\newcommand{\refundbalance}{\textsf{balance}}
\newcommand{\suicideset}{\textit{S}_\dagger}

\renewcommand{\exconf}[2]{#1}

\newcommand{\CALL}{\textsf{CALL}}
\newcommand{\STOP}{\textsf{STOP}}
\newcommand{\RETURN}{\textsf{RETURN}}
\newcommand{\CREATE}{\textsf{CREATE}}
\newcommand{\SHA}{\textsf{SHA3}}
\newcommand{\BLOCKHASH}{\textsf{BLOCKHASH}}
\newcommand{\JUMP}{\textsf{JUMP}}
\newcommand{\JUMPI}{\textsf{JUMPI}}
\newcommand{\EXTCODESIZE}{\textsf{EXTCODESIZE}}
\newcommand{\EXTCODECOPY}{\textsf{EXTCODECOPY}}

\newcommand{\curropcode}[2]{\omega_{#1, #2}}

\newcommand{\valu}{\textit{va}}
\newcommand{\io}{\textit{io}}
\newcommand{\is}{\textit{is}}
\newcommand{\oo}{\textit{oo}}
\newcommand{\os}{\textit{os}}
\newcommand{\aw}{\textit{aw}}
\newcommand{\addr}{\textit{addr}}
\newcommand{\freshaddress}{\rho}

\newcommand{\emptymemory}{\lam{x}{0}}
\newcommand{\emptystack}{\epsilon}
\newcommand{\emptyarray}{\epsilon}

\newcommand{\memext}[3]{M\,(#1, #2, #3)} 
\newcommand{\funL}[1]{\textit{L} \, ({#1})}
\newcommand{\basecosts}[2]{C_{\textit{base}} \, (#1, #2)}
\newcommand{\flag}{\textit{flag}}
\newcommand{\gascapacity}[4]{C_{\textit{gascap}} \, (#1, #2, #3, #4)}
\newcommand{\costmem}[2]{C_{\textit{mem}} \, (#1, #2)}

\newcommand{\simvalid}[3]{\textit{valid} \, (#1, #2, #3)}
\newcommand{\getfreshaddress}[2]{\textit{newAddress}\, (#1, #2)}

\newcommand{\costs}{\textit{c}}

\newcommandx{\pcheckpremscallhelp}[6]{\textsf{checkPrems}^{\textsf{\CALL}}_{\tempa} \, ({\tempb}, {\tempc}, {\tempd}, {\tempe}, {\tempf}, {\tempg}, {\temph}, {\tempi}, {#1}, {#2}, {#3}, {#4}, {#5}, {#6})}

\newcommand{\pos}{\textit{pos}}

\newcommand{\lpc}{\textit{pc}}
\newcommand{\code}{\textit{code}}
\newcommand{\stor}{\textit{stor}}
\newcommand{\siz}{\textit{size}}

\newcommand{\arrayinterval}[3]{{#1}[#2, #3]}

\newcommand{\accountnoncev}{\textit{nonce}}
\newcommand{\accountbalancev}{\textit{balance}}
\newcommand{\accountstorv}{\textit{stor}}
\newcommand{\accountcodev}{\textit{code}}
\newcommand{\refundbalancev}{\textit{bal}_r}
\newcommand{\datav}{d}
\newcommand{\logseqv}{L}
\newcommand{\mstatev}{\mu}
\newcommand{\gasv}{\textit{gas}}
\newcommand{\pcv}{\textit{pc}}
\newcommand{\memov}{m}
\newcommand{\stackv}{s}
\newcommand{\actwv}{i} 
\newcommand{\instv}{\omega}
\newcommand{\downv}{\textit{down}}
\newcommand{\upv}{\textit{up}}

\newcommand{\rlpencode}[1]{\textit{rlp}\,(#1)}

\newcommand{\bitstringtobytearray}[1]{{#1}_{\bytearray}}
\newcommand{\bytearraytobitstring}[1]{{#1}_\BB}

\newcommand{\ZZ}{\mathbb{Z}}

\newcommand{\letlet}{\textit{let}~}
\newcommand{\letin}{~\textit{in}~}

\newcommand{\gprice}{\textsf{price}}

\newcommand{\sinteger}[1]{\textit{Int}_{#1}}
\newcommand{\callstacksplain}{\callstacks_{\textit{plain}}}
\newcommand{\accountv}{\textit{acc}}

\newcommand{\binopv}{\textit{i}_\textit{bin}}
\newcommand{\binops}{\textit{Inst}_{\textit{bin}}}
\newcommand{\binopcost}[1]{\text{cost}_\textit{bin}(#1)}
\newcommand{\binopfun}[1]{\text{fun}_\textit{bin}(#1)}
\newcommand{\signed}[1]{{#1}^-}
\newcommand{\unsigned}[1]{{#1}^+}
\newcommand{\signof}[1]{\textit{sign}{(#1)}}
\newcommand{\bitand}{\&} 
\newcommand{\bitor}{\|}
\newcommand{\bitxor}{\oplus}
\newcommand{\bitneg}{\neg}
\newcommand{\keccak}[1]{\textit{Keccak}(#1)}

\newcommand{\funP}[3]{P\,(#1, #2, #3)}
\newcommand{\funD}[1]{D\,(#1)}
\newcommand{\funN}[2]{N\,(#1, #2)}
\newcommand{\funDhelp}[2]{D_H\,(#1, #2)}
\newcommand{\maxi}[2]{\textit{max}(#1, #2)}

\newcommand{\parentc}{\textsf{parent}}
\newcommand{\beneficiaryc}{\textsf{beneficiary}}
\newcommand{\difficultyc}{\textsf{difficulty}}  
\newcommand{\blocknumberc}{\textsf{number}}
\newcommand{\gaslimitc}{\textsf{gaslimit}}
\newcommand{\timestampc}{\textsf{timestamp}}
\newcommand{\suicidesetc}{\textsf{S}_{\dagger}}
\newcommand{\logseq}{\textsf{L}}

\newcommand{\SUB}{\textsf{SUB}}
\newcommand{\LT}{\textsf{LT}}
\newcommand{\GT}{\textsf{GT}}
\newcommand{\EQ}{\textsf{EQ}}
\newcommand{\AND}{\textsf{AND}}
\newcommand{\OR}{\textsf{OR}}
\newcommand{\XOR}{\textsf{XOR}}
\newcommand{\SLT}{\textsf{SLT}}
\newcommand{\SGT}{\textsf{SGT}}
\newcommand{\MUL}{\textsf{MUL}}
\newcommand{\DIV}{\textsf{DIV}}
\newcommand{\SDIV}{\textsf{SDIV}}
\newcommand{\MOD}{\textsf{MOD}}
\newcommand{\SMOD}{\textsf{SMOD}}
\newcommand{\SIGNEXTEND}{\textsf{SIGNEXTEND}}
\newcommand{\BYTE}{\textsf{BYTE}}
\newcommand{\ADDMOD}{\textsf{ADDMOD}}
\newcommand{\MULMOD}{\textsf{MULMOD}}
\newcommand{\NOT}{\textsf{NOT}}
\newcommand{\EXP}{\textsf{EXP}}
\newcommand{\ADDRESS}{\textsf{ADDRESS}}
\newcommand{\CALLER}{\textsf{CALLER}}
\newcommand{\CALLVALUE}{\textsf{CALLVALUE}}
\newcommand{\CODESIZE}{\textsf{CODESIZE}}
\newcommand{\BALANCE}{\textsf{BALANCE}}
\newcommand{\ORIGIN}{\textsf{ORIGIN}}
\newcommand{\CALLDATASIZE}{\textsf{CALLDATASIZE}}
\newcommand{\CALLDATALOAD}{\textsf{CALLDATALOAD}}
\newcommand{\CODECOPY}{\textsf{CODECOPY}}
\newcommand{\CALLDATACOPY}{\textsf{CALLDATACOPY}}
\newcommand{\GASPRICE}{\textsf{GASPRICE}}
\newcommand{\COINBASE}{\textsf{COINBASE}}
\newcommand{\TIMESTAMP}{\textsf{TIMESTAMP}}
\newcommand{\NUMBER}{\textsf{NUMBER}}
\newcommand{\DIFFICULTY}{\textsf{DIFFICULTY}}
\newcommand{\GASLIMIT}{\textsf{GASLIMIT}}
\newcommand{\POP}{\textsf{POP}}
\newcommand{\MLOAD}{\textsf{MLOAD}}
\newcommand{\MSTORE}{\textsf{MSTORE}}
\newcommand{\MSTOREByte}{\textsf{MSTORE8}}
\newcommand{\SLOAD}{\textsf{SLOAD}}
\newcommand{\SSTORE}{\textsf{SSTORE}}
\newcommand{\JUMPDEST}{\textsf{JUMPDEST}}
\newcommand{\PC}{\textsf{PC}}
\newcommand{\MSIZE}{\textsf{MSIZE}}
\newcommand{\GAS}{\textsf{GAS}}
\newcommand{\DUP}[1]{\textsf{DUP}{#1}}
\newcommand{\SWAP}[1]{\textsf{SWAP}{#1}}
\newcommand{\LOG}[1]{\textsf{LOG}{#1}}
\newcommand{\INVALID}{\textsf{INVALID}}

\begin{document}
\title{A  Semantic Framework for the Security Analysis of Ethereum smart contracts}
\author{Ilya Grishchenko \and Matteo Maffei \and Clara Schneidewind}
\institute{TU Wien
\\
\email{\{ilya.grishchenko,matteo.maffei,clara.schneidewind\}@tuwien.ac.at}
}
\maketitle

\begin{abstract}
Smart contracts are programs running on cryptocurrency (e.g.,  Ethe-reum) blockchains, whose popularity stem from the possibility to perform financial transactions, such as payments and auctions, in a distributed environment without need for any trusted third party. Given their financial nature, bugs or vulnerabilities in these programs may lead to catastrophic consequences, as witnessed by recent attacks. Unfortunately, programming smart contracts is a delicate task that requires strong expertise: Ethereum smart contracts are written in Solidity, a dedicated language resembling JavaScript, and shipped over the blockchain in the EVM bytecode format. In order to rigorously verify the security of smart contracts, it is of paramount importance to formalize their semantics as well as  the  security properties of interest, in particular at the level of the bytecode being executed.

In this paper, we present the first complete small-step semantics of EVM bytecode, which we formalize in the F* proof assistant, obtaining executable code that we successfully validate against the official Ethereum test suite. Furthermore, we formally define   for the first time a number of central security properties for smart contracts, such as call integrity, atomicity, and independence from miner controlled parameters.  This formalization relies on a combination of hyper- and safety properties. Along this work, we identified various mistakes and imprecisions in existing semantics and verification tools for Ethereum smart contracts, thereby demonstrating once more the importance of rigorous semantic foundations  for the design of security verification techniques. 
\end{abstract}

\section{Introduction}

One of the determining factors for the growing interest in blockchain technologies is the groundbreaking promise  of secure distributed computations even in absence of trusted third parties. Building on a distributed ledger that keeps track of previous transactions and the state of each account, whose functionality and security is ensured by a delicate combination of incentives and cryptography, software developers can implement sophisticated distributed, transactions-based computations by leveraging the scripting language offered by the underlying cryptocurrency. While many of these cryptocurrencies have an intentionally limited scripting language (e.g., Bitcoin~\cite{nakamoto2008bitcoin}), Ethereum was designed from the ground up with a quasi Turing-complete language\footnote{While the language itself is Turing complete, computations are associated with a bounded computational budget (called gas), which gets consumed by each instruction thereby enforcing termination.}. Ethereum programs, called \emph{smart contracts}, have thus found  a variety of appealing use cases, such as financial contracts~\cite{biryukov2017findel}, auctions~\cite{hahn2017smart}, elections~\cite{McCorrySH17}, data management systems~\cite{adhikari2017secure}, trading platforms~\cite{notheisen2017trading,mathieu2017blocktix}, permission management \cite{azaria2016medrec} and verifiable cloud computing~\cite{DWAMM::17}, just to mention a few. 
  Given their financial nature, bugs and vulnerabilities in smart contracts may lead to catastrophic consequences. For instance, the infamous DAO vulnerability~\cite{thedao} recently led to a 60M\$ financial loss and similar vulnerabilities occur  on a regular basis~\cite{paritya,parityb}. Furthermore, many smart contracts in the wild are  intentionally fraudulent, as highlighted in a recent survey~\cite{survey}. 
 
 A rigorous security analysis of smart contracts is thus crucial for the trust of the society in blockchain technologies and their widespread deployment. Unfortunately, this task is a quite challenging  for various reasons. First, Ethereum smart contracts are developed in an ad-hoc language,  called Solidity, which resembles JavaScript but features specific transaction-oriented mechanisms and a number of non-standard semantic behaviours, as further described in this paper. Second, smart contracts are uploaded on the blockchain in the form of Ethereum Virtual Machine (EVM) bytecode, a stack-based low-level code featuring dynamic code creation and invocation and, in general, very little static information, which  makes it extremely difficult to analyze. 

\myparagraph{Related Work} Recognizing the importance of solid semantic foundations for smart contracts, the Ethereum foundation published a yellow paper~\cite{yellowpaper} to describe the intended behaviour of smart contracts. This semantics, however, exhibits several under-specifications and does not follow any standard approach for the specification of program semantics, thereby hindering program verification. In order to provide a more precise characterization,  Hirai formalizes the EVM semantics in the proof assistant Isabelle/HOL and uses it for manually proving safety properties for concrete programs~\cite{eth-isabelle}.  This  semantics, however, constitutes just a sound over-approximation of the original semantics \cite{yellowpaper}. More specifically,  once a contract performs a call that is not a self-call, it is assumed that arbitrary code gets executed and consequently arbitrary changes to the account's state and to the global state can be performed. Consequently, this semantics can not serve as a general-purpose basis for static analysis techniques that might not rely on the same over-approximation. 

In a concurrent, unpublished work, Hildebrandt et al.~\cite{kevm} define the EVM semantics in the $\mathbb{K}$ framework~\cite{stefuanescu2016semantics} -- a language independent verification framework based on reachability logics. The authors leverage the power of the $\mathbb{K}$ framework in order to automatically derive analysis tools for the specified semantics,  presenting as an  example a gas analysis tool, a semantic debugger, and a program verifier based on reachability logics. The underlying semantics relies on  non-standard  local rewriting rules on the system configuration. Since  parts of the execution are treated in separation such as the exception behavior and the gas calculations, one small-step  consists of several rewriting steps, which makes this semantics harder to use as a basis for new static analysis techniques. This is relevant whenever the static analysis tools derivable by the $\mathbb{K}$ framework are not sufficient for the desired purposes: for instance, their analysis  requires the user to manually specify loop invariants, which is hardly doable for EVM bytecode and clearly does not scale to large programs. Furthermore, all these works concentrate on the semantics of EVM bytecode but do not study security properties for smart contracts. 

Sergey et al.~\cite{sergey2017concurrent} compare smart contracts on the blockchain with concurrent objects using shared memory and use this analogy to explain typical problems that arise when programming smart contracts in terms of concepts known from concurrency theory. They encourage the application of state-of-the art verification techniques for concurrent programs to smart contracts, but do not describe any specific analysis method applied to smart contracts themselves.  
Mavridou et al.~\cite{mavridoudesigning} define a high-level semantics for smart contracts that is based on finite state machines and aims at simplifying the development of smart contracts. They provide a translation of their state machine specification language to Solidity, a higher-order language for writing Ethereum smart contracts, and present design patterns that should help users to improve the security of their contracts. The translation to Solidity is not backed up by a correctness proof and the design patterns are not claimed to provide any security guarantees. 

Bhargavan et al.~\cite{Bhargavan:2016} introduce a framework  to analyze  Ethereum contracts by translation into F*, a functional programming language aimed at program verification and equipped with an interactive proof assistant. The translation supports only a fragment of the EVM bytecode and does not come with a justifying semantic argument. 

Luu et al. have recently presented Oyente \cite{oyente}, a state-of-the-art static analysis tool for EVM bytecode that relies on  symbolic execution. Oyente comes with a semantics of a simplified fragment of the  EVM bytecode and, in particular, misses several important commands related to contract calls and contract creation. Furthermore, it is affected by a major bug related to calls as well as several other minor ones which we discovered while formalizing our semantics, which is inspired by theirs.   Oyente supports a variety of security properties, such as  transaction order dependency, timestamp dependency, \TODOM{somewhere, probably in otherapproaches, we have to mention how these properties are captured by our definitions} and reentrancy,  but the security definitions are rather syntactic and described informally. As we show in this paper, the lack of solid semantic foundations causes several sources of unsoundness in Oyente.

\myparagraph{Our Contributions} This work lays the semantic foundations for Ethereum smart contracts. Specifically, we introduce

\begin{itemize}
\item The first complete small-step semantics for EVM bytecode;	
\item A formalization in F* of a large fragment of our semantics, which can serve as a foundation for verification techniques based on encoding into this language~\cite{Bhargavan:2016} as well as  machine-checked proofs for other analysis techniques (e.g., \cite{oyente}). By compiling F* in OCaml, we   could successfully validate  our semantics against the official Ethereum test suite;
\item  The first formal definitions of crucial security properties for smart contracts, such as call integrity, for which we devise a dedicated proof technique, atomicity, and  independence from miner controlled parameters.  Interestingly enough, the formalization of these properties requires hyper-properties, while existing static analysis techniques for smart contracts rely on  reachability properties and syntactic conditions; 
\item A collection of examples showing how the syntactic conditions employed in current analysis techniques are imprecise and, in several cases, unsound, thereby further motivating the need for solid semantic foundations and rigorous security definitions for smart contracts.
\end{itemize}

\noindent
The complete semantics as well as the formalization in F* are publicly available~\cite{fullversion}. 

\myparagraph{Outline} The remainder of this paper is organized as follows. \autoref{sec:ethereum} briefly overviews the Ethereum architecture,
\autoref{sec:semantics} introduces the Ethereum semantics and our formalization in F*,
\autoref{sec:definitions} formally defines various security properties for Ethereum smart contracts, 
and
\autoref{sec:conclusion} concludes highlighting interesting research directions.

\section{Background on  Ethereum}
\label{sec:ethereum}


\subsubsection{Ethereum}
Ethereum is a cryptographic currency system built on  top of a blockchain. 
Similar to Bitcoin, network participants publish transactions to the network that are then grouped into blocks by distinct nodes (the so called \emph{miners}) and appended to the blockchain using a proof of work (PoW) consensus mechanism. 
The state of the system -- that we will also refer to as \emph{global state} -- consists of the state of the different accounts populating it. An account can either be an external account (belonging to a user of the system) that carries information on its current balance or it can be a contract account that additionally obtains persistent storage and the contract's code. 
The account's balances are given in the subunit \emph{wei} of the virtual currency \emph{Ether}.\footnote{One Ether is equivalent to $10^{18}$ wei.} 

Transactions can alter the state of the system by either creating new contract accounts or by calling an existing account. Calls to external accounts can only transfer Ether to this account, but calls to contract accounts additionally execute the code associated to the contract. The contract execution might alter the storage of the account or might again perform transactions -- in this case we talk about \emph{internal transactions}. 

The execution model underlying the execution of contract code is described by a virtual state machine, the \emph{Ethereum Virtual Machine} (EVM). This is \emph{quasi Turing complete} as the otherwise Turing complete execution is restricted by the upfront defined resource \emph{gas} that effectively limits the number of execution steps.  
The originator of the transaction can specify the maximal gas that should be spent for the contract execution and also determines the gas prize (the amount of wei to pay for a unit of gas). Upfront, the originator pays for the gas limit according to the gas prize and in case of successful contract execution that did not spend the whole amount of gas dedicated to it, the originator gets reimbursed with gas that is left. The remaining wei paid for the used gas are given as a fee to a beneficiary address specified by the miner. 

\subsubsection{EVM Bytecode}
The code of contracts is written in \emph{EVM byte\REMOVEMfor{CY}{171125}{ }code} -- \TODOM{I would write bytecode...please make it consistent throughout the paper} an Assembler like bytecode language. As the core of the EVM is a stack\NEWMfor{CY}{171125}{-}based machine, the set of instructions in EVM bytecode consists mainly of standard instructions for stack operations, arithmetics, jumps and local memory access. The classical set of instructions is enriched with an opcode for the SHA3 hash and several opcodes for accessing the environment that the contract was called in.
In addition, there are opcodes for accessing and modifying the storage of the account currently running the code and distinct opcodes for performing internal call and create transactions. 
Another instruction particular to the block\REMOVEMfor{CY}{171125}{ }chain setting is the $\SUICIDE$ code that deletes the currently executed contract - but only after the successful execution of the external transaction. 

\paragraph{Gas and Exceptions}
The execution of  each instruction consumes a positive amount of gas. There is a gas limit set by the sender of the transaction. 
Exceeding the gas limit results in an exception that reverts the effects of the current transaction on the global state. 
In the case of nested transactions, the occurrence of an exception only reverts its own effects, but not those of the calling transaction. Instead, the failure of an internal transaction is only indicated by writing zero to the caller's stack. 

\subsubsection{Solidity}
\label{solidity}
In practice, most Ethereum smart contracts are not written in EVM bytecode directly, but in the high-level language Solidity which is developed by the Ethereum Foundation~\cite{solidity}. For understanding the typical problems that arise when writing smart contracts, it is important to consider the design of this high-level language.

Solidity is a so called ``contract-oriented" programming language that uses the concept of class from object-oriented languages for the representation of contracts. 
Similar to classes in object-oriented programming, contracts specify fields and methods for contract instances. Fields can be seen as persistent storage of a contract (instance) and contract methods can by default be invoked by any internal or external transaction. 
For interacting with another contract one either needs to create a new instance of this contract (in which case a new contract account with the functionality described in the contract class is created) or one can directly make transactions to a known contract address holding a contract of the required shape.
The syntax of Solidity resembles Java\REMOVEMfor{CY}{171125}{ }Script, enriched with additional primitives  accounting for the distributed setting of Ethereum. 
In particular, Solidity provides primitives for accessing the transaction and the block information, like \lstinline|msg.sender| for accessing the address of the account invoking the method or \lstinline|msg.value| for accessing the amount of $\wei$ transferred by the transaction that invoked the method. 

Solidity shows some particularities when it comes to transferring money to another contract especially using the provided low level functions \lstinline|send| and \lstinline|call|. 
A value transfer initiated using these functions is finally translated to an internal call transaction which implies that calling a contract might also execute code and in particular it can fail because the available gas is not sufficient for executing the code. In addition -- as in the EVM -- these kinds of calls do not enable exception propagation, so that the caller manually needs to checks for the return result. 
Another special feature of Solidity is that it allows for defining so called \emph{fallback functions} for contracts that get executed when a call via the \lstinline|send| function was performed or (using the \lstinline|call| function) an address is called that however does not properly specifies the concrete function of the contract to be called.

\section{Small-Step Semantics}
\label{sec:semantics}
We introduce a small-step semantics covering the full EVM bytecode, inspired by  the one presented by Luu et al. \cite{oyente}, which we substantially revise  in order to handle the missing instructions, in particular  contract calls and  call creation. In addition, while formalizing our semantics, we  found a major flaw related to calls and several minor ones (cf.  \autoref{sec:comparison-oyente}), which we fixed and reported to the authors.  Due to space constraints, we refer the interested reader to
\iffull Appendix~\ref{sec:formalization} and Appendix~\ref{sec:arules}
\else
the full version of the paper~\cite{fullversion}
\fi
for a formal account of the semantic rules and present below the most significant ones.

\subsection{Preliminaries}
In the following, we will use $\BB$ to denote the set $\{0,1\}$ of bits and accordingly $\BB^{x}$ for sets of bitstrings of size $x$. We further  let $\integer{x}$ denote the set of non-negative integers representable by $x$ bits and allow for implicit conversion between those two representations. In addition, we will use the notation $\arrayof{X}$ (resp. $\stackof{X}$)  for arrays (resp. lists) of  elements from the set $X$. We use standard notations for operations on arrays and lists. 

\subsection{Global state}
As mentioned before, the global state is a (partial) mapping from account addresses (that are bitstrings of size 160) to accounts. In the case that an account does not exist, we assume it to map to $\bot$.
Accounts, irrespectively of their type, are tuples of the form $\accountstate{n}{b}{\stor}{\code}$, with $n \in \integer{256}$ being the account's nonce that is incremented with every other account that the account creates, $b \in \integer{256}$ being the account's balance in $\wei$, $\stor \in \storagetype$ being the accounts persistent storage that is represented as a mapping from 256-bit words to 256-bit words and finally $\code \in \bytearray$ being the contract  that is an array of bytes. 
In contrast to contract accounts, external accounts have the empty bytearray as code. As only the execution of code in the context of the account can access and modify the account's storage, the fact that formally external accounts have persistent storage does not have any effect. 
In the following, we will denote the set of addresses with $\addresses$ and the set of global states with $\gstates$ and we will assume that $\gstate \in \gstates$.  

\subsection{Small-Step Relation}
In order to define the small-step semantics, we give a small-step relation $\sstep{\transenv}{\exconf{\callstack}{\transeffects}}{\exconf{\callstack'}{\transeffects'}}$ that specifies how a call stack $\callstack \in \callstacks$ representing the state of the execution evolves within one step under the transaction environment $\transenv \in \transenvs$. 

In Figure~\ref{fig:grammar} we give a full grammar for call stacks and transaction environments: 
\begin{figure*}[h]
\begin{mathpar}
\begin{array}{rlclll}
\text{Call stacks} & \callstacks & \ni & \callstack & \define & \cons{\excstate}{\callstackplain} ~|~  \cons{\haltstatefull{\gstate}{d}{g}{\transeffects}}{\callstackplain} ~|~ \callstackplain  \\
\text{Plain call stacks} & 
\callstacksplain & \ni & \callstackplain & \define & \cons{\regstatefull{\mstate}{\exenv}{\gstate}{\transeffects}}{\callstackplain} \\
\text{Machine states} & \mstates & \ni & \mstate & \define & \smstate{\lgas}{\lpc}{m}{i}{s} \\
\text{Execution environments} & \exenvs & \ni & \exenv & \define & \sexenv{\textit{actor}}{\textit{input}}{\textit{sender}}{\textit{value}}{\textit{code}}\\
\text{Global states} & \gstates  & \ni &\gstate & &  \\
\text{Account states} & \accounts & \ni & \accountv & \define & \accountstate{n}{b}{\textit{code}}{\textit{stor}} ~|~ \bot   \\ 
\text{Transaction effects} & \teffects  & \ni &\transeffects & \define & (b, L, \suicideset)  \\
\text{Transaction environments} & \transenvs & \ni & \transenv & \define & (o, \textit{prize}, H) \\
\\
\text{Notations:} 
&\multicolumn{5}{c}{
d \in \bytearray, \quad g \in \integer{256}, \quad \transeffects \in \teffects, \quad 
o \in \addresses, \quad \textit{prize} \in \integer{256}, \quad H \in \blockheaders} \\
&\multicolumn{5}{c}{
\lgas \in \integer{256}, \quad \lpc \in \integer{256}, \quad m \in \BB^{256}, \to \BB^8 \quad i \in \integer{256}, \quad s \in \stackof{\BB^{256}}}  \\
&\multicolumn{5}{c}{
\textit{sender} \in \addresses \quad \textit{input} \in \bytearray \quad \textit{sender} \in \addresses \quad \textit{value} \in \integer{256} \quad \textit{code} \in \bytearray} \\
&\multicolumn{5}{c}{
b \in \integer{256} \quad L \in \sequenceof{\logevents} \quad \suicideset \subseteq \addresses \quad 
\gstates = \addresses \to \accounts
}
\end{array}
\end{mathpar}
\caption{Grammar for call stacks and transaction environments}
\label{fig:grammar}
\end{figure*}

\subsubsection{Transaction Environments}
The transaction environment represents the static information of the block that the transaction is executed in and the immutable parameters given to the transaction as the gas prize or the gas limit. 
More specifically, the transaction environment $\transenv \in \transenvs = \addresses \times \integer{256} \times \blockheaders$ is a tuple of the form $(o, \gasprize, \blockheader)$ with $o \in \addresses$ being the address of the account that made the transaction, $\gasprize \in \integer{256}$ denoting amount of wei that needs to paid for a unit of gas in this transaction and $\blockheader \in \blockheaders$ being the header of the block that the transaction is part of. We do not specify the format of block headers here, but just assume a set $\blockheaders$ of block headers. 

\subsubsection{Callstacks}
A call stack $\callstack$ is a stack of execution states which represents the state of the execution within one internal transaction. We give a formal definition of the set of possible callstacks $\callstacks$ as follows:
\begin{align*}
\callstacks \define \{& \cons{\excstate}{\callstackplain}, ~\cons{\haltstatefull{\gstate}{\lgas}{d}{\transeffects}}{\callstackplain}, ~\callstackplain \\ ~|~ 
& \gstate \in \gstates, ~\lgas \in \NN, ~d \in \bytearray,~  \transeffects \in \teffects ~,\callstackplain \in \stackof{\mstates \times \exenvs \times \gstates \times \teffects} \}
\end{align*}
Syntactically, a call stack is a stack  of regular execution states of the form $\regstatefull{\mstate}{\exenv}{\gstate}{\transeffects}$ that can optionally be topped with a halting state $\haltstatefull{\gstate}{\lgas}{d}{\transeffects}$ or an exception state $\excstate$. We summarize these three types of states as execution states $\exstates$. 
Semantically, halting states indicate regular halting of an internal transaction, exception states indicate exceptional halting, and regular execution states describe the state of internal transactions in progress. 
Halting and exception states can only occur as top elements of the call stack as they represent terminated internal transactions. Exception states of the form $\excstate$ do not carry any information as in the case of an exception all effects of the terminated internal transaction are reverted and the caller state therefore stays  unaffected, except for the gas.  
Halting states instead are of the form $\haltstatefull{\gstate}{\lgas}{d}{\transeffects}$ specifying the global state $\gstate$ the execution halted in, the gas $\lgas \in \integer{256}$ remaining from the execution, the return data $d \in \bytearray$ and the additional transaction effects $\transeffects \in \teffects$ of the internal transaction. The additional transaction effects carry information that are accumulated during execution, but do not influence the small-step execution itself. Formally, the additional transaction effects are a triple of the form $(b, L, \suicideset) \in \teffects = \integer{256} \times \sequenceof{\logevents} \times \setof{\addresses}$ with $b \in \integer{256}$ being the refund balance that is increased by account storage operations and will finally be paid to the transaction's beneficiary, $L \in \sequenceof{\logevents}$ being the sequence of log events that the bytecode execution invoked during execution and $\suicideset \subseteq \addresses$ being the so called suicide set -- the set of account addresses that executed the $\SUICIDE$ command and therefore registered their account for deletion. 
The information held by the halting state is  carried over to the calling state. 
 
The state of a non-terminated internal transaction is described by a regular execution state  of the form $\regstatefull{\mstate}{\exenv}{\gstate}{\transeffects}$. The state is determined by the current global state $\gstate$ of the system as well as the execution environment $\exenv \in \exenvs$ that specifies the parameters of the current transaction (including inputs and the code to be executed), the local state $\mstate \in \mstates$ of the stack machine, and the transaction effects $\transeffects \in \teffects$ collected during execution so far. 

\subsubsection{Execution Environment}
The execution environment $\exenv$ of an internal transaction specifies the static parameters of the transaction. It is a  tuple of the form $\sexenv{\textit{actor}}{\textit{input}}{\textit{sender}}{\textit{value}}{\textit{code}} \in \exenvs = \addresses \times \bytearray \times \addresses \times \integer{256} \times \bytearray$ with the following components: 
\begin{itemize}
\item $\textit{actor} \in \addresses$ is the address of the account currently executing;
\item $\textit{input} \in \bytearray$ is the data given as an input to the internal transaction;
\item $\textit{sender} \in \addresses$ is the address of the account that initiated the internal transaction;
\item $\textit{value} \in \integer{256}$ is the value transferred by the internal transaction;
\item $\textit{code} \in \bytearray$ is the code currently executed.
\end{itemize}
This information is determined at the beginning of an internal transaction execution and it can be accessed, but not altered during the execution.

\subsubsection{Machine State}
The local machine state $\mstate$ represents the state of the underlying state machine used for execution and is a tuple of the form $\smstate{\lgas}{\textit{pc}}{\textit{m}}{\textit{i}}{\textit{s}}$ where 
\begin{itemize}
\item $\lgas \in \integer{256}$ is the current amount of gas still available for execution;
\item $\textit{pc}\in \integer{256}$ is the current program counter;
\item $\textit{m} \in \memorytype$ is a mapping from 256-bit words to bytes that represents the local memory;
\item $\textit{i} \in \integer{256}$ is the current number of active words in memory;
\item $\textit{s} \in \stackof{\word}$ is the local 256-bit word stack of the stack machine.
\end{itemize}
The execution of each internal transaction starts in a fresh machine state, with an empty stack, memory initialized to all zeros, and program counter and active words in memory set to zero. Only the gas is instantiated with the gas value available for the execution.   

\subsection{Small-Step Rules}
In the following, we will present a selection of interesting small-step rules in order to illustrate the most important features of the semantics. 

For demonstrating the overall design of the semantics, we start with the example of the arithmetic expression $\ADD$ performing addition of two values on the machine stack. Note that as the word size of the stack machine is $256$, all arithmetic operations are performed modulo $2^{256}$. 
\begin{mathpar}
\small\infer{
\arraypos{\access{\exenv}{\code}}{\access{\mstate}{\pc}}= \ADD \\
\access{\mstate}{\stack} = \cons{a}{\cons{b}{s}} \\
\access{\mstate}{\gas} \geq 3 \\
\mstate'= \dec{\inc{\update{\mstate}{\stack}{\cons{(a+b)}{s}}}{\pc}{1}}{\gas}{3}}
{\sstep{\transenv}{\cons{\regstatefull{\mstate}{\exenv}{\gstate}{\transeffects}}{\callstack}}{\cons{\regstatefull{\mstate'}{\exenv}{\gstate}{\transeffects}}{\callstack}}}

\infer{
\arraypos{\access{\exenv}{\code}}{\access{\mstate}{\pc}}= \ADD \\
(\size{\access{\mstate}{\stack}} < 2 \lor \access{\mstate}{\gas} < 3)}
{\sstep{\transenv}{\cons{\regstatefull{\mstate}{\exenv}{\gstate}{\transeffects}}{\callstack}}{\cons{\excstate}{\callstack}}}
\end{mathpar}
We use a dot notation, in order to access components of the different state parameters. We name the components with the variable names introduced for these components in the last section written in sans-serif-style. 
In addition, we use the usual notation for updating components: $\update{\textit{t}}{\textsf{c}}{v}$ denotes that the component $\textsf{c}$ of tuple $\textit{t}$ is updated with value $v$. For expressing incremental updates in a simpler way, we additionally use the notation $\inc{t}{\textsf{c}}{v}$ to denote that the (numerical) component of $\textsf{c}$ is incremented by $v$ and similarly $\dec{t}{\textsf{c}}{v}$ for decrementing a component $\textsf{c}$ of $t$. 

The execution of the arithmetic instruction $\ADD$ only performs local changes in the machine state affecting the local stack, the program counter, and the gas budget. For deciding upon the correct instruction to execute, the currently executed code (that is part of the execution environment) is accessed at the position of the current program counter. The cost of an $\ADD$ instruction is constantly three units of gas that get subtracted from the gas budget in the machine state. 
As every other instruction, $\ADD$ can fail due to lacking gas or due to underflows on the machine stack. In this case, the exception state is entered and the execution of the current internal transaction is terminated. 
For better readability, we use here the slightly sloppy $\lor$ notation for combining the two error cases in one inference rule. 

A more interesting example of a semantic rule is the one of the $\CALL$ instruction that initiates an internal call transaction. In the case of calling, several corner cases need to be treated which results in several inference rules for this case. Here, we only present one rule for illustrating the main functionality. More precisely, we present the case in that the account that should be called exists, the call stack limit of $1024$ is not reached yet, and the account initiating the transaction has a sufficiently large balance for sending the specified amount of wei to the called account. 
\begin{mathpar}\small
\infer{
\arraypos{\access{\exenv}{\code}}{\access{\mstate}{\pc}}=\CALL  \\
\access{\mstate}{\stack}=\cons{g}{\cons{\recipient}{\cons{\valu}{\cons{\io}{\cons{\is}{\cons{\oo}{\cons{\os}{s}}}}}}}   \\
\gstate(\recipient) \neq \none\\ 
\size{A} + 1 < 1024 \\
\access{\getaccount{\gstate}{\access{\exenv}{\activeaccount}}}{\balance} \geq \valu \\
\aw = \memext{\memext{\access{\mstate}{\actw}}{\io}{\is}}{\oo}{\os} \\ 
c_{\textit{call}} = \gascapacity{\valu}{1}{g}{\access{\mstate}{\gas}} \\
c = \basecosts{\valu}{1} + \costmem{\access{\mstate}{\actw}}{\aw} +c_{\textit{call}} \\
\access{\mstate}{\gas} \geq c \\
\gstate' = \updategstate{\updategstate{\gstate}{\recipient}{\inc{\getaccount{\gstate}{\recipient}}{\balance}{\valu}}}{\access{\exenv}{\activeaccount}}{\dec{\getaccount{\gstate}{\access{\exenv}{\activeaccount}}}{\balance}{\valu}} \\
d =\getinterval{\access{\mstate}{\memo}}{\io}{\io + \is -1}  \\ 
\mstate' =\smstate{c_{\textit{call}}}{0}{\emptymemory}{0}{\emptystack}  \\ 
\exenv' =\update{\update{\update{\update{\update{\exenv}{\sender}{\access{\exenv}{\activeaccount}}}{\activeaccount}{\recipient}}{\tvalue}{\valu}}{\inputdata}{d}}{\activecode}{\access{\getaccount{\gstate}{\recipient}}{\accountcode}}\\
} 
{\sstep{\transenv}{\cons{\regstatefull{\mstate}{\exenv}{\gstate}{\transeffects}}{\callstack}}{\cons{\regstatefull{\mstate'}{\exenv'}{\gstate'}{\transeffects}}{\cons{\regstatefull{\mstate}{\exenv}{\gstate}{\transeffects}}{\callstack}}}}
\end{mathpar}
For performing a call, the parameters to this call need to be specified on the machine stack. These are the amount of gas $g$ that should be given as budget to the call, the recipient $\recipient$ of the call and the amount $\valu$ of wei to be transferred with the call. In addition, the caller needs to specify the input data that should be given to the transaction and the place in memory where the return data of the call should be written after successful execution. To this end, the remaining arguments specify the offset and size of the memory fragment that input data should be read from (determined by $\io$ and $\is$) and return data should be written to (determined by $\oo$ and $\os$).

Calculating the cost in terms of gas for the execution is quite complicated in the case of $\CALL$ as it is influenced by several factors including the arguments given to the call and the current machine state. First of all, the gas that should be given to the call (here denoted by $c_\textit{call}$) needs to be determined. This value is not necessarily equal to the value $g$ specified on the stack, but also depends on the value $\valu$ transferred by the call and the currently available gas.
In addition, as the memory needs to be accessed for reading the input value and writing the return value, the number of active words in memory might be increased. This effect is captured by the memory extension function $M$. As accessing additional words in memory costs gas, this cost needs to be taken into account in the overall cost. The costs resulting from an increase in the number of active words is calculated by the function $C_\textit{mem}$. 
Finally, there is also a base cost charged for the call that depends on the value $\valu$. 
As the cost also depends on the specific case for calling that is considered, the cost calculation functions receive a flag (here $1$) as arguments. These technical details are spelled out in
\iffull
Appendix~\ref{sec:arules}.
\else
the full version~\cite{fullversion}.
\fi 

The call itself then has several effects: 
First, it transfers the balance from the executing state ($\textit{actor}$ in the execution environment) to the recipient ($\recipient$). To this end, the global state is updated. Here we use a special notation for the functional update on the global state using $\langle \rangle$ instead of $[]$. 
Second, for initializing the execution of the initiated internal transaction, a new regular execution state is placed on top of the execution stack. The internal transaction starts in a fresh machine state at program counter zero. This means that the initial memory is initialized to all zeros and consequently the number of active words in memory is zero as well and additionally the initial stack is empty. The gas budget given to the internal transaction is $c_\textit{call}$ calculated before. The transaction environment of the new call records the call parameters. This includes the sender that is the currently executing account $\textit{actor}$, the new active account that is now the called account $\recipient$ as well as the value $\valu$ sent and the input data given to the call. To this end the input data is extracted from the memory using the offset $\io$ and the size $\is$. We use an interval notation here to denote that a part of the memory is extracted. Finally, the code in the execution environment of the new internal transaction is the code of the called account. 

Note that the execution state of the caller stays completely unaffected at this stage of the execution. This is a conscious design decision in order to simplify the expression of security properties and to make the semantics more suitable to abstractions. 

Besides $\CALL$ there are two different instructions for initiating internal call transactions that implement slight variations of the simple $\CALL$ instruction. These variations are called $\CALLCODE$ and $\DELEGATECALL$, which both allow for executing another's account code in the context of the caller. The difference is that in the case of $\CALLCODE$ a new internal transaction is started and the currently executed account is registered as the sender of this transaction while in the case of $\DELEGATECALL$ an existing call is really forwarded in the sense that the sender and the value of the initiating transaction are propagated to the new internal transaction. 

Analogously to the instructions for initiating internal call transactions, there is also one instruction $\CREATE$ that allows for the creation of a new account. The semantics of this instruction is similar to the one of $\CALL$, with the exception  that a fresh account is created, which gets the specified transferred value, and that the input provided to this internal transaction, which is again specified in the local memory, is interpreted as the initialization code to be executed in order to produce the newly created account's code as output.  
In contrast to the call transaction, a create transaction does not await a return value, but only an indication of success or failure. 

For discussing how to return from an internal transaction, we show the rule for returning from a successful internal call transaction. 
\begin{mathpar}\small
\infer{
\arraypos{\access{\exenv}{\code}}{\access{\mstate}{\pc}}=\CALL  \\
\access{\mstate}{\stack}=\cons{g}{\cons{\recipient}{\cons{\valu}{\cons{\io}{\cons{\is}{\cons{\oo}{\cons{\os}{s}}}}}}}   \\
\flag = \cond{\gstate(\recipient)= \none}{0}{1} \\  
\aw = \memext{\memext{\access{\mstate}{\actw}}{\io}{\is}}{\oo}{\os} \\
c_{\textit{call}} = \gascapacity{\valu}{\flag}{g}{\access{\mstate}{\gas}} \\
c = \basecosts{\valu}{\flag} + \costmem{\access{\mstate}{\actw}}{\aw} +c_{\textit{call}} \\
\mstate' =\update{\inc{\inc{\update{\update{\mstate}{\actw}{\aw}}{\stack}{\cons{1}{s}}}{\pc}{1}}{\gas}{\lgas - c}}{\memo}{\updateinterval{\access{\mstate}{\memo}}{\oo}{\oo + s -1}{d}}
}
{\sstep{\transenv}{\cons{\haltstatefull{\gstate'}{\lgas}{d}{\transeffects'}}{\cons{\regstatefull{\mstate}{\exenv}{\gstate}{\transeffects}}{\callstack}}}{\cons{\regstatefull{\mstate'}{\exenv}{\gstate'}{\transeffects'}}{\callstack}}}
\end{mathpar}
Leaving the caller state unchanged at the point of calling has the negative side effect that the cost calculation needs to be redone at this point in order to determine the  new gas value of the caller state. 
But besides this, the rule is  straightforward: the program counter is incremented as usual and the number of active words in memory is adjusted as memory accesses for reading the input and return data have been made. The gas is decreased, meaning that  the overall amount of gas $c$ allocated for the execution is subtracted. However, as this cost already includes the gas budget given to the internal transaction, the gas $\lgas$ that is left after the execution is refunded again. In addition, the return data $d$ is written to the local memory of the caller at the place specified by $\oo$ and $\os$. Finally, the value one is written to the caller's stack in order to indicate the success of the internal call transaction. 
As the execution was successful, as indicated by the halting state, the global state and the transaction effects of the callee are adopted by the caller. 


%

EVM bytecode offers several instructions for explicitly halting (internal) transaction execution. Besides the standard instructions $\STOP$ and $\RETURN$, there is the $\SUICIDE$ instruction that is very particular to the blockchain setting. 
The $\STOP$ instruction causes regular halting of the internal transaction without returning data to the caller. In contrast, the $\RETURN$ instruction allows one to specify the memory fragment containing the return data that will be handed to the caller. 

Finally, the $\SUICIDE$ instruction halts the execution and lists the currently execution account for later deletion. More precisely, this means that this account will be deleted when finalizing the external transaction, but its behavior during the ongoing small-step execution is not affected. Additionally, the whole balance of the deleted account is transferred to some beneficiary specified on the machine stack. 

We show the small-step rules depicting the main functionality of $\SUICIDE$. As for $\CALL$, capturing the whole functionality of $\SUICIDE$ would require to consider several corner cases. Here we consider the case where the beneficiary exists, the stack does not underflow and the available amount of gas is sufficient. 

\begin{mathpar} \small 
\infer{
\curropcode{\mstate}{\exenv} = \SUICIDE \\
\access{\mstate}{\stack} = \cons{a_\textit{ben}}{s} \\
a = a_\textit{ben} \mod  2^{160} \\
\gstate(a) \neq \bot \\
\access{\mstate}{\gas} \geq 5000 \\
g = \access{\mstate}{\gas} - 5000 \\
\gstate' = \updategstate{\updategstate{\gstate}{\access{\exenv}{\activeaccount}}{\update{\gstate(\access{\exenv}{\activeaccount})}{\accountbalance}{0}}}{a}{\inc{\gstate(a)}{\accountbalance}{\access{\access{\gstate}{(\access{\exenv}{\activeaccount})}}{\accountbalance}}} \\ 
r = \cond{(\access{\exenv}{\activeaccount} \in \access{\transenv}{\suicideset})}{0}{24000} \\
\transeffects' = \inc{\update{\transeffects}{\suicidesetc}{\access{\transeffects}{\suicidesetc} \cup \{ \access{\exenv}{\activeaccount} \} }}{\refundbalance}{r}
}
{\sstep{\transenv}{\cons{\regstatefull{\mstate}{\exenv}{\gstate}{\transeffects}}{\callstack}}{\cons{\haltstatefull{\gstate'}{g}{\emptyarray}{\transeffects'}}{\callstack}}}
\end{mathpar}

The $\SUICIDE$ command takes one argument $a_{\textit{ben}}$ from the stack specifying the address of the beneficiary that should get the balance of the account that is destructed. If all preconditions are satisfied, the balance of the executing account ($\access{\exenv}{\activeaccount}$) is transferred to the beneficiary address and the current internal transaction execution enters a halting state. Additionally, the transaction effects are extended by adding $\access{\exenv}{\activeaccount}$ to the suicide set and by possibly increasing the refund balance. The refund balance is only increased in case that $\access{\exenv}{\activeaccount}$ is not already scheduled for deletion. The halting state captures the global state $\gstate$ after the money transfer, the remaining gas $g$ after executing the $\SUICIDE$ and the updated transaction effects $\transeffects'$. As no return data is handed to the caller, the empty bytearray $\emptyarray$ is specified as return data in the halting state. 

Note that $\SUICIDE$ deletes the currently executing account $\access{\exenv}{\activeaccount}$ which is not necessarily the same account as the one owning the code $\access{\exenv}{\activecode}$. This might be due a previous execution of $\DELEGATECALL$ or $\CALLCODE$. 

\subsection{Transaction Execution}
The outcome of an external transaction execution does not only consist of the result of the EVM bytecode execution.
Before executing the bytecode, the transaction environment and the execution environment are determined from the transaction information and the block header. 
In the following we assume $\transactions$ to denote the set of transactions. 
An (external) transaction $T \in \transactions$, similar to the internal transactions, specifies a gas limit, a recipient and a value to be transferred. In addition, it also contains the originator and the gas prize that will be recorded in the transaction environment. Finally, it specifies an input to the transaction and the transaction type that can either be a call or a create transaction. The transaction type determines whether the input will be interpreted as input data to a call transaction or as initialization code for a create transaction. 
In addition to the transaction of the environment initialization, some initial changes on the global state and validity checks are performed. For the sake of presentation we assume in the following a function $\inittrans{\cdot}{\cdot}{\cdot} \in \transactions \times \blockheaders \times \gstates \to (\transenvs \times \exstates) \cup \{ \bot \}$ performing the initialization phase and returning a transaction environment and initial execution state in the case of a valid transaction and $\bot$ otherwise. 
Similarly, we assume a function $\finalizetrans{\cdot}{\cdot}{\cdot}{\cdot} \in \transaction \times \exstates \times \teffects \times \gstates$ that given the final global state of the execution, the accumulated transaction effects and the transaction, computes the final effects on the global state. These include for example the deletion of the contracts from the suicide set and the payout to the beneficiary of the transaction.

Formally we can define the execution of a transaction $\transaction \in \transactions$ in a block with header $\blockheader \in \blockheaders$ as follows: 
\begin{mathpar}\small
\infer
{(\transenv, \exstate) = \inittrans{\transaction}{\blockheader}{\gstate} \\
\ssteps{\transenv}{\exconf{\cons{\exstate}{\nil}}{\emptytranseffects}}{\exconf{\cons{\exstate'}{\nil}}{\transeffects'}} \\
\finalstate{\exstate'} \\ 
\gstate' = \finalizetrans{\exstate'}{\transeffects'}{\transaction}{}}
{\transstep{\transaction}{\blockheader}{\gstate}{\gstate'}}
\end{mathpar}
where $\rightarrow^*$ denotes the reflexive and transitive closure of the  small-step relation and the predicate $\finalstate{\cdot}$ characterizes a state that cannot be further reduced using the small-step relation.

\subsection{Formalization in F*}
We provide a formalization of a large fragment of our small-step semantics in the proof assistant F* \cite{fstar}. At the time of writing, we are formalizing the remaining part, which only consists  of straightforward local operations, such as bitwise operators and opcodes to write code to (resp. read code from) the memory. F* is an ML-dialect that is optimized for program verification and allows for performing manual proofs as well as automated proofs leveraging the power of SMT solvers. 

Our formalization strictly follows the small-step semantics as presented in this paper. The core functionality is implemented by the function \lstinline|step| that describes how an execution stack evolves within one execution state. To this end it has two possible outcomes: either it performs an execution step and returns the new callstack or -- in the case that a final configuration is reached (which is a stack containing only one element that is either a halting or an exception state) -- it reports the final state. In order to provide a total function for the step relation, we needed to introduce a third execution outcome that signalizes that a problem occurred due to an inconsistent state. When running the semantics from a valid initial configuration this result, however, should never be produced. 
For running the semantics, the function \lstinline|execution| is defined that subsequently performs execution steps using \lstinline|step| until reaching the final state and reports it. 

The current implementation encompasses approximately thousand lines of code. Since F* code can be compiled into OCaml, we validate our semantics against the official EVM test suite \cite{evmtests}. Our semantics passes 304 out of 624 tests, failing only in those involving any of the missing functionalities.

We make the formalization in F* publicly available~\cite{fullversion}  in order to facilitate the design of static analysis techniques for EVM bytecode as well as their soundness proofs. 

\subsection{Comparison with the Semantics by Luu et al.~\cite{oyente}}
The small-step semantics defined by Luu et al.~\cite{oyente} encompasses only a variation of a subset of EVM bytecode instructions (called EtherLite) and assumes a heavily simplified execution configuration. The instructions covered span simple stack operations for pushing and popping values, conditional branches, binary operations, instructions for accessing and altering local memory and account storage, as well as as the ones for calling, returning and destructing the account. Essential instructions as $\CREATE$ and those for accessing the transaction and block information are omitted. The authors represent a configuration as a tuple of a call stack of activation records and the global state. An activation record contains the code to be executed, the program counter, the local memory and the machine stack. The global state is modelled as mapping from addresses to accounts, with the latter  consisting of code, balance and persistent storage. 

The overall abstraction contains a conceptual flaw, as not including the global state in the activation records of the call stack does not allow for  modelling that, in the case of an exception in the execution of the callee, the global state is rolled back to the one of the caller at the point of calling. 
In addition, the model cannot be easily extended with further instructions -- such as further call instructions or instructions accessing the environment -- without major changes in the abstraction as a lot of information, e.g., the one captured in our small-step semantics in the transaction and the execution environment, are missing. 

\label{sec:comparison-oyente}

\section{Security Definitions}
\label{sec:definitions}
In the following, we introduce the  semantic characterization of the  most significant  security properties for smart contracts, motivating them with typical vulnerabilities recurring in the wild.  

For selecting those properties, we inspected the classification of bugs performed in~\cite{oyente} and~ \cite{survey}. To our knowledge, these are the only works published so far that aim at systematically summarizing bugs in Ethereum smart contracts.

For the presented bugs, we synthesized the semantic security properties that were violated. In this process we realized that some bugs share the same underlying property violation and that other bugs can not be captured by such generic properties -- either because they are of a purely syntactic nature or because they constitute a derivation from a desired behavior that is particular to a specific contract.

\myparagraph{Preliminary Notations}
Formally, we represent a contract as a tuple of the form $\contract{\contractaddress}{\contractcode}$ where $\contractaddress \in \addresses$ denotes the address of the contract and $\contractcode \in \arrayof{\BB}$  denotes the contract's code. We denote the set of contracts by $\contracts$ and assume functions $\getcontractaddress{\cdot}$ and $\getcontractcode{\cdot}$ that extract the contract address and code respectively. 

As we will argue about contracts being called in an arbitrary setting, we additionally introduce the notion of \emph{reachable configuration}. Intuitively, a pair $(\transenv, \callstack)$ of a transaction environment $\transenv$ and a call stack $\callstack$  is reachable  if there exists a state $s$ such that $\callstack,s$ are the result of  \inittrans{$T$}{$H$}{$\sigma$}, for some transaction $T$, block header $H$, a global state $\sigma$, and $\callstack$ is reachable from $s$. 


\begin{definition}[Reachable Configuration]
	The pair $(\transenv, A) \in \transenvs \times \exstates$ is a reachable configuration if for some transaction $T \in \transactions$, some block header $H \in \blockheaders$ and some global state $\gstate \in\addresses\rightarrow \accounts$ of the blockchain it holds that 
	\begin{align*}
	(\transenv, s) = \inittrans{\transaction}{\blockheader}{\gstate} \wedge \ssteps{\transenv}{\cons{s}{\nil}}{\callstack}
	\end{align*}
\end{definition}

In order to give concise security definitions, we further introduce, and assume throughout the paper, an annotation to the small step semantics in order to highlight the contract $c$ that is currently executed. 
In the case of  initialization code being executed, we use $\bot$. 
Specifically, we let
\begin{align*}
\annotatedcallstacks \define 
\{& \cons{\annotate{\excstate}{c}}{\callstackplain}, ~\cons{\annotate{\haltstatefull{\gstate}{\lgas}{\transeffects}{d}}{c}}{\callstackplain}, ~\callstackplain \\ ~|~ 
& \gstate \in \gstates, ~\lgas \in \NN, ~d \in \bytearray,~ \transeffects \in \teffects,~\callstackplain \in \stackof{(\mstates \times \exenvs \times \gstates \times \teffects) \times \contracts} \}
\end{align*}
where $c \in \contracts \cup \{ \bot \} = \contractsbot$. 


Next, we introduce the notion of  execution trace for smart contract execution. Intuitively, a trace is a sequence of actions. In our setting, the actions to be recorded are composed of an opcode, the address of the executing contract, and a sequence of arguments to the opcode. We denote the set of actions with $\actions$. Accordingly, every small step produces a trace consisting of a single action. Again, we lift the resulting trace semantics to multiple execution steps that then produce sequences of actions $\pi \in \sequenceof{\actions}$. 
We only report the trace semantics definition for the $\textsf{CALL}$ case here, referring to 
\iffull
Appendix~\ref{sec:arules} for further details.
\else
the full version of the paper for the details~\cite{fullversion}.
\fi

\begin{mathpar}\small
	\infer{
		\arraypos{\access{\exenv}{\code}}{\access{\mstate}{\pc}}=\CALL \\
		\access{\mstate}{\stack}=\cons{g}{\cons{\recipient}{\cons{\valu}{\cons{\io}{\cons{\is}{\cons{\oo}{\cons{\os}{s}}}}}}}   \\
		\cdots \\
		\mstate' = \cdots \\
		\exenv' = \cdots \\
		\gstate' = \cdots
	} 
	{\ssteptrace{\transenv}{\cons{\annotate{\regstate{\mstate}{\exenv}{\gstate}}{c}}{\callstack}}{\cons{\annotate{\regstate{\mstate'}{\exenv'}{\gstate'}}{\recipient}}{\cons{\annotate{\regstate{\mstate}{\exenv}{\gstate}}{c}}{\callstack}}}{\CALL_{c}(g, \recipient, \value, \io, \is, \oo, \os)}}
\end{mathpar}
We will write $\project{\pi}{\filtercallscreates{c}}$ to denote the projection of $\pi$ to calls performed by contract $c$, i.e., actions   of the form  $\CALL_{c}(g, \recipient, \valu, \io, \is, \oo, \os)$,
$\CREATE_{c}(\valu, \io, \is)$, \\
 $\CALLCODE_{c}(g, \recipient, \valu, \io, \is, \oo, \os)$, and $\DELEGATECALL_{c}(g, \recipient,  \io, \is, \oo, \os)$. 

\subsection{Call Integrity}

\myparagraph{Dependency on Attacker Code}
One of the most famous bugs of Ethereum's history is the so called DAO bug that led to a loss of 60 million dollars in June 2016 \cite{thedao}. This bug is in the literature classified as reentrancy bug \cite{survey,oyente} as the affected contract was drained out of money by subsequently reentering it and performing transactions to the attacker on behalf of the contract. More generally, the problem of this contract was that malicious code was able to affect the outgoing money flows of the contract.  
The cause of such bugs mostly roots in the developer's misunderstanding of the semantics of Solidity's call primitives. 
In general, calling a contract can invoke two kinds of actions: 
	Transferring Ether to the contract's account or 
	Executing (parts of) a contracts code.
In particular, the \lstinline|call| construct invokes the called contract's fallback function when no particular function of the contract is specified (\ref{solidity}). 
Consequently, the developer may expect an atomic value transfer where potentially another contract's code is executed. For illustrating how to exploit this sort of bug, we consider the following contracts: 

\begin{minipage}[t]{0.5 \textwidth}
\begin{lstlisting}
contract Bob{
 bool sent = false;
  function ping(address c){
    if (!sent) { c.call.value(2)();
      		       sent = true; }}}
\end{lstlisting}
\end{minipage}
\begin{minipage}[t]{0.5 \textwidth}
\begin{lstlisting}
contract Mallory{
  function(){
    Bob(msg.sender).ping(this);}} 
\end{lstlisting}
\end{minipage}

The function \lstinline|ping| of contract \lstinline|Bob| sends an amount of $2$ {\wei} to the address specified in the argument. However, this should only be possible once, which is potentially ensured by the \lstinline|sent| variable that is set after the successful money transfer. 
Instead, it turns out that invoking the \lstinline|call.value| function on a contract's address invokes the contract's fallback function as well. 

Given a second contract \lstinline|Mallory|, it is possible to transfer more money than the intended $2$ {\wei} to the account of \lstinline|Mallory|. 
By invoking \lstinline|Bob|'s function \lstinline|ping| with the address of \lstinline|Mallory|'s account, $2$ {\wei} are transferred to \lstinline|Mallory|'s account and additionally the fallback function of \lstinline|Mallory| is invoked. 
As the fallback function again calls the \lstinline|ping| function with \lstinline|Mallory|'s address another $2$ {\wei} are transferred before the variable \lstinline|sent| of contract \lstinline|Bob| was set. 
This looping goes on until all gas of the initial call is consumed or the callstack limit is reached. In this case, only the last transfer of {\wei} is reverted and the effects of all former calls stay in place. Consequently the intended restriction on contract \lstinline|Bob|'s \lstinline|ping| function (namely to only transfer $2$ {\wei} once) is circumvented.

\myparagraph{Call Integrity}
In order to protect from this  class of bugs, it is crucial to secure the code against being reentered \TODOM{to be picky, re-entering or reentering? we are inconsistent} before regaining control over the control flow. From a security perspective, the fundamental problem is that the contract behaviour  depends on untrusted code, even though this was not intended by the developer.  We capture this intuition through a  hyperproperty, \TODOM{same here, hyper-property or hyperproperty?} which we name \emph{call integrity}. The idea is that no matter how the attacker can schedule $c$ (callstacks $\callstack$ and $\callstack'$ in the definition), the  calls of $c$ (traces $\pi$, $\pi'$) cannot be controlled by the attacker, even if $c$ hands over the control to the attacker. \TODOM{With regards to consistency, we should also go through the bibliography and make that consistent}


\begin{definition}[Call Integrity]
	A contract $c \in \contracts$ satisfies  call integrity  for a set of addresses $\untrustedaddresses \subseteq \addresses$ if for all reachable configurations $(\transenv,\cons{\annotate{s}{c}}{\callstack}),(\transenv,\cons{\annotate{s'}{c}}{\callstack'})$ with $s, s'$ differing only in the code with address in $\untrustedaddresses$, it holds that for all $t,t'$
	\begin{align*}
	\sstepstrace{\transenv}{\cons{\annotate{s}{c}}{\callstack}}{\cons{\annotate{t}{c}}{\callstack}}{\pi} ~\land~ \finalstate{\annotate{t}{c}} 
	~\land ~
	\sstepstrace{\transenv}{\cons{\annotate{s'}{c}}{\callstack'}}{\cons{\annotate{t'}{c}}{\callstack'}}{\pi'} ~\land~ \finalstate{\annotate{t'}{c}} \\
	\implies \project{\pi}{\filtercallscreates{c}} = \project{\pi'}{\filtercallscreates{c}}
	\end{align*}
\end{definition}


\subsection{Proof Technique for Call Integrity}
We now  establish a proof technique for call integrity, based on local properties that are arguably easier to verify and that we show to  imply call integrity. 
As  a first observation, we identify   the different ways in which external contracts can influence the execution of a smart contract $c$ and introduce corresponding  security properties : 
\begin{description}
\item[Code Dependency] The contract $c$ might access (information on) the untrusted contracts code via the $\EXTCODECOPY$ or the $\EXTCODESIZE$ instructions and make his behaviour  depend on those values;
\item[Effect Dependency] The contract $c$ might call the untrusted contract and might depend on its execution effects and return value;
\item[Re-entrancy] The contract $c$ might call the untrusted contract, with the latter influencing the behaviour of the former by performing changes to the global state itself or \REPLACEMfor{CY}{171125}{'}{``}on behalf\REPLACEMfor{CY}{171125}{'}{''} of $c$ by reentering it and thereby potentially decreasing the balance of $c$.
\end{description}


The first two of these properties can be seen  as value dependencies and therefore can be formalized as hyperproperties. The first property says that the calls performed by a contract should not be affected by the effects on the execution state produced by adversarial contracts. Technically, we consider a contract $c$ calling an adversarial contract $c'$ (captured as $\sstep{\transenv}{\cons{\annotate{s}{c}}{\callstack}}{\cons{\annotate{s''}{c'}}{\cons{\annotate{s}{c}}{\callstack}}}$ in the premise), which we let terminate in two arbitrary   states $s',t'$: we require that  $c$'s continuation code performs the same calls in both states. 

\begin{definition}[$\untrustedaddresses$-effect Independence]
A contract $c \in \contracts$ is $\untrustedaddresses$-effect independent of for a set of addresses $\untrustedaddresses \subseteq \addresses$ if for all reachable configurations $(\transenv, \cons{\annotate{s}{c}}{\callstack})$ such that $\sstep{\transenv}{\cons{\annotate{s}{c}}{\callstack}}{\cons{\annotate{s''}{c'}}{\cons{\annotate{s}{c}}{\callstack}}}$ for some $s''$ and $\getcontractaddress{c'} \in \untrustedaddresses$, it holds that for all final states $s', t'$ whose global state might differ in all components but the code from the global state of $s$, 
\begin{align*}
\sstepstrace{\transenvinit}{\cons{\annotate{s'}{c'}}{\cons{\annotate{s}{c}}{\callstack}}}{\cons{\annotate{s''}{c}}{\callstack}}{\pi} 
~\land~ \finalstate{s''}  \\
~\land~ \sstepstrace{\transenvinit}{\cons{\annotate{t'}{c'}}{\cons{\annotate{s}{c}}{\callstack}}}{\cons{\annotate{t''}{c}}{\callstack}}{\pi'} 
\land \finalstate{t''}  \\
\implies \project{\pi}{\filtercallscreates{c}} = \project{\pi'}{\filtercallscreates{c}}
\end{align*}
\end{definition}

The second property says that the calls of a contract should not be affected by the code read from the blockchain (e.g., the code does not branch on  code read from the blockchain). 
To this end we introduce the notation $\sstepslocalupdatetrace{\transenv}{\cons{s}{\callstack}}{\cons{s'}{\callstack}}{f}{\pi}$ to denote that the local small-step execution of state $s$ on stack $\callstack$ under $\transenv$ results in several steps in state $s'$ producing trace $\pi$ given that in the local execution steps of $\EXTCODECOPY$ and $\EXTCODESIZE$, which are the operations used to access  the code on the global state,  the code returned by these functions is determined by the partial function $f \in \addresses \pto \arrayof{\BB}$ as opposed to the global state. In other words, we consider in the premise a contract $c$ reading two different codes from the blockchain and terminating in both runs (captured as $\sstepslocalupdatetrace{\transenv}{\cons{\annotate{s}{c}}{\callstack}}{\cons{\annotate{s'}{c}}{\callstack}}{f}{\pi}$ and $\sstepslocalupdatetrace{\transenv}{\cons{\annotate{s}{c}}{\callstack}}{\cons{\annotate{s''}{c}}{\callstack}}{f'}{\pi'}$), and we require that $c$ performs the same calls in both runs. 
\begin{definition}[$\untrustedaddresses$-code Independence]
A contract $c \in \contracts$ is $\untrustedaddresses$-code independent for a set of addresses $\untrustedaddresses \subseteq \addresses$
if for all reachable configurations $(\transenv, \cons{\annotate{s}{c}}{\callstack})$
it holds for all local code updates $f, f' \in \addresses \pto \arrayof{\BB}$ on $\untrustedaddresses$ that
\begin{align*}
\sstepslocalupdatetrace{\transenv}{\cons{\annotate{s}{c}}{\callstack}}{\cons{\annotate{s'}{c}}{\callstack}}{f}{\pi}
~\land~ \finalstate{s'} 
~\land~ \sstepslocalupdatetrace{\transenv}{\cons{\annotate{s}{c}}{\callstack}}{\cons{\annotate{s''}{c}}{\callstack}}{f'}{\pi'}
~\land~ \finalstate{s''} \\
\implies \project{\pi}{\filtercallscreates{c}} = \project{\pi'}{\filtercallscreates{c}}
\end{align*}
\end{definition}

Both these independence properties can be overapproximated by static analysis techniques based on program dependence graphs~\cite{Hammer:2009:FCO}, as done by Joana to verify non-interference in Java~\cite{joana14it}.  The idea is to traverse the  dependence graph in order to detect dependencies between the sensitive sources, in our case the data controlled by the adversary and returned to the contract, and the observable sinks, in our case the local contract calls. 

The last property constitutes a safety  property. Specifically, single-entrancy states that it cannot happen that when reentering the contract $c$ another call is performed before returning (i.e., after reentrancy, which we capture in the call stack as two distinct states with the same running contract $c$, the call stack cannot further increase).

\begin{definition}[Single-entrancy]
	A contract $c \in \contracts$ is single-entrant if for all reachable configurations $(\transenv,\cons{\annotate{\exstate}{c}}{\callstack})$, it holds for all $s', s'',\callstack' $ that 
	\begin{align*}
	\ssteps{\transenv}{\cons{\annotate{\exstate}{c}}{\callstack}}{\concatstack{\cons{\annotate{\exstate'}{c}}{\callstack'}}{\cons{\annotate{\exstate}{c}}{\callstack}}} 
	 \\ 
	\implies 
	\neg \exists \exstate'' \in \exstates, c' \in \contractsbot . \, 
	\ssteps{\transenv}{\concatstack{\cons{\annotate{\exstate'}{c}}{\callstack'}}{\cons{\annotate{\exstate}{c}}{\callstack}}}{\cons{\annotate{s''}{c'}}{\concatstack{\cons{\annotate{\exstate'}{c}}{\callstack'}}{\cons{\annotate{\exstate}{c}}{\callstack}}}} 
	\end{align*}
\end{definition}
This safety property can be easily overapproximated by  syntactic conditions, as for instance done in the Oyente analyzer~\cite{oyente}.


Finally, the next theorem proves the soundness of our proof technique, i.e., the two independence properties and the single-entrancy property together entail call integrity.

\begin{theorem}
Let $c \in \contracts$ be a contract and $\untrustedaddresses \subseteq \addresses$ be a set of untrusted addresses. 
If $c$ is $\untrustedaddresses$-local independent, $c$ is $\untrustedaddresses$-effect independent, and $c$ is single-entrant then $c$ provides call integrity for $\untrustedaddresses$.
\end{theorem}
\noindent
\textit{Proof Sketch.} 
Let $(\transenv,\cons{\annotate{s}{c}}{\callstack}),(\transenv,\cons{\annotate{s'}{c}}{\callstack'})$ be reachable configurations such that $s,s'$ differ only in the code with address in $\untrustedaddresses$. 
We now compare the two small-step runs of those configurations. 
Due to $\untrustedaddresses$-code independence, the execution until the first call to an address $a \in \untrustedaddresses$ produces the same partial trace until the call to $a$. Indeed, we can express the runs under different address mappings through the code update from the $\untrustedaddresses$-code independence property,  as long as no call to one of the updated addresses is performed. When a first call to $a \in \untrustedaddresses$ is performed, we know due to single-entrancy that the following call cannot produce any partial execution trace for any of the runs as this would imply that contract $c$ is reentered and a call out of the contract is performed. 
Due to  $\untrustedaddresses$-code independence and $\untrustedaddresses$-effect independence \TODOM{update consistently!}, the traces after returning must coincide till the next call to  an address in $\untrustedaddresses$. This argument can be iteratively applied until reaching the final state of the execution of $c$. 
%

\subsection{Atomicity}
\subsubsection{Exception Handling} 
As discussed in section \ref{solidity}, the way exceptions are propagated varies with the way contracts are called. In particular, in the case of \lstinline|call| and \lstinline|send|, exceptions are not propagated, but a manual check for the successful completion of the called function's execution is required. This behavior reflects the way exceptions are reported during bytecode execution: Instead of propagating up through the call stack, the callee reports the exception to the caller by writing zero to the stack. In the context of Ethereum, the issue of exception handling is particularly delicate as due to the gas restriction, it might always happen that a call fails simply because it ran out of gas. Intuitively, a user would expect a contract not to depend on the concrete gas value that is given to it, with the exception that a contract might always fail completely (and consequently does not perform any changes on the global state). Such a behavior would prevent contracts from entering an inconsistent state as the one presented in the following excerpt of a simple banking contract: 

\begin{lstlisting}
contract SimpleBank{mapping(address => uint) balances;
  function withdraw(){ msg.sender.send(balances[msg.sender]));
    balances[msg.sender] = 0;}}
\end{lstlisting}


The contract keeps a record of the user balances and provides a function that allows a user to withdraw its own balance -- which results in an update of the record. A developer might not expect that the \lstinline|send| might fail, but as it is on the bytecode level represented by a $\CALL$ instruction, additional to the Ether transfer, code might be executed that runs out of gas. As a consequence, the contract would end up in a state where the money was not transferred (as all effects of the call are reverted in case of an exception), but still the internal balance record of the contract was updated and consequently the money cannot be withdrawn by the owner anymore. 

Inspired by such situations where an inconsistent state is entered by a contract due to mishandled gas exceptions, we introduce the notion of \emph{atomicity} of a contract. Intuitively, atomicity requires that  the effects of the execution on the global state do not depend on the amount of gas available -- except when an exception is triggered,  in which case the overall execution should have no effect at all. The last condition is captured by requiring that the final global state is the same as the initial one for at least one of the two executions (intuitively, the one causing the exception).  

\begin{definition}
A contract $c \in \contracts$ satisfies atomicity 
if for all reachable configurations $(\transenv, \callstack')$ such that 
$\sstep{\transenv}{\callstack'}{\cons{\annotate{s}{c}}{\callstack}}$, it holds for all gas values $g , g' \in \integer{256}$ that 
\begin{align*}
\ssteps{\transenv}{\cons{\update{\annotate{s}{c}}{\access{\mstate}{\gas}}{g}}{\callstack}}{\cons{\annotate{s'}{c}}{\callstack}}
~\land~ \finalstate{s'} \\
~\land~ \ssteps{\transenv}{\cons{\update{\annotate{s}{c}}{\access{\mstate}{\gas}}{g'}}{\callstack}}{\cons{\annotate{s''}{c}}{\callstack}}
~\land~ \finalstate{s''} \\
\implies \access{s'}{\gstate} = \access{s''}{\gstate} \lor  \access{s}{\gstate} = \access{s'}{\gstate} \lor \access{s}{\gstate} = \access{s''}{\gstate} 
\end{align*}
\end{definition}

\subsection{Independence of Miner controlled Parameters}
\label{statedependency}
Another particularity of the distributed blockchain environment is that users while performing transactions cannot make assumptions on large parts of the context their transaction will be executed in. 
A part of this is due to the asynchronous nature of the system: it can always be that another transaction that alters the context was performed first.  Actually, the situation is even more delicate as transactions are not processed in a first-come-first-serve manner, but miners have a big influence on the execution context of transactions. They can decide upon the order of the transactions in a block (and also sneak their own transactions in first) and in addition they can even control some parameters as the block timestamp within a certain range. 
Consequently, contracts whose (outgoing) money flows depend either on miner controlled block information or on state information (as the state of their storage or their balance) that might be changed by other transactions are prone to manipulations by miners. A typical example adduced in the literature is the use of block timestamps as source of randomness \cite{survey,oyente}. In a classical lottery implementation that randomly pays out to one of the participants and uses the block timestamp as source of randomness, a malicious miner can easily influence the result in his favor by selecting a beneficial timestamp. 

We capture the absence of the miner's influence by two definitions, one saying that the outgoing Ether flows of a contract should not be influenced by components of the transaction environment that can be (within a certain range) set by miners and the other one saying that the Ether flows should not depend on those parts of the contract state that might have been influenced by previously executed transactions. The first definition rules out what is in the literature often described as timestamp dependency \cite{survey,oyente}. 

First, we define \emph{independence of} (parts of) \emph{the transaction environment}.
To this end, we assume $\transenvcomponents$ to be the set of components of the transaction environment and write $\transenv \equalupto{\transenvcomponent} \transenv'$ to denote that the transaction environments $\transenv, \transenv'$ are equal up to component $\transenvcomponent$. 


\begin{definition}[Independence of  the Transaction Environment]
A contract $c \in \contracts$ is independent of a subset $I \subseteq \transenvcomponents$ of components of the transaction environment if for all $\transenvcomponent \in I$ and all reachable configurations $(\transenv, \cons{\annotate{s}{c}}{\callstack})$ it holds for all $\transenv'$ that 
\begin{align*}
& \transenvcomponent(\transenv) \neq \transenvcomponent(\transenv') 
\land \transenv \equalupto{\transenvcomponent} \transenv' \\ 
&\land \sstepstrace{\transenv}{\cons{\annotate{\exstate}{c}}{\callstack}}{\cons{\annotate{\exstate'}{c}}{\callstack}}{\pi}
~\land~\finalstate{\exstate'} 
~\land~ \sstepstrace{\transenv'}{\cons{\annotate{\exstate}{c}}{\callstack}}{\cons{\annotate{\exstate''}{c}}{\callstack}}{\pi'} ~\land~ \finalstate{\exstate''} \\
&\implies \project{\pi}{\filtercallscreates{c}} = \project{\pi'}{\filtercallscreates{c}}
\end{align*}
\end{definition}
%
%
%

Next, we define the notion of \emph{independence of the account state}.  Formally, we capture this property by requiring that the outgoing Ether flows of the contract under consideration should not be affected by those parameters of the contract that might have been changed by previous executions which are the balance, the account's nonce, and the account's persistent storage. 

\begin{definition}[Independence of Mutable Account State]
A contract $c \in \contracts$ is independent of the account state if for all reachable configurations $(\transenv,\cons{\annotate{s}{c}}{\callstack}),(\transenv,\cons{\annotate{s}{c}}{\callstack'})$ with $s, s'$ differing only in the nonce, balance and storage for $\getcontractaddress{c}$, it holds that
	\begin{align*}
	\sstepstrace{\transenv}{\cons{\annotate{s}{c}}{\callstack}}{\cons{\annotate{s'}{c}}{\callstack}}{\pi} ~\land~ \finalstate{\annotate{s'}{c}} 
	~\land ~
	\sstepstrace{\transenv}{\cons{\annotate{s}{c}}{\callstack'}}{\cons{\annotate{s''}{c}}{\callstack}}{\pi'} ~\land~ \finalstate{\annotate{s''}{c}} \\
	\implies \project{\pi}{\filtercallscreates{c}} = \project{\pi'}{\filtercallscreates{c}}
	\end{align*}
\end{definition}

As far the other independence properties, both these properties can be statically verified using program dependence graphs.

\subsection{Classification of Bugs}

The previously presented security definitions are motivated by the bugs that were observed in real Ethereum smart contracts and studied in~\cite{oyente} and~\cite{survey}. 
Table~\ref{tab:properties} gives an overview on the bugs from the literature that are ruled out by our security properties. 

\begin{table}[b]
\centering
\caption{Bugs from~\cite{oyente} and~ \cite{survey} ruled out by the security properties}
\label{tab:properties}
\begin{tabular}{cp{5.75cm}}
\multicolumn{1}{p{1.4cm}}{\parbox{4cm}{\centering\textbf{Security Property}}} & \multicolumn{1}{c}{\textbf{Bug}} \\[7pt]
\toprule
	\multicolumn{1}{c}{\multirow{2}{*}{\parbox{4cm}{\centering Call Integrity}}}	 
	& Reentrancy~\cite{survey,oyente} \\
	& Call to the Unknown~\cite{survey} \\
\midrule 
	\multicolumn{1}{c}{\multirow{1}{*}{\parbox{4cm}{\centering Atomicity}}}	 
	& Mishandled Exceptions~\cite{survey,oyente}\\
\midrule 
	\multicolumn{1}{c}{\multirow{2}{*}{\parbox{4cm}{\centering Independence of Mutable Account State}}}	 
	& Transaction Order Dependency~\cite{oyente}  \\ & Unpredictable State~\cite{survey} \\
\midrule 
	\multicolumn{1}{c}{\multirow{3}{*}{\parbox{4cm}{\centering Independence of Transaction Environment}}}	 
	& Timestamp Dependancy~\cite{oyente} \\
	& Time Constraints~\cite{survey}\\
	& Generating Randomness~\cite{survey} \\
\bottomrule
\end{tabular}
\end{table}

Our security properties do not cover all bugs described by Atzei et al.~\cite{survey}, as some of the bugs do not constitute violations of general security properties, i.e., properties that are not specific to the particular contract implementation. There are two classes of bugs that we do not consider: 
The first class deals with the occurrence of unexpected exceptions (such as the Gasless Send and the Call stack Limit bug) and the second class encompasses bugs caused by the Solidity semantics deviating from the programmer's intuitions (such as the Keeping Secrets, Type Cast and Exception Disorders bugs). 

The first class of bugs encompasses runtime exceptions that are hard to predict for the developer and that are consequently not handled correctly. Of course, it would be possible to formalize the absence of those particular kinds of exceptions as simple reachability properties using the small-step semantics. Still, such properties would not give any insight about the security of a contract: the fact that a particular exception occurs can be unproblematic in the case that proper exception handling is in place. In general, the notion of a correct exception handling highly depends on the specific contract's intended behavior. For the special case of out-of-gas exceptions, we could introduce the notion of atomicity in order to capture a generic goal of proper exception handling. But such a notion is not necessarily sufficient for characterizing reasonable ways of dealing with other kinds of runtime exceptions. 

The second class of bugs are introduced on the Solidity level and are similarly hard to account for by using generic security properties. Even though these bugs might all originate from similar idiosyncrasies of the Solidity semantics, the impact of the bugs on the contract's semantics might deviate a lot. This might result in violations of the security properties discussed before, but also in violating the contract's functional correctness. Consequently, catching those bugs might require the introduction of contract-specific correctness properties. 

Finally, Atzei et al.~\cite{survey} discuss the Ether Lost in Transfer bug. This bug is introduced by sending Ether to addresses that do not belong to any contract or user, so called orphan addresses. We could easily formalize a reachability property stating that no valid contract execution should ever send Ether to such an address. We omit such a definition here as it is quite straightforward and at the same time it is not a property that directly affects the security of an individual contract: Sending Ether to such an orphan address might have negative impacts on the overall system as money is effectively lost. For the specific contract sending this money, this bug can be seen as a corner case of sending Ether to an unintended address which rather constitutes a correctness violation. 


\subsection{Discussion}
\label{sec:limitations}

\newcommand\newsubcap[1]{\phantomcaption%
       \caption*{\thefigure.\thesubfigure: #1}}

As previously discussed, we are not aware of any prior formal security definitions of smart contracts. Nevertheless, we  compared our definitions with the verification conditions used in Oyente \cite{oyente}. Our investigation shows  that the verification conditions adopted in this tool are neither sound nor complete.

For detecting mishandled exceptions, it is checked whether each $\CALL$ instruction in the contract code is directly followed by the $\ISZERO$ instruction that checks whether the top element of the stack is zero. Unfortunately, Oyente (although stated in the paper) does not implement this check, so that we needed to manually inspect the bytecodes for determining the outcomes of the syntactic check. 
As shown in Figure~\ref{exc_fn} a check for the caller returning zero does not necessarily imply a proper exception handling and therefore atomicity of the contract.
This excerpt of a simple banking contract that keeps track of the users' balances and allows users to withdraw their balances using the function \lstinline|withdraw| checks for the success of the performed call, but still does not react accordingly. It only makes sure that the number of successes is updated consistently, but does not perform the update on the user's balance record according to the call outcome. 

On the other hand, not performing the desired check does not imply the absence of atomicity as illustrated in Figure~\ref{exc_fp}. 
Writing the outcome in some variable before checking it, satisfies the negative pattern, but still correct exception handling is performed. 
For detecting timestamp dependency, Oyente checks whether the contract has a symbolic execution path with the timestamp (that is represented as own symbolic variable) being included in one of its constraints. 
This definition however, does not capture the case shown in Figure~\ref{time_fn}. 
\begin{figure}[t]
\begin{subfigure}[b]{0.55\textwidth}
\begin{lstlisting}
contract SimpleBank{
 mapping(address => uint) bal;
 uint successes; 
  function withdraw(){
   if (msg.sender.send(bal[msg.sender])) 
    {  successes++; }
   bal[msg.sender] = 0;}}
\end{lstlisting}
\newsubcap{Exception handling: False negative}
\label{exc_fn}
\end{subfigure}
\begin{subfigure}[b]{0.45\textwidth}
\begin{lstlisting}
contract SimpleBank{
 mapping(address => uint) bal;
  function withdraw(){
   bool b = 
    msg.sender.send(bal[msg.sender]); 
    if (b) bal[msg.sender] = 0;}}
  
\end{lstlisting}
\newsubcap{Exception handling: False positive}
\label{exc_fp}
\end{subfigure}

\bigskip
\begin{subfigure}[b]{.5\textwidth}
\centering
\begin{lstlisting}
contract Test{ 
 uint time = block.timestamp;
  function pay (){ 
   if (time % 2 == 1){ 
    msg.sender.send(100);}}}
\end{lstlisting}
\newsubcap{Timestamp dependency: False negative}
\label{time_fn}
\end{subfigure}
\begin{subfigure}[b]{.5\textwidth}
\centering
\begin{lstlisting}
contract Test { 
 function pay (){ 
  if (block.timestamp % 2 == 1 ||
   block.timestamp % 2 == 0){ 
    msg.sender.send(100);}}}
\end{lstlisting}
\newsubcap{Timestamp dependency: False positive}
\label{time_fp}
\end{subfigure}

\bigskip
\begin{subfigure}[b]{0.6 \textwidth}
\begin{lstlisting}
contract Fund{ 
 mapping(address => uint) shares;
  function withdraw(){ 
   if (msg.sender.send(shares[msg.sender]))
    shares[msg.sender] = 0;}}
\end{lstlisting}
\newsubcap{Reentrancy: False negative}
\label{reentrant_fn}
\end{subfigure}
\begin{subfigure}[b]{0.4 \textwidth}
\begin{lstlisting}
contract Bob{ 
 bool sent = false;
  function ping(address c){
   if (!sent) { 
    sent = true; 
    c.call.value(2)();}}}
\end{lstlisting}
\newsubcap{Reentrancy: False positive}
\label{reentrant_fp}
\end{subfigure}
\end{figure}

This contract is clearly timestamp dependent as  whether or not the function \lstinline|pay| pays out some money to the sender depends on the timestamp set when creating the contract. A malicious miner could consequently manipulate the block timestamp for a transaction that creates such a contract in a way that money is paid out and then subsequently query it for draining it out. This is however, not captured by the characterization of the property in Oyente as they only capture the local execution paths of the contract. 

On the other hand, using the block timestamp in path constraints does not imply a dependency as can easily be seen by the example in Figure~\ref{time_fp}. 

For the transaction order dependency and the reentrancy property, we were unfortunately not able to  reconcile the property characterization provided in the paper with the implementation of Oyente. 

For checking reentrancy according to the paper, it should be checked whether the constraints on the path leading to a $\CALL$ instruction can still be satisfied after performing the updates on the path (e.g. changing the storage). If so, the contract is flagged as reentrant. According to our understanding, this approach should not flag contracts that correctly guard their calls as reentrant. Still, by the version of Oyente provided with the paper the contract in Figure~\ref{reentrant_fp} is tagged as reentrant. 

There exists an updated version of Oyente~\cite{newo} that is able to precisely tag this contract as not reentrant, but we could not find any concrete information on the criteria used for checking this property. Still, we found out that the underlying characterization can not be sufficient for detecting reentrancy as the contract in Figure~\ref{reentrant_fn} is classified not to exhibit a reentrancy vulnerability even though it should as the \lstinline|send| command also executes the recipient's callback function (even though with limited gas). 
The example is taken from the Solidity documentation \cite{solidity} where it is listed as negative example.
For transaction order dependency, Oyente should check whether execution traces exhibiting different Ether flows exists. But it turned out that not even a simple example of a transaction dependent contract can be detected by any of the versions of Oyente. 


\section{Conclusions}
\label{sec:conclusion}
We presented the first complete small-step semantics of EVM bytecode and formalized a large fragment thereof in the F* proof assistant, successfully validating it   against the official Ethereum test suite. We further defined for the first time a number of salient security properties for smart contracts, relying on a combination of hyper- and safety properties. Our framework is available to the academic community in order to facilitate future research on rigorous security analysis of smart contracts. 

In particular, this work opens up a number of interesting research directions. First, it would be interesting to formalize in F* the semantics of  Solidity code and a compiler from Solidity into EVM, formally proving  its soundness against our semantics. This would allow us to provide software developers with a tool to verify the security of their code, from which they could obtain bytecode that is secure by construction. Second, we intend  to design an efficient  static analysis technique for EVM bytecode and to formally prove its soundness   against our semantics. 

\paragraph{Acknowledgments.} This work has been partially supported by the European Research Council (ERC) under the European Union's Horizon 2020 research (grant agreement  No 771527-BROWSEC).


\bibliographystyle{splncs}
\bibliography{biblio}

\iffull
\newpage
\appendix 
\section{Formalization}

\label{sec:formalization}
\subsection{Notations}
\label{subsec:formalization_notations}
In the following, we will use $\BB$ to denote the set $\{0,1\}$ of bits and accordingly $\BB^{x}$ for sets of bitstrings of size $x$. We further  let $\integer{x}$ denote the set of non-negative integers representable by $x$ bits and allow for implicit conversion between those two representations (assuming bitstrings to represent a big-endian encoding of natural numbers). In addition, we will use the notation $\arrayof{X}$ (resp. $\stackof{X}$)  for arrays (resp. lists) of  elements from the set $X$. We use standard notations for operations on arrays and lists. In particular we write $\arraypos{a}{\pos}$ to access position $\pos \in [1, \size{a} - 1]$ of array $a \in \arrayof{X}$ and $\arrayinterval{a}{\downv}{\upv}$ to access the subarray of size $\upv - \downv$ from position $\downv \in  [1, \size{a} - 1]$ to $\upv \in  [1, \size{a} - 1]$. In case that $\downv > \upv$ this operation results in the empty array $\emptyarray$. 
In addition, we write $\concat{a_1}{a_2}$ for the concatenation of two arrays $a_1, a_2 \in \arrayof{X}$. 

In the following formalization, we will make use of bytearrays $b \in \bytearray$. To this end, we will assume functions $\bitstringtobytearray{(\cdot)} \in \BB^x \to \bytearray$ and $\bytearraytobitstring{(\cdot)} \in \bytearray \to \BB^x$ to chunk bitstrings with size dividable by $8$ to bytearrays and vice versa.
To denote the zero byte, we write $0^8$ and accordingly for an array of zero bytes of size n, we write $0^{8\cdot n}$.  

For lists, we denote the empty list by $\nil$ and write $\cons{x}{\textit{xs}}$ for placing element $x \in X$ on top of list $\textit{xs} \in \stackof{X}$. 
In addition, we write $\concatstack{\textit{xs}}{\textit{ys}}$ for concatenating lists $\textit{xs}, \textit{ys} \in \stackof{X}$. 

We let $\addresses$ denote the set of $160$-bit addresses ($\BB^{160}$).

\subsection{Configurations}

The global state of the system is defined by the accounts that are existing and their current state, including their balances and their codes. Formally, the global state is a (partial) mapping from account addresses to accounts:
\[\gstate \in \gstates = \addresses \to \optionof{(\integer{256}\times \integer{256} \times (\BB^{256} \to \BB^{256}) \times \arrayof{\BB^8})}\]
An account \account{\accountnoncev}{\accountbalancev}{\accountstorv}{\accountcodev} is described by the account's balance $\accountbalancev \in \integer{256}$, the state of its persistent storage $\accountstorv \in \word \to \word$, its nonce $\accountnoncev \in \integer{256}$ and the account's code $\accountcodev \in \arrayof{\byte}$. 

A configuration $\exconf{\callstack}{\transeffects}$ of the execution consists of the stack $\callstack$ of execution states .
The call stack $\callstack$ keeps track of the calls made during execution. To this end it consists of execution states of one of the following forms: 
\begin{itemize}
\item $\excstate$ denotes an exceptional halting state and can only occur as top element. It expresses that the execution of the current call ended with an exception. 
\item $\haltstatefull{\gstate}{\lgas}{\datav}{\transeffects}$ denotes regular halting and can only occur as top element. It expresses that the execution of the current call halted in global state $\gstate \in \gstate$ with transaction effects $\transeffects \in \teffects$ and with an amount $\lgas \in \integer{256}$ of remaining gas and return data $\datav \in \bytearray$
\item $\regstatefull{\mstate}{\exenv}{\gstate}{\transeffects}$ denotes a regular execution state and represents the state of the execution of the current call. A regular execution state includes the local state of the stack machine $\mstate \in \mstates$, the execution environment $\exenv \in \exenvs$ that contains the parameters given to the call and the current global state $\gstate \in \gstates$ and the transaction effects $\transeffects \in \teffects$
\end{itemize}

The reason to make the global state part of the call stack is that it does not change linearly during the execution. In the case of an exception, all effects of the call's execution on the global state are reverted and the execution continues in the global state of the caller. 
The same holds for the transaction effects. 

Formally, we give the syntax of call stacks as follows: 
\begin{align*}
\callstacks \define \{& \cons{\excstate}{\callstackplain}, ~\cons{\haltstatefull{\gstate}{\lgas}{d}{\transeffects}}{\callstackplain}, ~\callstackplain \\ ~|~ 
& \gstate \in \gstates, ~\lgas \in \NN, ~d \in \bytearray,~  \transeffects \in \teffects ~,\callstackplain \in \stackof{\mstates \times \exenvs \times \gstates \times \teffects} \}
\end{align*}

In Figure~\ref{fig:grammar'} we give a full grammar for call stacks: 
\begin{figure*}[h]
\begin{mathpar}
\begin{array}{rlclll}
\text{Call stacks} & \callstacks & \ni & \callstack & \define & \cons{\excstate}{\callstackplain} ~|~  \cons{\haltstatefull{\gstate}{d}{g}{\transeffects}}{\callstackplain} ~|~ \callstackplain  \\
\text{Plain call stacks} & 
\callstacksplain & \ni & \callstackplain & \define & \cons{\regstatefull{\mstate}{\exenv}{\gstate}{\transeffects}}{\callstackplain} \\
\text{Machine states} & \mstates & \ni & \mstate & \define & \smstate{\lgas}{\lpc}{m}{i}{s} \\
\text{Execution environments} & \exenvs & \ni & \exenv & \define & \sexenv{\textit{actor}}{\textit{input}}{\textit{sender}}{\textit{value}}{\textit{code}}\\
\text{Global states} & \gstates  & \ni &\gstate & &  \\
\text{Account states} & \accounts & \ni & \accountv & \define & \accountstate{n}{b}{\textit{code}}{\textit{stor}} ~|~ \bot   \\ 
\text{Transaction effects} & \teffects  & \ni &\transeffects & \define & (b, L, \suicideset)  \\
\text{Transaction environments} & \transenvs & \ni & \transenv & \define & (o, \textit{prize}, H) \\
\\
\text{Notations:} 
&\multicolumn{5}{c}{
d \in \bytearray, \quad g \in \integer{256}, \quad \transeffects \in \teffects, \quad 
o \in \addresses, \quad \textit{prize} \in \integer{256}, \quad H \in \blockheaders} \\
&\multicolumn{5}{c}{
\lgas \in \integer{256}, \quad \lpc \in \integer{256}, \quad m \in \BB^{256}, \to \BB^8 \quad i \in \integer{256}, \quad s \in \stackof{\BB^{256}}}  \\
&\multicolumn{5}{c}{
\textit{sender} \in \addresses \quad \textit{input} \in \bytearray \quad \textit{sender} \in \addresses \quad \textit{value} \in \integer{256} \quad \textit{code} \in \bytearray} \\
&\multicolumn{5}{c}{
b \in \integer{256} \quad L \in \sequenceof{\logevents} \quad \suicideset \subseteq \addresses \quad 
\gstates = \addresses \to \accounts
}
\end{array}
\end{mathpar}
\caption{Grammar for calls stacks and transaction environments}
\label{fig:grammar'}
\end{figure*}

\subsubsection{Regular execution states}
In the following we give a detailed description of the components of regular executions state. 

\paragraph{Local machine state}
The local machine state $\mstate \in \mstates = \integer{256} \times \integer{256} \times (\word \to \byte) \times \integer{256} \times \stackof{\word}$ represents the state of the underlying state machine used for execution consists of the following components: 
\begin{itemize}
\item $\lgas \in \integer{256}$ is the current amount of gas still available for execution;
\item $\textit{pc}\in \integer{256}$ is the current program counter;
\item $\textit{m} \in \memorytype$ is a mapping from 256-bit words to bytes that represents the local memory;
\item $\textit{i} \in \integer{256}$ is the current number of active words in memory;
\item $\textit{s} \in \stackof{\word}$ is the local 256-bit word stack of the stack machine.
\end{itemize}
The execution of each internal transaction starts in a fresh machine state, with an empty stack, memory initialized to all zeros, and program counter and active words in memory set to zero. Only the gas is instantiated with the gas value available for the execution.   

\paragraph{Execution environment}
The execution environment $\exenv$ of an internal transaction specifies the static parameters of the transaction. It is a tuple of the form $\sexenv{\textit{actor}}{\textit{input}}{\textit{sender}}{\textit{value}}{\textit{code}} \in \exenvs = \addresses \times \bytearray \times \addresses \times \integer{256} \times \bytearray$ with the following components: 
\begin{itemize}
\item $\textit{actor} \in \addresses$ is the address of the account currently executing;
\item $\textit{input} \in \bytearray$ is the data given as an input to the internal transaction;
\item $\textit{sender} \in \addresses$ is the address of the account that initiated the internal transaction;
\item $\textit{value} \in \integer{256}$ is the value transferred by the internal transaction;
\item $\textit{code} \in \bytearray$ is the code currently executed.
\end{itemize}
This information is determined at the beginning of an internal transaction execution and it can be accessed, but not altered during the execution.

\paragraph{Transaction effects}
The transaction effects $\transeffects \in \teffects= \integer{256} \times \stackof{\logevents} \times \setof{\addresses}$ collect information on changes that will be applied to the global state after the transaction's execution. They do not effect the code execution itself. 
In particular, the transaction effects contain the following components: 
\begin{itemize}
\item $\refundbalancev \in \integer{256}$ is the refund balance that is increased by memory operations and will finally be paid to the transaction's beneficiary
\item $\logseqv \in \stackof{\logevents}$ is the sequence of log events performed during executions. A log event is a tuple of the address of the currently executing a count, a tuple with zero to four components specified when executing a logging instruction and finally a fraction of the local memory. Consequently, $\logevents = \addresses \times (\{() \} \cup \word \cup (\word)^2 \cup (\word)^3 \cup (\word)^4) \times \bytearray$.
\item 
$\suicideset \subseteq \addresses$ is the suicide set that keeps track of the contracts that destroyed themselves (using the $\SUICIDE$ command) during the execution (of the external transaction). These contracts are recorded in $\suicideset$ and only removed from the global state after the end of the execution. 
\end{itemize}

\subsection{Transaction environment}
The transaction environment represents the static information of the block that the transaction is executed in and the immutable parameters given to the transaction as the gas prize or the gas limit. 
More specifically, the transaction environment $\transenv \in \transenvs = \addresses \times \integer{256} \times \blockheaders$ is a tuple of the form $(o, \gasprize, \blockheader)$ with the following components: 
\begin{itemize}
\item 
$o \in \addresses$ is the address of the account that made the transaction
\item $\gasprize \in \integer{256}$ denotes the amount of wei that needs to paid for a unit of gas in this transaction 
\item $\blockheader \in \blockheaders = \integer{256} \times \addresses \times \integer{256} \times \integer{256} \times \integer{256} \times \integer{256}$ is the header of the block that the transaction is part. A block header is of the form  $(\parent,$ $\beneficiary,$ $\difficulty,$  $\blocknumber,$ $\gaslimit,$ $\timestamp)$. 
Where $\parent \in \integer{256}$ identifies the header of the block's parent block, $\beneficiary \in \addresses$ is the address of the beneficiary of the transaction, $\difficulty \in \integer{256}$ is a measure of the difficulty of solving the proof of work puzzle required to mine the block, $\blocknumber \integer{256}$ is the number of ancestor blocks, $\gaslimit \in \integer{256}$ is the maximum amount of gas that might be consumed when executing the blocks transactions and $\timestamp \in \integer{256}$ is the Unix time stamp at the block's inception.
Note that this is a simplified version of the block header described in the yellow paper~\cite{yellowpaper} that only contains those components needed for transaction execution. 
\end{itemize}

\section{Small step semantics}
\label{sec:arules}

We define a small step relation $\rightarrow$. We write $\sstep{\transenv}{
\exconf{\callstack}{}}{\exconf{\callstack'}{}}$ to denote that the call stack $\callstack \in \callstacks$ evolves under the transaction environment $\transenv \in \transenvs$ to the call stack $\callstack' \in \callstacks$.
The transaction environment contains information concerning the block or transaction the current code is executed in and that does not change over code execution. 

\subsection{Notations}
In order to present the small-step rules in a concise fashion we introduce some notations for accessing and updating state. 


For the global state we use a slightly different notation for accessing and updating. As the global state is a mapping from addresses to account, the account's state can be accessed by applying the address to the global state. For updating we introduce a simplifying notation: 
\begin{align*}
\updategstate{\gstate}{\addr}{s} & \define \lam{a}{\cond{a = \addr}{s}{\getaccount{\sigma}{a}}}
\end{align*}

For accessing memory fragments we use the following notation: 
\begin{align*}
\getinterval{\memo}{o}{s} & \define [\memo(o),\memo(o + 1), \dots, \memo(o+s -1)]
\end{align*}
Correspondingly, we define updates for memory fragments. Let $o, s \in \integer{256}$ and $v \in \bytearray$: 
\begin{align*}
\updateinterval{\memo}{o}{s}{v} & \define \lam{x}{\cond{(x \geq o \land x < o + \mini{s}{\size{v}})}{\arraypos{v}{x- o}}{\memo(x)}}
\end{align*}
Similarly to accessing arrays, we write $\extract{v}{\textit{down}}{\textit{up}}$ to extract the bitvector's bits from position $\textit{down}$ until position $\textit{up}$ (where we require $\textit{down} \leq \textit{up}$). Additionally, we assume a concatenation function for bitvectors and write $\concat{b_1}{b_2}$ for concatenating bit vectors $b_1$ and $b_2$. 

Most of the state components used in the formalization of the EVM execution configurations consist of tuples. For sake of better readability, instead of accessing tuple components using projection, we name the components according to the variable names we used in the description in Section~\ref{sec:formalization} and use a dot notation for accessing them. To differentiate component names from variable names, we typeset components in sans serifs font. 
For example, given $\mstatev \in \mstates$, we write $\access{\mstatev}{\gas}$ to access the first component of the tuple $\mstatev$. 
Similarly, we use a simple update notation for components. E.g., instead of writing $\letlet \mstatev = (\gasv, \pcv, \memov, \actwv, \stackv) \letin (\gasv, \pcv + 1, \memov, \actwv, \stackv)$, we write $\update{\mstatev}{\pc}{\access{\mstate}{\pc} + 1}$. For the case of incrementing or decrementing numerical values we use the usual short cuts $+=$ and $-=$ and would for example write the example shown before as $\inc{\mstatev}{\pc}{1}$. 

As mentioned in section~\ref{subsec:formalization_notations}, we use the notions of $\BB^{x}$ and $\integer{x}$ interchangeably as we interpret bitvectos usually as unsigned integers. As some operations however are performed on the signed interpretation of the machine words, we assume functions $\signed{(\cdot)}: \BB^{x} \to \sinteger{x}$ and $\signed{(\cdot)}: \integer{x} \to \sinteger{x}$ that output the signed interpretation of a bitvector or unsigned integer respectively. Note the that $\sinteger{x}$ denotes the set of unsigned integers re presentable with $x$ bits. Accordingly, we assume a functions $\unsigned{(\cdot)}: \sinteger{x} \to \BB^{x}$ and $\unsigned{(\cdot)}: \sinteger{x} \to \integer{x}$ for converting signed integers back to their unsigned interpretation. 

\subsection{Auxiliary definitions}
\paragraph{Accessing bytecode}
For extracting the command that is currently executed, the instruction at position $\access{\mstate}{\pc}$ of the code $\activecode$ provided in the execution environment needs to be accessed. For sake of presentation, we define a function doing so: 
\begin{definition}[Currently executed command]
The currently executed command in the machine state $\mstate$ and execution environment $\exenv$ is denoted by $\curropcode{\mstate}{\exenv}$ and  
defined as follows: 
\begin{align*}
\curropcode{\mstate}{\exenv} := \begin{cases}
\arraypos{\access{\exenv}{\activecode}}{\access{\mstate}{\pc}} & \access{\mstate}{\pc} < \size{\access{\exenv}{\activecode}} \\
\STOP & \text{otherwise}
\end{cases}
\end{align*}
\end{definition}

All EVM instructions have in common that running out of gas as well as over and under flows of the local machine stack cause an exception. 
We define a function $\simvalid{\cdot}{\cdot}{\cdot}: \integer{256} \times \integer{256} \times \NN \to \BB$ that given the available gas, the instruction cost and the new stack size determines whether one of the conditions mentioned above is satisfied. We do not check for stack underflows as this is realized by pattern matching in the individual small step rules. 
\begin{align*}
\simvalid{g}{c}{s} \define 
\begin{cases}
1 & g \geq c \land s < 1024 \\
0 & \textit{otherwise}
\end{cases}
\end{align*}
We also write $\simvalid{g}{c}{s}$ for $\simvalid{g}{c}{s} = 1$ and $\neg \simvalid{g}{c}{s}$ for $\simvalid{g}{c}{s} = 0$. 

In EVM bytecode jump potential destinations are explicitly marked by the distinct $\JUMPDEST$ instruction. Jumps to other destination cause an exception. For simplifying this check, we define the set of valid jump destinations as follows: 

\begin{definition}{Valid jump destinations \cite{yellowpaper}.}
$\funD{\cdot}: \bytearray \to \setof{\NN}$ determines the set of valid jump destinations given the code $\code \in \bytearray$, that is being run. It is defined as any position in the code occupied by a $\JUMPDEST$ instruction. Formally $\funD{c} = \funDhelp{c}{0}$, where:
\begin{align*}
\funDhelp{\cdot}{\cdot}: \bytearray \times \NN \to \setof{\NN} \\
\funDhelp{c}{i} := \begin{cases}
\emptyset &  i \geq \size{c} \\
\{i\} \cup \funDhelp{c}{\funN{i}{c[i]}} & \arraypos{c}{i} = \JUMPDEST \\
\funDhelp{c}{\funN{i}{\arraypos{c}{i}}} & \text{otherwise}
\end{cases}
\end{align*}

where $\funN{\cdot}{\cdot}: \NN \times \byte \to \NN$ is the next valid instruction position in the code, skipping the data of a $\PUSH{n}$ instruction, if any:
\begin{align*}
\funN{i}{\instv} := 
\begin{cases}
    i + n + 1 & \instv = \PUSH{n}\\
    i + n & \text{otherwise}
    \end{cases}
\end{align*}
\end{definition}
\paragraph{Memory Consumption}

The execution tracks the number of active words in memory and charges fees for memory that is used. 
The active words in memory are those words that are accessed either
for reading or writing. If a command increases the number of active
words, it needs to pay accordingly to the amount of words that became active. 

To model the increasing number of active words in memory we define a memory expansion function as done in \cite{yellowpaper} that determines the number of active words in memory given the number of active memory words so far as well as the offset and the size of the memory fraction accessed. 
\begin{align*}
\memext{i}{o}{s} := 
\begin{cases}
i & \text{if $s = 0$} \\
\maxi{i}{\left \lceil \frac{(o +s)}{32} \right \rceil} &  \text{otherwise}
\end{cases} 
\end{align*}

According to the amount of additional words in memory that are used by the execution of an instruction, additional execution costs are charged. 
For describing the cost that occur due to memory consumption, we use a function $\costmem{\cdot}{\cdot}: \NN \times \NN \to \ZZ$ that given the number of active words in memory before and after the command execution, outputs the corresponding costs. 
\begin{align*}
\costmem{\aw}{\aw'} & \define 3 \cdot (\aw' - \aw) + \left \lfloor \frac{aw'^2}{512} \right \rfloor - \left \lfloor \frac{aw^2}{512} \right \rfloor
\end{align*}

\paragraph{Creating new account addresses}

We define  a function $\getfreshaddress{\cdot}{\cdot}: \addresses \times \NN \to \addresses$ that given an address and a nonce provides a fresh address. 
\begin{align*}
\getfreshaddress{a}{n} = \keccak{\rlpencode{(a, n-1)}}[96, 255]
\end{align*}

where $\rlpencode{\cdot}$ is the RLP encoding function. The RLP encoding is a canonical way of transforming different structures such as tuples to a sequence of bytes. We will not comment on this in detail, but refer to the reader to the Ethereum yellow paper~\cite{yellowpaper}. 

Note that the $\getfreshaddress{\cdot}{\cdot}$ function is assumed to be collision resistant.

\subsection{Small-step rules}

\paragraph{Binary stack operations} 
We start by giving the rules for arithmetic operations. As all of these instructions alter only the local stack and gas and their only difference consists of the operations performed and the (constant) amount of gas computed, we assume assume a set $\binops$ of binary operations and functions $\binopcost{\cdot}: \binops \to \integer{256}$ and $\binopfun{\cdot}: \binops \to (\word \times \word \to \word)$ that map the binary operations to their costs and functionality. 

For all binary operations $\binopv \in \binops$, we create rules of the following form 
\begin{mathpar}
\infer{
\curropcode{\mstate}{\exenv}= \binopv \\
\simvalid{\access{\mstate}{\gas}}{\binopcost{\binopv}}{\size{s} +1} \\
\access{\mstate}{\stack} = \cons{a}{\cons{b}{s}} \\
\mstate'= \dec{\inc{\update{\mstate}{\stack}{\cons{(\binopfun{\binopv})}{s}}}{\pc}{1}}{\gas}{\binopcost{\binopv}}}
{\sstep{\transenv}{\cons{\regstatefull{\mstate}{\exenv}{\gstate}{\transeffects}}{\callstack}}{\cons{\regstatefull{\mstate'}{\exenv}{\gstate}{\transeffects}}{\callstack}}}
\end{mathpar}

\begin{mathpar}
\infer{
\curropcode{\mstate}{\exenv}= \binopv  \\
(\neg \simvalid{\access{\mstate}{\gas}}{\binopcost{\binopv}}{\size{s} +1} 
\lor \size{\access{\mstate}{\stack}} < 2)}
{\sstep{\transenv}{\cons{\regstatefull{\mstate}{\exenv}{\gstate}{\transeffects}}{\callstack}}{\cons{\excstate}{\callstack}}}
\end{mathpar}

We define 
\begin{align*}
\binops \define \{ \ADD,  \SUB, \LT,  \GT,  \EQ, \AND, \OR, \XOR, \SLT,  \SGT, \MUL, \DIV, \SDIV, \\
\MOD, \SMOD, \SIGNEXTEND, \BYTE \} 
\end{align*}
and 
\begin{align*}
\binopcost{\binopv} = 
\begin{cases}
3 & \binopv \in  \{\ADD, \SUB, \LT, \GT, \SLT, \SGT, \EQ, \AND, \OR, \XOR, \BYTE \}  \\
5 &\binopv \in \{\MUL, \DIV, \SDIV, \MOD, \SMOD, \SIGNEXTEND \} \\
\end{cases}
\end{align*}
and 
\begin{align*}
\binopfun{\binopv} = 
\begin{cases}
\lambda (a,b).\, a + b \mod 2^{256} & \binopv = \ADD \\
\lambda (a,b).\,a - b \mod 2^{256} & \binopv = \SUB \\
\lambda (a,b).\, \cond{a < b}{1}{0} & \binopv = \LT \\
\lambda (a,b). \, \cond{a > b}{1}{0} & \binopv = \GT \\
\lambda (a,b). \, \cond{\signed{a} < \signed{b}}{1}{0} & \binopv = \SLT \\
\lambda (a,b). \, \cond{\signed{a} > \signed{b}}{1}{0} & \binopv = \SGT \\
\lambda (a,b). \, \cond{a = b}{1}{0} & \binopv = \EQ \\
\lambda (a,b).\, a \bitand b & \binopv = \AND\\
\lambda (a,b).\, a \bitor b & \binopv = \OR \\
\lambda (a,b).\, a \bitxor b & \binopv = \XOR \\
\lambda (a,b).\, a \cdot b \mod 2^{256} & \binopv = \MUL \\
\lambda (a,b).\, \cond{(b = 0)}{0}{\lfloor a \div b \rfloor}  & \binopv = \DIV\\
\lambda (a,b).\, \cond{(b=0)}{0}{a \mod b}  & \binopv = \MOD\\
\lambda (a,b).\, (b=0)?~0~:~(a= 2^{255}\land \signed{b} = -1)?~2^{256}~: \\
~\textit{let}~x = \signed{a} \div \signed{b} ~\textit{in}~ \unsigned{(\signof{x} \cdot \lfloor | x| \rfloor)}  & \binopv = \SDIV\\
\lambda (a,b).\, \cond{(b=0)}{0}{\unsigned{(\signof{a} \cdot |a| \mod |b|)}} &\binopv = \SMOD\\
\lambda (o, b). \, \cond{(o \geq 32)}{0}{\concat{\extract{b}{8 \cdot o}{8 \cdot o + 7}}{0^{248}}} & \binopv = \BYTE \\ 
\lambda (a, b). \, \textit{let}~x = 256-8(a+1) ~\textit{in} \\
~\textit{let}~ s = \arraypos{b}{x} ~\textit{in}~ \concat{s^x}{\extract{b}{x}{255}} & \binopv = \SIGNEXTEND 
\end{cases}
\end{align*}

where $\signof{\cdot}: \sinteger{x} \to \{-1, 1\}$ is defined as 
\begin{align*}
\signof{x} = 
\begin{cases}
1 & x \geq 0 \\
0 & \text{otherwise}
\end{cases}
\end{align*}

and $\bitand$, $\bitor$ and $\bitxor$ are bitwise and, or and xor, respectively.

Exceptions to the normal binary operations are the exponentiation as this instruction uses non-constant costs and the computation of the Keccack-256 hash. 

\begin{mathpar}
\infer{
\curropcode{\mstate}{\exenv}= \EXP \\
\simvalid{\access{\mstate}{\gas}}{\costs}{\size{s} +1} \\
\access{\mstate}{\stack} = \cons{a}{\cons{b}{s}} \\
\costs = \cond{(b = 0)}{10}{10 + 10* (1 + \left \lfloor \log_{256}{b} \right \rfloor)} \\
x = (a^b) \mod 2^{256} \\
\mstate'= \dec{\inc{\update{\mstate}{\stack}{\cons{x}{s}}}{\pc}{1}}{\gas}{\costs}}
{\sstep{\transenv}{\cons{\regstatefull{\mstate}{\exenv}{\gstate}{\transeffects}}{\callstack}}{\cons{\regstatefull{\mstate'}{\exenv}{\gstate}{\transeffects}}{\callstack}}}
\end{mathpar}

\begin{mathpar}
\infer{
\curropcode{\mstate}{\exenv}= \EXP  \\
\costs = \cond{(b = 0)}{10}{10 + 10* (1 + \left \lfloor \log_{256}{b} \right \rfloor)} \\
\access{\mstate}{\stack} = \cons{a}{\cons{b}{s}} \\
\neg \simvalid{\access{\mstate}{\gas}}{\costs}{\size{s} +1}}
{\sstep{\transenv}{\cons{\regstatefull{\mstate}{\exenv}{\gstate}{\transeffects}}{\callstack}}{\cons{\excstate}{\callstack}}}
\end{mathpar}

\begin{mathpar}
\infer{
\curropcode{\mstate}{\exenv}= \SHA \\
\simvalid{\access{\mstate}{\gas}}{\costs}{\size{s} +1} \\
\access{\mstate}{\stack} = \cons{\pos}{\cons{\siz}{s}} \\
\aw = \memext{\access{\mstate}{\actw}}{\pos}{\siz}\\ 
\costs = \costmem{\access{\mstate}{\actw}}{\aw} + 30 + 6 \cdot \left \lceil \frac{\siz}{32} \right \rceil\\
v =  \arraypos{\access{\mstate}{\memo}}{\pos, \pos + \siz -1}\\
h = \keccak{v}\\ 
\mstate'= \update{\dec{\inc{\update{\mstate}{\stack}{\cons{h}{s}}}{\pc}{1}}{\gas}{\costs}}{\actw}{\aw}}
{\sstep{\transenv}{\cons{\regstatefull{\mstate}{\exenv}{\gstate}{\transeffects}}{\callstack}}{\cons{\regstatefull{\mstate'}{\exenv}{\gstate}{\transeffects}}{\callstack}}}
\end{mathpar}
where $\keccak{x}$ is the Keccak-256 hash of $x$. ()

\begin{mathpar}
\infer{
\curropcode{\mstate}{\exenv}= \SHA \\
\access{\mstate}{\stack} = \cons{\pos}{\cons{\siz}{s}} \\
\access{\mstate}{\stack} = \cons{\pos}{\cons{\siz}{s}} \\
\aw = \memext{\access{\mstate}{\actw}}{\pos}{\siz}\\ 
\costs = \costmem{\access{\mstate}{\actw}}{\aw} + 30 + 6 \cdot \left \lceil \frac{\siz}{32} \right \rceil\\
\access{\mstate}{\stack} = \cons{a}{\cons{b}{s}} \\
\neg \simvalid{\access{\mstate}{\gas}}{\costs}{\size{s} +1}}
{\sstep{\transenv}{\cons{\regstatefull{\mstate}{\exenv}{\gstate}{\transeffects}}{\callstack}}{\cons{\excstate}{\callstack}}}
\end{mathpar}

\begin{mathpar}
\infer{
(\curropcode{\mstate}{\exenv}= \EXP  \lor \curropcode{\mstate}{\exenv}= \SHA) \\
\size{\access{\mstate}{\stack}} < 2} 
{\sstep{\transenv}{\cons{\regstatefull{\mstate}{\exenv}{\gstate}{\transeffects}}{\callstack}}{\cons{\excstate}{\callstack}}}
\end{mathpar}

\paragraph{Unary stack operations}

\begin{mathpar}
\infer{
\curropcode{\mstate}{\exenv}= \ISZERO \\
\simvalid{\access{\mstate}{\gas}}{3}{\size{s} +1} \\
\access{\mstate}{\stack} = \cons{a}{s} \\
x = \cond{(a=0)}{1}{0} \\
\mstate'= \dec{\inc{\update{\mstate}{\stack}{\cons{x}{s}}}{\pc}{1}}{\gas}{3}}
{\sstep{\transenv}{\cons{\regstatefull{\mstate}{\exenv}{\gstate}{\transeffects}}{\callstack}}{\cons{\regstatefull{\mstate'}{\exenv}{\gstate}{\transeffects}}{\callstack}}}
\end{mathpar}

\begin{mathpar}
\infer{
\curropcode{\mstate}{\exenv}= \NOT \\
\simvalid{\access{\mstate}{\gas}}{3}{\size{s} +1} \\
\access{\mstate}{\stack} = \cons{a}{s} \\
x = \bitneg a \\
\mstate'= \dec{\inc{\update{\mstate}{\stack}{\cons{x}{s}}}{\pc}{1}}{\gas}{3}}
{\sstep{\transenv}{\cons{\regstatefull{\mstate}{\exenv}{\gstate}{\transeffects}}{\callstack}}{\cons{\regstatefull{\mstate'}{\exenv}{\gstate}{\transeffects}}{\callstack}}}
\end{mathpar}

where $\bitneg$ is bitwise negation. 

\begin{mathpar}
\infer{
(\curropcode{\mstate}{\exenv}= \ISZERO \lor \curropcode{\mstate}{\exenv}= \NOT) \\
(\neg \simvalid{\access{\mstate}{\gas}}{3}{\size{s} +1} 
\lor \size{\access{\mstate}{\stack}} < 1)}
{\sstep{\transenv}{\cons{\regstatefull{\mstate}{\exenv}{\gstate}{\transeffects}}{\callstack}}{\cons{\excstate}{\callstack}}}
\end{mathpar}

\paragraph{Ternary stack operations}

\begin{mathpar}
\infer{
\curropcode{\mstate}{\exenv}= \ADDMOD \\
\simvalid{\access{\mstate}{\gas}}{8}{\size{s} +1} \\
\access{\mstate}{\stack} = \cons{a}{\cons{b}{\cons{c}{s}}} \\
x = \cond{(c=0)}{0}{(a + b) \mod c} \\
\mstate'= \dec{\inc{\update{\mstate}{\stack}{\cons{x}{s}}}{\pc}{1}}{\gas}{8}}
{\sstep{\transenv}{\cons{\regstatefull{\mstate}{\exenv}{\gstate}{\transeffects}}{\callstack}}{\cons{\regstatefull{\mstate'}{\exenv}{\gstate}{\transeffects}}{\callstack}}}
\end{mathpar}

\begin{mathpar}
\infer{
\curropcode{\mstate}{\exenv}= \MULMOD \\
\simvalid{\access{\mstate}{\gas}}{8}{\size{s} +1} \\
\access{\mstate}{\stack} = \cons{a}{\cons{b}{\cons{c}{s}}} \\
x = \cond{(c=0)}{0}{(a \cdot b) \mod c} \\
\mstate'= \dec{\inc{\update{\mstate}{\stack}{\cons{x}{s}}}{\pc}{1}}{\gas}{8}}
{\sstep{\transenv}{\cons{\regstatefull{\mstate}{\exenv}{\gstate}{\transeffects}}{\callstack}}{\cons{\regstatefull{\mstate'}{\exenv}{\gstate}{\transeffects}}{\callstack}}}
\end{mathpar}

\begin{mathpar}
\infer{
(\curropcode{\mstate}{\exenv}= \ADDMOD \lor \curropcode{\mstate}{\exenv}= \MULMOD) \\
(\neg \simvalid{\access{\mstate}{\gas}}{8}{\size{\access{\mstate}{\stack}} -2} 
\lor \size{\access{\mstate}{\stack}} < 3)}
{\sstep{\transenv}{\cons{\regstatefull{\mstate}{\exenv}{\gstate}{\transeffects}}{\callstack}}{\cons{\excstate}{\callstack}}}
\end{mathpar}

\paragraph{Accessing the execution environment}

There are some simple access operations for accessing parts of the execution environment such as the addresses of the executing account and the caller, the value given to the internal transaction and the sizes of the executed code and the data given as input to the call. 

\begin{mathpar}
\infer{
\curropcode{\mstate}{\exenv}= \ADDRESS \\
\simvalid{\access{\mstate}{\gas}}{2}{\size{\access{\mstate}{\stack}} +1} \\
\mstate'= \dec{\inc{\update{\mstate}{\stack}{\cons{\access{\exenv}{\activeaccount}}{\access{\mstate}{\stack}}}}{\pc}{1}}{\gas}{2}}
{\sstep{\transenv}{\cons{\regstatefull{\mstate}{\exenv}{\gstate}{\transeffects}}{\callstack}}{\cons{\regstatefull{\mstate'}{\exenv}{\gstate}{\transeffects}}{\callstack}}}
\end{mathpar}

\begin{mathpar}
\infer{
\curropcode{\mstate}{\exenv}= \CALLER \\
\simvalid{\access{\mstate}{\gas}}{2}{\size{\access{\mstate}{\stack}} +1} \\
\mstate'= \dec{\inc{\update{\mstate}{\stack}{\cons{\access{\exenv}{\sender}}{\access{\mstate}{\stack}}}}{\pc}{1}}{\gas}{2}}
{\sstep{\transenv}{\cons{\regstatefull{\mstate}{\exenv}{\gstate}{\transeffects}}{\callstack}}{\cons{\regstatefull{\mstate'}{\exenv}{\gstate}{\transeffects}}{\callstack}}}
\end{mathpar}

\begin{mathpar}
\infer{
\curropcode{\mstate}{\exenv}= \CALLVALUE\\
\simvalid{\access{\mstate}{\gas}}{2}{\size{\access{\mstate}{\stack}} +1} \\
\mstate'= \dec{\inc{\update{\mstate}{\stack}{\cons{\access{\exenv}{\tvalue}}{\access{\mstate}{\stack}}}}{\pc}{1}}{\gas}{2}}
{\sstep{\transenv}{\cons{\regstatefull{\mstate}{\exenv}{\gstate}{\transeffects}}{\callstack}}{\cons{\regstatefull{\mstate'}{\exenv}{\gstate}{\transeffects}}{\callstack}}}
\end{mathpar}

\begin{mathpar}
\infer{
\curropcode{\mstate}{\exenv}= \CODESIZE\\
\simvalid{\access{\mstate}{\gas}}{2}{\size{\access{\mstate}{\stack}} +1} \\
\mstate'= \dec{\inc{\update{\mstate}{\stack}{\cons{\size{\access{\exenv}{\activecode}}}{\access{\mstate}{\stack}}}}{\pc}{1}}{\gas}{2}}
{\sstep{\transenv}{\cons{\regstatefull{\mstate}{\exenv}{\gstate}{\transeffects}}{\callstack}}{\cons{\regstatefull{\mstate'}{\exenv}{\gstate}{\transeffects}}{\callstack}}}
\end{mathpar}

\begin{mathpar}
\infer{
\curropcode{\mstate}{\exenv}= \CALLDATASIZE\\
\simvalid{\access{\mstate}{\gas}}{2}{\size{\access{\mstate}{\stack}} +1} \\
\mstate'= \dec{\inc{\update{\mstate}{\stack}{\cons{\size{\access{\exenv}{\inputdata}}}{\access{\mstate}{\stack}}}}{\pc}{1}}{\gas}{2}}
{\sstep{\transenv}{\cons{\regstatefull{\mstate}{\exenv}{\gstate}{\transeffects}}{\callstack}}{\cons{\regstatefull{\mstate'}{\exenv}{\gstate}{\transeffects}}{\callstack}}}
\end{mathpar}

\begin{mathpar}
\infer{
(\curropcode{\mstate}{\exenv}= \ADDRESS \lor \curropcode{\mstate}{\exenv}= \CALLER \lor \curropcode{\mstate}{\exenv}= \CALLVALUE \\
\lor \curropcode{\mstate}{\exenv}= \CODESIZE \lor \curropcode{\mstate}{\exenv}= \CALLDATASIZE)\\
\neg \simvalid{\access{\mstate}{\gas}}{2}{\size{\access{\mstate}{\stack}} +1}}
{\sstep{\transenv}{\cons{\regstatefull{\mstate}{\exenv}{\gstate}{\transeffects}}{\callstack}}{\cons{\excstate}{\callstack}}}
\end{mathpar}

Accessing the code and the input data in the execution environment is more involved. 

The $\CALLDATALOAD$ instruction writes the (first 256 bit of) data given as input to the current call to the stack. 
\begin{mathpar}
\infer{
	\curropcode{\mstate}{\exenv}= \CALLDATALOAD\\
	\access{\mstate}{\stack} = \cons{a}{s}\\
	\simvalid{\access{\mstate}{\gas}}{3}{\size{\access{\mstate}{\stack}}} \\
	k = \cond{(\size{\access{\exenv}{\data}} - a < 0)}{0} {\mini{\size{\access{\exenv}{\data}} - a}{32}}\\
	v' = \arraypos{\access{\exenv}{\data}}{a, a + k - 1}\\
	v = \concat{v'}{0^{256 - k\cdot 8}} \\
	\mstate'= \dec{\inc{\update{\mstate}{\stack}{\cons{v}{s}}}{\pc}{1}}{\gas}{3}
}
{\sstep{\transenv}{\cons{\regstatefull{\mstate}{\exenv}{\gstate}{\transeffects}}{\callstack}}{\cons{\regstatefull{\mstate'}{\exenv}{\gstate}{\transeffects}}{\callstack}}}
\end{mathpar}

\begin{mathpar}
\infer{
\curropcode{\mstate}{\exenv}= \CALLDATALOAD \\
\neg \simvalid{\access{\mstate}{\gas}}{3}{\size{\access{\mstate}{\stack}}}}
{\sstep{\transenv}{\cons{\regstatefull{\mstate}{\exenv}{\gstate}{\transeffects}}{\callstack}}{\cons{\excstate}{\callstack}}}
\end{mathpar}

The $\CALLDATACOPY$ instruction copies the data that was given as input to the current call to the memory. 

\begin{mathpar}
\infer{
	\curropcode{\mstate}{\exenv}= \CALLDATACOPY\\
	\access{\mstate}{\stack} = \cons{\pos_{\memo}}{\cons{\pos_\data}{\cons{\siz}{s}}}\\
	\aw = \memext{\access{\mstate}{\actw}}{\pos_\memo}{\siz} \\
	\costs = \costmem{\access{\mstate}{\actw}}{\aw}+3 + 3 \cdot \left \lceil \frac{\siz}{32} \right \rceil\\ 
	\simvalid{\access{\mstate}{\gas}}{\costs}{\size{\access{\mstate}{\stack}}} \\
	k = \cond{(\size{\access{\exenv}{\inputdata})} - \pos_\data < 0}{0}{\mini{\size{\access{\exenv}{\inputdata}}- \pos_\data}{\siz}}\\
	d' = \arraypos{\access{\exenv}{\inputdata}}{\pos_\data, \pos_\data + k - 1}\\
	d = \concat{d'}{0^{8 \cdot(\siz - k)}}\\
	\mstate'= \update{\update{\dec{\inc{\update{\mstate}{\stack}{s}}{\pc}{1}}{\gas}{\costs}}{\memo}{\update{\memo}{[\pos_\memo, \pos_\memo + \siz - 1]}{d}}}{\actw}{\aw}
}
{\sstep{\transenv}{\cons{\regstatefull{\mstate}{\exenv}{\gstate}{\transeffects}}{\callstack}}{\cons{\regstatefull{\mstate'}{\exenv}{\gstate}{\transeffects}}{\callstack}}}
\end{mathpar}

The $\CODECOPY$ instruction copies the code that is currently executed to the memory. 
\begin{mathpar}
\infer{
	\curropcode{\mstate}{\exenv}= \CODECOPY  \\
	\access{\mstate}{\stack} = \cons{\pos_{\memo}}{\cons{\pos_\code}{\cons{\siz}{s}}}\\
	\aw = \memext{\access{\mstate}{\actw}}{\pos_\memo}{\siz} \\
	\costs = \costmem{\access{\mstate}{\actw}}{\aw}+3 + 3 \cdot \left \lceil \frac{\siz}{32} \right \rceil\\ 
	\simvalid{\access{\mstate}{\gas}}{\costs}{\size{\access{\mstate}{\stack}}} \\
	k = \cond{(\size{\access{\exenv}{\activecode})} - \pos_\code < 0}{0}{\mini{\size{\access{\exenv}{\code}}- \pos_\code}{\siz}}\\
	d' = \arraypos{\access{\exenv}{\activecode}}{\pos_\code, \pos_\code + k - 1}\\
	d = \concat{d'}{\STOP^{\siz - k}}\\
	\mstate'= \update{\update{\dec{\inc{\update{\mstate}{\stack}{s}}{\pc}{1}}{\gas}{\costs}}{\memo}{\update{\memo}{[\pos_\memo, \pos_\memo + \siz - 1]}{d}}}{\actw}{\aw}
}
{\sstep{\transenv}{\cons{\regstatefull{\mstate}{\exenv}{\gstate}{\transeffects}}{\callstack}}{\cons{\regstatefull{\mstate'}{\exenv}{\gstate}{\transeffects}}{\callstack}}}
\end{mathpar}

\begin{mathpar}
\infer{
(\curropcode{\mstate}{\exenv}= \CODECOPY \lor \curropcode{\mstate}{\exenv} = \CALLDATACOPY)\\
\size{\access{\mstate}{\stack}} < 3} 
{\sstep{\transenv}{\cons{\regstatefull{\mstate}{\exenv}{\gstate}{\transeffects}}{\callstack}}{\cons{\excstate}{\callstack}}}
\end{mathpar}

\begin{mathpar}
\infer{
(\curropcode{\mstate}{\exenv}= \CODECOPY \lor \curropcode{\mstate}{\exenv}= \CALLDATACOPY)\\
	\access{\mstate}{\stack} = \cons{\pos_{\memo}}{\cons{\siz}{\cons{\pos_\code}{s}}}\\
	\aw = \memext{\access{\mstate}{\actw}}{\pos_\memo}{\pos_\code} \\
	\costs = \costmem{\access{\mstate}{\actw}}{\aw}+3 + 3 \cdot \left \lceil \frac{\pos_\code}{32} \right \rceil\\ 
	\neg \simvalid{\access{\mstate}{\gas}}{\costs}{\size{\access{\mstate}{\stack}}}} 
{\sstep{\transenv}{\cons{\regstatefull{\mstate}{\exenv}{\gstate}{\transeffects}}{\callstack}}{\cons{\excstate}{\callstack}}}
\end{mathpar}

\paragraph{Accessing the transaction environment}

\begin{mathpar}
\infer{
\curropcode{\mstate}{\exenv}= \ORIGIN \\
\simvalid{\access{\mstate}{\gas}}{2}{\size{\access{\mstate}{\stack}} +1} \\
\mstate'= \dec{\inc{\update{\mstate}{\stack}{\cons{\access{\transenv}{\originator}}{\access{\mstate}{\stack}}}}{\pc}{1}}{\gas}{2}}
{\sstep{\transenv}{\cons{\regstatefull{\mstate}{\exenv}{\gstate}{\transeffects}}{\callstack}}{\cons{\regstatefull{\mstate'}{\exenv}{\gstate}{\transeffects}}{\callstack}}}
\end{mathpar}

\begin{mathpar}
\infer{
\curropcode{\mstate}{\exenv}= \GASPRICE \\
\simvalid{\access{\mstate}{\gas}}{2}{\size{\access{\mstate}{\stack}} +1} \\
\mstate'= \dec{\inc{\update{\mstate}{\stack}{\cons{\access{\transenv}{\gprice}}{\access{\mstate}{\stack}}}}{\pc}{1}}{\gas}{2}}
{\sstep{\transenv}{\cons{\regstatefull{\mstate}{\exenv}{\gstate}{\transeffects}}{\callstack}}{\cons{\regstatefull{\mstate'}{\exenv}{\gstate}{\transeffects}}{\callstack}}}
\end{mathpar}

\begin{mathpar}
\infer{
(\curropcode{\mstate}{\exenv}= \ORIGIN \lor \curropcode{\mstate}{\exenv}= \GASPRICE) \\
\neg \simvalid{\access{\mstate}{\gas}}{2}{\size{\access{\mstate}{\stack}} +1}}
{\sstep{\transenv}{\cons{\regstatefull{\mstate}{\exenv}{\gstate}{\transeffects}}{\callstack}}{\cons{\excstate}{\callstack}}}
\end{mathpar}

The $\BLOCKHASH$ command writes the hash of one of the 256 most recently completed block (that is specified on the stack) to the stack: 

\begin{mathpar}
\infer{
\curropcode{\mstate}{\exenv}= \BLOCKHASH \\
\simvalid{\access{\mstate}{\gas}}{20}{\size{\access{\mstate}{\stack}}} \\
\access{\mstate}{\stack} = \cons{n}{s} \\ 
h = \funP{\access{\exenv}{\parentc}}{n}{0} \\
\mstate'= \dec{\inc{\update{\mstate}{\stack}{\cons{h}{\access{\mstate}{\stack}}}}{\pc}{1}}{\gas}{20}}
{\sstep{\transenv}{\cons{\regstatefull{\mstate}{\exenv}{\gstate}{\transeffects}}{\callstack}}{\cons{\regstatefull{\mstate'}{\exenv}{\gstate}{\transeffects}}{\callstack}}}
\end{mathpar}

\begin{mathpar}
\infer{
\curropcode{\mstate}{\exenv}= \BLOCKHASH \\
(\neg \simvalid{\access{\mstate}{\gas}}{20}{\size{\access{\mstate}{\stack}}}
\lor \size{\access{\mstate}{\stack}} < 1)}
{\sstep{\transenv}{\cons{\regstatefull{\mstate}{\exenv}{\gstate}{\transeffects}}{\callstack}}{\cons{\excstate}{\callstack}}}
\end{mathpar}

where the function $\funP{h}{n}{a}$ tries to access the block with number $n$ by traversing the block chain starting from $h$ until the counter $a$ reaches the limit of $256$ or the genesis block is reached. 
\begin{align*}
\funP{h}{n}{a} \define 
\begin{cases}
0 & n > \access{h}{\blocknumberc} \lor a = 256 \lor h = 0 \\
h & n = \access{h}{\blocknumberc} \\
\funP{\access{h}{\parentc}}{n}{a + 1} & \text{otherwise}
\end{cases}
\end{align*}

\begin{mathpar}
\infer{
\curropcode{\mstate}{\exenv}= \COINBASE \\
\simvalid{\access{\mstate}{\gas}}{2}{\size{\access{\mstate}{\stack}} +1} \\
\mstate'= \dec{\inc{\update{\mstate}{\stack}{\cons{\access{(\access{\transenv}{\blockheader})}{\beneficiaryc}}{\access{\mstate}{\stack}}}}{\pc}{1}}{\gas}{2}}
{\sstep{\transenv}{\cons{\regstatefull{\mstate}{\exenv}{\gstate}{\transeffects}}{\callstack}}{\cons{\regstatefull{\mstate'}{\exenv}{\gstate}{\transeffects}}{\callstack}}}
\end{mathpar}

\begin{mathpar}
\infer{
\curropcode{\mstate}{\exenv}= \TIMESTAMP\\
\simvalid{\access{\mstate}{\gas}}{2}{\size{\access{\mstate}{\stack}} +1} \\
\mstate'= \dec{\inc{\update{\mstate}{\stack}{\cons{\access{(\access{\transenv}{\blockheader})}{\timestampc}}{\access{\mstate}{\stack}}}}{\pc}{1}}{\gas}{2}}
{\sstep{\transenv}{\cons{\regstatefull{\mstate}{\exenv}{\gstate}{\transeffects}}{\callstack}}{\cons{\regstatefull{\mstate'}{\exenv}{\gstate}{\transeffects}}{\callstack}}}
\end{mathpar}

\begin{mathpar}
\infer{
\curropcode{\mstate}{\exenv}= \NUMBER\\
\simvalid{\access{\mstate}{\gas}}{2}{\size{\access{\mstate}{\stack}} +1} \\
\mstate'= \dec{\inc{\update{\mstate}{\stack}{\cons{\access{(\access{\transenv}{\blockheader})}{\blocknumberc}}{\access{\mstate}{\stack}}}}{\pc}{1}}{\gas}{2}}
{\sstep{\transenv}{\cons{\regstatefull{\mstate}{\exenv}{\gstate}{\transeffects}}{\callstack}}{\cons{\regstatefull{\mstate'}{\exenv}{\gstate}{\transeffects}}{\callstack}}}
\end{mathpar}

\begin{mathpar}
\infer{
\curropcode{\mstate}{\exenv}= \DIFFICULTY\\
\simvalid{\access{\mstate}{\gas}}{2}{\size{\access{\mstate}{\stack}} +1} \\
\mstate'= \dec{\inc{\update{\mstate}{\stack}{\cons{\access{(\access{\transenv}{\blockheader})}{\difficultyc}}{\access{\mstate}{\stack}}}}{\pc}{1}}{\gas}{2}}
{\sstep{\transenv}{\cons{\regstatefull{\mstate}{\exenv}{\gstate}{\transeffects}}{\callstack}}{\cons{\regstatefull{\mstate'}{\exenv}{\gstate}{\transeffects}}{\callstack}}}
\end{mathpar}

\begin{mathpar}
\infer{
\curropcode{\mstate}{\exenv}= \GASLIMIT\\
\simvalid{\access{\mstate}{\gas}}{2}{\size{\access{\mstate}{\stack}} +1} \\
\mstate'= \dec{\inc{\update{\mstate}{\stack}{\cons{\access{(\access{\transenv}{\blockheader})}{\gaslimitc}}{\access{\mstate}{\stack}}}}{\pc}{1}}{\gas}{2}}
{\sstep{\transenv}{\cons{\regstatefull{\mstate}{\exenv}{\gstate}{\transeffects}}{\callstack}}{\cons{\regstatefull{\mstate'}{\exenv}{\gstate}{\transeffects}}{\callstack}}}
\end{mathpar}

\begin{mathpar}
\infer{
(\curropcode{\mstate}{\exenv}= \COINBASE \lor \curropcode{\mstate}{\exenv}= \TIMESTAMP \lor \curropcode{\mstate}{\exenv}= \NUMBER \\
\lor \curropcode{\mstate}{\exenv}= \DIFFICULTY \lor \curropcode{\mstate}{\exenv}= \GASLIMIT)\\
\neg \simvalid{\access{\mstate}{\gas}}{2}{\size{\access{\mstate}{\stack}} +1}}
{\sstep{\transenv}{\cons{\regstatefull{\mstate}{\exenv}{\gstate}{\transeffects}}{\callstack}}{\cons{\excstate}{\callstack}}}
\end{mathpar}

\paragraph{Accessing the global state}

\begin{mathpar}
\infer{
\curropcode{\mstate}{\exenv}= \BALANCE\\
\access{\mstate}{\stack} = \cons{a}{s} \\
\simvalid{\access{\mstate}{\gas}}{400}{\size{s} +1} \\
b = \cond{(\gstate(a \mod 2^{160}) = \accountstate{\accountnoncev}{\accountbalancev}{\accountstorv}{\accountcodev})}{\accountbalancev}{0}\\
\mstate'= \dec{\inc{\update{\mstate}{\stack}{\cons{b}{\access{\mstate}{\stack}}}}{\pc}{1}}{\gas}{400}}
{\sstep{\transenv}{\cons{\regstatefull{\mstate}{\exenv}{\gstate}{\transeffects}}{\callstack}}{\cons{\regstatefull{\mstate'}{\exenv}{\gstate}{\transeffects}}{\callstack}}}
\end{mathpar}

\begin{mathpar}
\infer{
\curropcode{\mstate}{\exenv}= \BALANCE  \\
(\neg \simvalid{\access{\mstate}{\gas}}{400}{\size{\access{\mstate}{\stack}}} 
\lor \size{\access{\mstate}{\stack}} < 1)}
{\sstep{\transenv}{\cons{\regstatefull{\mstate}{\exenv}{\gstate}{\transeffects}}{\callstack}}{\cons{\excstate}{\callstack}}}
\end{mathpar}

\begin{mathpar}
\infer{
\curropcode{\mstate}{\exenv}= \EXTCODESIZE\\
\access{\mstate}{\stack} = \cons{a}{s} \\
\simvalid{\access{\mstate}{\gas}}{700}{\size{s} +1} \\
\siz = \size{\access{\gstate(a \mod 2^{160})}{\accountcode}} \\
\mstate'= \dec{\inc{\update{\mstate}{\stack}{\cons{s}{\access{\mstate}{\stack}}}}{\pc}{1}}{\gas}{700}}
{\sstep{\transenv}{\cons{\regstatefull{\mstate}{\exenv}{\gstate}{\transeffects}}{\callstack}}{\cons{\regstatefull{\mstate'}{\exenv}{\gstate}{\transeffects}}{\callstack}}}
\end{mathpar}

\begin{mathpar}
\infer{
\curropcode{\mstate}{\exenv}= \EXTCODESIZE  \\
(\neg \simvalid{\access{\mstate}{\gas}}{700}{\size{\access{\mstate}{\stack}}} 
\lor \size{\access{\mstate}{\stack}} < 1)}
{\sstep{\transenv}{\cons{\regstatefull{\mstate}{\exenv}{\gstate}{\transeffects}}{\callstack}}{\cons{\excstate}{\callstack}}}
\end{mathpar}

\begin{mathpar}
\infer{
	\curropcode{\mstate}{\exenv}= \EXTCODECOPY  \\
	\access{\mstate}{\stack} = \cons{a}{\cons{\pos_{\memo}}{\cons{\pos_\code}{\cons{\siz}{s}}}}\\
	\code = \access{\gstate(a \mod 2^{160})}{\accountcode} \\
	\aw = \memext{\access{\mstate}{\actw}}{\pos_\memo}{\siz} \\
	\costs = \costmem{\access{\mstate}{\actw}}{\aw}+700 + 3 \cdot \left \lceil \frac{\siz}{32} \right \rceil\\ 
	\simvalid{\access{\mstate}{\gas}}{\costs}{\size{\access{\mstate}{\stack}}} \\
	k = \cond{(\size{\code)} - \pos_\code < 0}{0}{\mini{\size{\access{\exenv}{\code}}- \pos_\code}{\siz}}\\
	d' = \arraypos{\code}{\pos_\code, \pos_\code + k - 1}\\
	d = \concat{d'}{\STOP^{\siz - k}}\\
	\mstate'= \update{\update{\dec{\inc{\update{\mstate}{\stack}{s}}{\pc}{1}}{\gas}{\costs}}{\memo}{\update{\memo}{[\pos_\memo, \pos_\memo + \siz - 1]}{d}}}{\actw}{\aw}
}
{\sstep{\transenv}{\cons{\regstatefull{\mstate}{\exenv}{\gstate}{\transeffects}}{\callstack}}{\cons{\regstatefull{\mstate'}{\exenv}{\gstate}{\transeffects}}{\callstack}}}
\end{mathpar}

\begin{mathpar}
\infer{
\curropcode{\mstate}{\exenv}= \EXTCODECOPY \\
\size{\access{\mstate}{\stack}} < 4} 
{\sstep{\transenv}{\cons{\regstatefull{\mstate}{\exenv}{\gstate}{\transeffects}}{\callstack}}{\cons{\excstate}{\callstack}}}
\end{mathpar}

\begin{mathpar}
\infer{
\curropcode{\mstate}{\exenv}= \EXTCODECOPY \\
	\access{\mstate}{\stack} = \cons{a}{\cons{\pos_{\memo}}{\cons{\siz}{\cons{\pos_\code}{s}}}}\\
	\aw = \memext{\access{\mstate}{\actw}}{\pos_\memo}{\pos_\code} \\
	\costs = \costmem{\access{\mstate}{\actw}}{\aw}+ 700 + 3 \cdot \left \lceil \frac{\pos_\code}{32} \right \rceil\\ 
	\neg \simvalid{\access{\mstate}{\gas}}{\costs}{\size{\access{\mstate}{\stack}}}} 
{\sstep{\transenv}{\cons{\regstatefull{\mstate}{\exenv}{\gstate}{\transeffects}}{\callstack}}{\cons{\excstate}{\callstack}}}
\end{mathpar}

\paragraph{Stack operations}

\begin{mathpar}
\infer{
\curropcode{\mstate}{\exenv}= \POP \\
\simvalid{\access{\mstate}{\gas}}{2}{\size{s}} \\
\access{\mstate}{\stack} = \cons{a}{s} \\
\mstate'= \dec{\inc{\update{\mstate}{\stack}{s}}{\pc}{1}}{\gas}{2}}
{\sstep{\transenv}{\cons{\regstatefull{\mstate}{\exenv}{\gstate}{\transeffects}}{\callstack}}{\cons{\regstatefull{\mstate'}{\exenv}{\gstate}{\transeffects}}{\callstack}}}
\end{mathpar}

\begin{mathpar}
\infer{
\curropcode{\mstate}{\exenv}= \POP \\
(\neg \simvalid{\access{\mstate}{\gas}}{2}{\size{s}} \lor \size{\access{\mstate}{\stack}} < 1)}
{\sstep{\transenv}{\cons{\regstatefull{\mstate}{\exenv}{\gstate}{\transeffects}}{\callstack}}{\cons{\excstate}{\callstack}}}
\end{mathpar}

There are $32$ instructions for pushing values to the stack. We summarize the behavior of all these instructions with the following rules by parameterising the instruction with number of following bytecodes that are pushed to the stack. 
The $\PUSH{n}$ (with $m \in [1, 32]$) command pushes the bytecodes at the next $n$ program counter position to the stack. 

\begin{mathpar}
\infer{
\curropcode{\mstate}{\exenv}= \PUSH{n} \\ 
k = \mini{\size{\access{\exenv}{\code}}}{\access{\mstate}{\pc} + x} \\
\simvalid{\access{\mstate}{\gas}}{3}{\size{\access{\mstate}{\stack}} + 1} \\
d = \arraypos{\access{\exenv}{\code}}{\access{\mstate}{\pc} + 1, k} \\
d' = \concat{d}{0^{8 \cdot (32 - (k - \access{\mstate}{\pc} ))}} \\
\mstate'= \dec{\inc{\update{\mstate}{\stack}{\cons{d'}{\access{\mstate}{\stack}}}}{\pc}{(x+1)}}{\gas}{3}}
{\sstep{\transenv}{\cons{\regstatefull{\mstate}{\exenv}{\gstate}{\transeffects}}{\callstack}}{\cons{\regstatefull{\mstate'}{\exenv}{\gstate}{\transeffects}}{\callstack}}}
\end{mathpar}

\begin{mathpar}
\infer{
\curropcode{\mstate}{\exenv}= \PUSH{n} \\
\neg \simvalid{\access{\mstate}{\gas}}{3}{\size{s} + 1}}
{\sstep{\transenv}{\cons{\regstatefull{\mstate}{\exenv}{\gstate}{\transeffects}}{\callstack}}{\cons{\excstate}{\callstack}}}
\end{mathpar}

The $\DUP{n}$ instructions (with $n \in [1, 16]$) duplicate the $n$th  stack element:  

\begin{mathpar}
\infer{
\curropcode{\mstate}{\exenv}= \DUP{n} \\
\simvalid{\access{\mstate}{\gas}}{3}{\size{\access{\mstate}{\stack}} +1} \\
\access{\mstate}{\stack} = \concatstack{s_1}{(\cons{x_n}{s_2})}\\
\size{s_1} = n-1 \\
\mstate'= \dec{\inc{\update{\mstate}{\stack}{\cons{x_n}{\access{\mstate}{\stack}}}}{\pc}{1}}{\gas}{3}
}
{\sstep{\transenv}{\cons{\regstatefull{\mstate}{\exenv}{\gstate}{\transeffects}}{\callstack}}{\cons{\regstatefull{\mstate'}{\exenv}{\gstate}{\transeffects}}{\callstack}}}
\end{mathpar}

\begin{mathpar}
\infer{
\curropcode{\mstate}{\exenv}= \DUP{n} \\
( \neg \simvalid{\access{\mstate}{\gas}}{3}{\size{\access{\mstate}{\stack}} +1} 
\lor  \size{\access{\mstate}{\stack}} < n)}
{\sstep{\transenv}{\cons{\regstatefull{\mstate}{\exenv}{\gstate}{\transeffects}}{\callstack}}{\cons{\excstate}{\callstack}}}
\end{mathpar}

The $\SWAP{n}$ instructions (with $n \in [1, 16]$) swap the first and the $n$th  stack element: 

\begin{mathpar}
\infer{
\curropcode{\mstate}{\exenv}= \SWAP{n} \\
\simvalid{\access{\mstate}{\gas}}{3}{\size{\access{\mstate}{\stack}}} \\
\access{\mstate}{\stack} = \cons{y}{(\concatstack{s_1}{(\cons{x_n}{s_2})})} \\
\size{s_1} = n-1 \\
\mstate'= \dec{\inc{\update{\mstate}{\stack}{\cons{x_n}{(\concatstack{s_1}{(\cons{y}{s_2})})}}}{\pc}{1}}{\gas}{3}}
{\sstep{\transenv}{\cons{\regstatefull{\mstate}{\exenv}{\gstate}{\transeffects}}{\callstack}}{\cons{\regstatefull{\mstate'}{\exenv}{\gstate}{\transeffects}}{\callstack}}}
\end{mathpar}

\begin{mathpar}
\infer{
\curropcode{\mstate}{\exenv}= \SWAP{n} \\
( \neg \simvalid{\access{\mstate}{\gas}}{3}{\size{\access{\mstate}{\stack}}} 
\lor  \size{\access{\mstate}{\stack}} < n + 1)}
{\sstep{\transenv}{\cons{\regstatefull{\mstate}{\exenv}{\gstate}{\transeffects}}{\callstack}}{\cons{\excstate}{\callstack}}}
\end{mathpar}

\paragraph{Jumps}
The $\JUMP$ command updates the program counter to $i$ (specified in the stack) if $i$ is a valid jump destination. 

\begin{mathpar}
\infer
{\curropcode{\mstate}{\exenv} = \JUMP \\
\simvalid{\access{\mstate}{\gas}}{8}{\size{s}} \\
\access{\mstate}{\stack} = \cons{i}{s} \\
i \in \funD{\access{\exenv}{\code}} \\
\mstate' = \dec{\update{\update{\mstate}{\stack}{s}}{\pc}{i}}{\gas}{8}}
{\sstep{\transenv}{\cons{\regstatefull{\mstate}{\exenv}{\gstate}{\transeffects}}{\callstack}}{\cons{\regstatefull{\mstate'}{\exenv}{\gstate}{\transeffects}}{\callstack}}}
\end{mathpar}

\begin{mathpar}
\infer
{\curropcode{\mstate}{\exenv} = \JUMP \\
\access{\mstate}{\stack} = \cons{i}{s} \\
(i \not \in \funD{\access{\exenv}{\code}} \lor 
\neg \simvalid{\access{\mstate}{\gas}}{8}{\size{s}})}
{\sstep{\transenv}{\cons{\regstatefull{\mstate}{\exenv}{\gstate}{\transeffects}}{\callstack}}{\cons{\excstate}{\callstack}}}
\end{mathpar}

\begin{mathpar}
\infer{
\curropcode{\mstate}{\exenv}= \JUMP \\
 \size{\access{\mstate}{\stack}} < 1}
{\sstep{\transenv}{\cons{\regstatefull{\mstate}{\exenv}{\gstate}{\transeffects}}{\callstack}}{\cons{\excstate}{\callstack}}}
\end{mathpar}

The conditional jump command $\JUMPI$ conditionally jumps to position $i$ depending on $b$. 

\begin{mathpar}
\infer
{\curropcode{\mstate}{\exenv} = \JUMPI \\
\simvalid{\access{\mstate}{\gas}}{10}{\size{s}} \\
\access{\mstate}{\stack} = \cons{i}{\cons{b}{s}} \\
i \in \funD{\access{\exenv}{\code}} \\
j = \cond{(b=0)}{\access{\mstate}{\pc}+1}{i} \\
\mstate' = \dec{\update{\update{\mstate}{\stack}{s}}{\pc}{j}}{\gas}{10}}
{\sstep{\transenv}{\cons{\regstatefull{\mstate}{\exenv}{\gstate}{\transeffects}}{\callstack}}{\cons{\regstatefull{\mstate'}{\exenv}{\gstate}{\transeffects}}{\callstack}}}
\end{mathpar}

\begin{mathpar}
\infer
{\curropcode{\mstate}{\exenv} = \JUMPI \\
\access{\mstate}{\stack} = \cons{i}{\cons{b}{s}} \\
(i \not \in \funD{\access{\exenv}{\code}} \lor 
\neg \simvalid{\access{\mstate}{\gas}}{10}{\size{s}})}
{\sstep{\transenv}{\cons{\regstatefull{\mstate}{\exenv}{\gstate}{\transeffects}}{\callstack}}{\cons{\excstate}{\callstack}}}
\end{mathpar}

\begin{mathpar}
\infer{
\curropcode{\mstate}{\exenv}= \JUMPI \\
 \size{\access{\mstate}{\stack}} < 2}
{\sstep{\transenv}{\cons{\regstatefull{\mstate}{\exenv}{\gstate}{\transeffects}}{\callstack}}{\cons{\excstate}{\callstack}}}
\end{mathpar}

The $\JUMPDEST$ command marks a valid jump destination. It does not trigger any execution and consequently the only effect of the command is the increasing of the program counter and charging the fee for the command execution.  

\begin{mathpar}
\infer
{\curropcode{\mstate}{\exenv} = \JUMPDEST \\
\simvalid{\access{\mstate}{\gas}}{1}{\size{\access{\mstate}{\stack}}} \\
\mstate' = \dec{\inc{\mstate}{\pc}{1}}{\gas}{1}
}
{\sstep{\transenv}{\cons{\regstatefull{\mstate}{\exenv}{\gstate}{\transeffects}}{\callstack}}{\cons{\regstatefull{\mstate'}{\exenv}{\gstate}{\transeffects}}{\callstack}}}
\end{mathpar}

\begin{mathpar}
\infer{
\curropcode{\mstate}{\exenv}= \JUMPDEST \\
\neg \simvalid{\access{\mstate}{\gas}}{1}{\size{\access{\mstate}{\stack}}}}
{\sstep{\transenv}{\cons{\regstatefull{\mstate}{\exenv}{\gstate}{\transeffects}}{\callstack}}{\cons{\excstate}{\callstack}}}
\end{mathpar}

\paragraph{Local memory operations}
The $\MLOAD$ command reads a fraction of the local memory specified by $a$ and pushes it to the stack. Note that this increases the number of active words in memory and therefore causes additional cost. 

\begin{mathpar}
\infer{
\curropcode{\mstate}{\exenv}= \MLOAD \\
c = \costmem{\access{\mstate}{\actw}}{\aw} + 3 \\
\simvalid{\access{\mstate}{\gas}}{c}{\size{s} + 1} \\ 
\access{\mstate}{\stack} = \cons{a}{s} \\
v = \access{\mstate}{\memo}[a, a +31]\\
\aw = \memext{\access{\mstate}{\actw}}{a}{32} \\
\mstate'= \dec{\inc{\update{\update{\mstate}{\actw}{\aw}}{\stack}{\cons{v}{s}}}{\pc}{1}}{\gas}{c}}
{\sstep{\transenv}{\cons{\regstatefull{\mstate}{\exenv}{\gstate}{\transeffects}}{\callstack}}{\cons{\regstatefull{\mstate'}{\exenv}{\gstate}{\transeffects}}{\callstack}}}
\end{mathpar}

\begin{mathpar}
\infer{
\curropcode{\mstate}{\exenv}= \MLOAD \\
 \size{\access{\mstate}{\stack}} < 1}
{\sstep{\transenv}{\cons{\regstatefull{\mstate}{\exenv}{\gstate}{\transeffects}}{\callstack}}{\cons{\excstate}{\callstack}}}
\end{mathpar}

\begin{mathpar}
\infer{
\curropcode{\mstate}{\exenv}= \MLOAD \\
\access{\mstate}{\stack} = \cons{a}{s} \\
\aw = \memext{\access{\mstate}{\actw}}{a}{32} \\
c = \costmem{\access{\mstate}{\actw}}{\aw} + 3 \\
\neg \simvalid{\access{\mstate}{\gas}}{c}{\size{s} + 1}}
{\sstep{\transenv}{\cons{\regstatefull{\mstate}{\exenv}{\gstate}{\transeffects}}{\callstack}}{\cons{\excstate}{\callstack}}}
\end{mathpar}

The $\MSTORE$ command writes a value $b$ given at the stack to address $a$ in the local memory. 
\textit{Notice that we abuse the update-notation here slightly to update whole intervals of the local memory}

\begin{mathpar}
\infer{
\curropcode{\mstate}{\exenv}= \MSTORE \\
c = \costmem{\access{\mstate}{\actw}}{\aw} + 3 \quad 
\access{\mstate}{\stack} = \cons{a}{\cons{b}{s}} \\
\simvalid{\access{\mstate}{\gas}}{c}{\size{s}} \\
\aw = \memext{\access{\mstate}{\actw}}{a}{32} \\
\mstate'= \dec{\inc{\update{\update{\update{\mstate}{\memo}{\update{\access{\mstate}{\memo}}{[a, a+31]}{\bitstringtobytearray{b}}}}{\actw}{\aw}}{\stack}{s}}{\pc}{1}}{\gas}{c}}
{\sstep{\transenv}{\cons{\regstatefull{\mstate}{\exenv}{\gstate}{\transeffects}}{\callstack}}{\cons{\regstatefull{\mstate'}{\exenv}{\gstate}{\transeffects}}{\callstack}}}
\end{mathpar}

\begin{mathpar}
\infer{
\curropcode{\mstate}{\exenv}= \MSTORE \\
\access{\mstate}{\stack} = \cons{a}{\cons{b}{s}} \\
\aw = \memext{\access{\mstate}{\actw}}{a}{32} \\
c = \costmem{\access{\mstate}{\actw}}{\aw} + 3 \\
\neg \simvalid{\access{\mstate}{\gas}}{c}{\size{s}}}
{\sstep{\transenv}{\cons{\regstatefull{\mstate}{\exenv}{\gstate}{\transeffects}}{\callstack}}{\cons{\excstate}{\callstack}}}
\end{mathpar}

\begin{mathpar}
\infer{
\curropcode{\mstate}{\exenv}= \MSTOREByte \\
c = \costmem{\access{\mstate}{\actw}}{\aw} + 3 \quad 
\access{\mstate}{\stack} = \cons{a}{\cons{b}{s}} \\
\simvalid{\access{\mstate}{\gas}}{c}{\size{s}} \\
\aw = \memext{\access{\mstate}{\actw}}{a}{1} \\
\mstate'= \dec{\inc{\update{\update{\update{\mstate}{\memo}{\update{\access{\mstate}{\memo}}{a}{b \mod 256}}}{\actw}{\aw}}{\stack}{s}}{\pc}{1}}{\gas}{c}}
{\sstep{\transenv}{\cons{\regstatefull{\mstate}{\exenv}{\gstate}{\transeffects}}{\callstack}}{\cons{\regstatefull{\mstate'}{\exenv}{\gstate}{\transeffects}}{\callstack}}}
\end{mathpar}

\begin{mathpar}
\infer{
\curropcode{\mstate}{\exenv}= \MSTOREByte \\
\access{\mstate}{\stack} = \cons{a}{\cons{b}{s}} \\
\aw = \memext{\access{\mstate}{\actw}}{a}{1} \\
c = \costmem{\access{\mstate}{\actw}}{\aw} + 3 \\
\neg \simvalid{\access{\mstate}{\gas}}{c}{\size{s}}}
{\sstep{\transenv}{\cons{\regstatefull{\mstate}{\exenv}{\gstate}{\transeffects}}{\callstack}}{\cons{\excstate}{\callstack}}}
\end{mathpar}

\begin{mathpar}
\infer{
(\curropcode{\mstate}{\exenv}= \MSTORE 
\lor  \curropcode{\mstate}{\exenv}= \MSTOREByte)\\
 \size{\access{\mstate}{\stack}} < 2}
{\sstep{\transenv}{\cons{\regstatefull{\mstate}{\exenv}{\gstate}{\transeffects}}{\callstack}}{\cons{\excstate}{\callstack}}}
\end{mathpar}

\paragraph{Persistent storage operations}

The $\SLOAD$ command reads the executing account's persistent storage at position $a$. 
\begin{mathpar}
\infer{
\curropcode{\mstate}{\exenv}= \SLOAD \\
\simvalid{\access{\mstate}{\gas}}{200}{\size{s} + 1} \\
\access{\mstate}{\stack} = \cons{a}{s} \\
\mstate' = \dec{\inc{\update{\mstate}{\stack}{\cons{(\access{\gstate(\access{\exenv}{\addr})}{\accountstor})(a)}{s}}}{\pc}{1}}{\gas}{200}
}
{\sstep{\transenv}{\cons{\regstatefull{\mstate}{\exenv}{\gstate}{\transeffects}}{\callstack}}{\cons{\regstatefull{\mstate'}{\exenv}{\gstate}{\transeffects}}{\callstack}}}
\end{mathpar}

\begin{mathpar}
\infer{
\curropcode{\mstate}{\exenv}= \SLOAD \\
(\neg \simvalid{\access{\mstate}{\gas}}{200}{\size{s} + 1} \lor \size{\access{\mstate}{\stack}} < 1)}
{\sstep{\transenv}{\cons{\regstatefull{\mstate}{\exenv}{\gstate}{\transeffects}}{\callstack}}{\cons{\excstate}{\callstack}}}
\end{mathpar}

The $\SSTORE$ command stores the value $b$ in the executing account's persistent storage at position $a$. 
\begin{mathpar}
\infer{
\curropcode{\mstate}{\exenv}= \SSTORE \\
c = \cond{(b \neq 0 \land (\access{\gstate(\access{\exenv}{\addr})}{\accountstor})(a) = 0)}{20000}{5000} \\
\simvalid{\access{\mstate}{\gas}}{c}{\size{s}} \\
\access{\mstate}{\stack} = \cons{a}{\cons{b}{s}} \\
\mstate'  = \dec{\inc{\update{\mstate}{\stack}{s}}{\pc}{1}}{\gas}{c} \\
\gstate'  = \updategstate{\gstate}{\access{\exenv}{\addr}}{\update{\access{\exenv}{\addr}}{\accountstor}{\update{\access{\gstate(\access{\exenv}{\addr})}{\accountstor}}{a}{b}}} \\
r = \cond{(b = 0 \land (\access{\gstate(\access{\exenv}{\addr})}{\accountstor})(a) \neq 0)}{15000}{0}  \\
\transeffects' = \inc{\transeffects}{\refundbalance}{r}}
{\sstep{\transenv}{\cons{\regstatefull{\mstate}{\exenv}{\gstate}{\transeffects}}{\callstack}}{\cons{\regstatefull{\mstate'}{\exenv}{\gstate'}{\transeffects'}}{\callstack}}}
\end{mathpar}

\begin{mathpar}
\infer{
\curropcode{\mstate}{\exenv}= \SSTORE \\
\access{\mstate}{\stack} = \cons{a}{\cons{b}{s}} \\
c = \cond{(b \neq 0 \land (\access{\gstate(\access{\exenv}{\addr})}{\accountstor})(a) = 0)}{20000}{5000} \\
\neg \simvalid{\access{\mstate}{\gas}}{c}{\size{s}}}
{\sstep{\transenv}{\cons{\regstatefull{\mstate}{\exenv}{\gstate}{\transeffects}}{\callstack}}{\cons{\excstate}{\callstack}}}
\end{mathpar}

\begin{mathpar}
\infer{
\curropcode{\mstate}{\exenv}= \SSTORE \\
 \size{\access{\mstate}{\stack}} < 2}
{\sstep{\transenv}{\cons{\regstatefull{\mstate}{\exenv}{\gstate}{\transeffects}}{\callstack}}{\cons{\excstate}{\callstack}}}
\end{mathpar}

\paragraph{Accessing the machine state}

\begin{mathpar}
\infer{
\curropcode{\mstate}{\exenv}= \PC \\
\simvalid{\access{\mstate}{\gas}}{2}{\size{\access{\mstate}{\stack}} +1} \\
\mstate'= \dec{\inc{\update{\mstate}{\stack}{\cons{\access{\mstate}{\pc}}{\access{\mstate}{\stack}}}}{\pc}{1}}{\gas}{2}}
{\sstep{\transenv}{\cons{\regstatefull{\mstate}{\exenv}{\gstate}{\transeffects}}{\callstack}}{\cons{\regstatefull{\mstate'}{\exenv}{\gstate}{\transeffects}}{\callstack}}}
\end{mathpar}

\begin{mathpar}
\infer{
\curropcode{\mstate}{\exenv}= \MSIZE \\
\simvalid{\access{\mstate}{\gas}}{2}{\size{\access{\mstate}{\stack}} +1} \\
\mstate'= \dec{\inc{\update{\mstate}{\stack}{\cons{32 \cdot \access{\mstate}{\actw}}{\access{\mstate}{\stack}}}}{\pc}{1}}{\gas}{2}}
{\sstep{\transenv}{\cons{\regstatefull{\mstate}{\exenv}{\gstate}{\transeffects}}{\callstack}}{\cons{\regstatefull{\mstate'}{\exenv}{\gstate}{\transeffects}}{\callstack}}}
\end{mathpar}

\begin{mathpar}
\infer{
\curropcode{\mstate}{\exenv}= \GAS \\
\simvalid{\access{\mstate}{\gas}}{2}{\size{\access{\mstate}{\stack}} +1} \\
\mstate'= \dec{\inc{\update{\mstate}{\stack}{\cons{\access{\mstate}{\gas}}{\access{\mstate}{\stack}}}}{\pc}{1}}{\gas}{2}}
{\sstep{\transenv}{\cons{\regstatefull{\mstate}{\exenv}{\gstate}{\transeffects}}{\callstack}}{\cons{\regstatefull{\mstate'}{\exenv}{\gstate}{\transeffects}}{\callstack}}}
\end{mathpar}

\begin{mathpar}
\infer{
(\curropcode{\mstate}{\exenv}= \PC \lor \curropcode{\mstate}{\exenv}= \MSIZE \lor \curropcode{\mstate}{\exenv}= \GAS) \\
\neg \simvalid{\access{\mstate}{\gas}}{2}{\size{\access{\mstate}{\stack}} +1}}
{\sstep{\transenv}{\cons{\regstatefull{\mstate}{\exenv}{\gstate}{\transeffects}}{\callstack}}{\cons{\excstate}{\callstack}}}
\end{mathpar}

\paragraph{Logging instructions}

The logging operation allows to append a new log entry to the log series. The log series keeps track of archived and indexable ‘checkpoints’ in the execution of Ethereum byte code. The motivation of the log series is to allow external observers to track the program execution. 
A log entry consists of the address of the currently executing account, up to for 'topics' (specified on stack) and a fraction of the memory. 
There are four logging instructions, but as seen before we describe their effects using common rules parameterising the instruction by the amount of log information read from the stack. 

\begin{mathpar}
\infer{
\curropcode{\mstate}{\exenv}= \LOG{n} \\
\access{\mstate}{\stack} = \cons{\pos_\memo}{\cons{\siz}{(\concatstack{s_1}{s_2})}}\\
\size{s_1} = n \\
\aw = \memext{\access{\mstate}{\actw}}{\pos_\memo}{\siz} \\ 
\costs = \costmem{\access{\mstate}{\actw}}{\aw} + 375 + 8 \cdot \siz + n  \cdot 375 \\ 
\simvalid{\access{\mstate}{\gas}}{\costs}{\size{\access{\mstate}{\stack}}} \\
\mstate'= \update{\dec{\inc{\update{\mstate}{\stack}{s}}{\pc}{1}}{\gas}{\costs}}{\actw}{\aw} \\
d= \access{\mstate}{\memo}[\pos_\memo, \pos_\memo + \siz -1] \\
\transeffects' = \update{\transeffects}{\logseq}{\concatstack{\access{\transeffects}{\logseq}}{[(\access{\exenv}{\activeaccount}, s_1, d)]}}}
{\sstep{\transenv}{\cons{\regstatefull{\mstate}{\exenv}{\gstate}{\transeffects}}{\callstack}}{\cons{\regstatefull{\mstate'}{\exenv}{\gstate}{\transeffects'}}{\callstack}}}
\end{mathpar}

\begin{mathpar}
\infer{
\curropcode{\mstate}{\exenv}= \LOG{n} \\
\access{\mstate}{\stack} = \cons{\pos_\memo}{\cons{\siz}{(\concatstack{s_1}{s_2})}}\\
\size{s_1} = n \\
\aw = \memext{\access{\mstate}{\actw}}{\pos_\memo}{\siz} \\ 
\costs = \costmem{\access{\mstate}{\actw}}{\aw} + 375 + 8 \cdot \siz + n  \cdot 375 \\ 
\neg \simvalid{\access{\mstate}{\gas}}{\costs}{\size{\access{\mstate}{\stack}}} }
{\sstep{\transenv}{\cons{\regstatefull{\mstate}{\exenv}{\gstate}{\transeffects}}{\callstack}}{\cons{\excstate}{\callstack}}}
\end{mathpar}

\begin{mathpar}
\infer{
\curropcode{\mstate}{\exenv}= \LOG{n} \\
\size{\access{\mstate}{\stack}} < n + 2 }
{\sstep{\transenv}{\cons{\regstatefull{\mstate}{\exenv}{\gstate}{\transeffects}}{\callstack}}{\cons{\excstate}{\callstack}}}
\end{mathpar}

\paragraph{Halting instructions}
The execution of a $\RETURN$ command requires to read data from the local memory. Consequently the cost for memory consumption is charged. Additionally the read data is recorded in the halting state in order to potentially propagate it to the caller. 

\begin{mathpar}
\infer
{\curropcode{\mstate}{\exenv}= \RETURN \\
\access{\mstate}{\stack} = \cons{\io}{\cons{\is}{s}} \\
\aw = \memext{\access{\mstate}{\actw}}{io}{is} \\
\costs = \costmem{\access{\mstate}{\actw}}{aw}  \\
\simvalid{\access{\mstate}{\gas}}{c}{\size{s}} \\
d = \access{\mstate}{\memo}[\io, \io+ \is + 1]  \\
g = \access{\mstate}{\gas} - \costs}
{\sstep{\transenv}{\cons{\regstatefull{\mstate}{\exenv}{\gstate}{\transeffects}}{\callstack}}{\cons{\haltstatefull{\gstate}{g}{d}{\transeffects}}{\callstack}}}
\end{mathpar}

\begin{mathpar}
\infer
{\curropcode{\mstate}{\exenv}= \RETURN \\
\access{\mstate}{\stack} = \cons{\io}{\cons{\is}{s}} \\
\aw = \memext{\access{\mstate}{\actw}}{io}{is} \\
\costs = \costmem{\access{\mstate}{\actw}}{aw}  \\
\neg \simvalid{\access{\mstate}{\gas}}{c}{\size{s}}}
{\sstep{\transenv}{\cons{\regstatefull{\mstate}{\exenv}{\gstate}{\transeffects}}{\callstack}}{\cons{\excstate}{\callstack}}}
\end{mathpar}

\begin{mathpar}
\infer{
\curropcode{\mstate}{\exenv}= \RETURN \\
\size{\access{\mstate}{\stack}} < 2 }
{\sstep{\transenv}{\cons{\regstatefull{\mstate}{\exenv}{\gstate}{\transeffects}}{\callstack}}{\cons{\excstate}{\callstack}}}
\end{mathpar}

The execution of a $\STOP$ command halts execution without propagating any data to the caller. 

\begin{mathpar}
\infer{
\curropcode{\mstate}{\exenv} = \STOP \\
g = \access{\mstate}{\gas}}
{\sstep{\transenv}{\cons{\regstatefull{\mstate}{\exenv}{\gstate}{\transeffects}}{\callstack}}{\cons{\haltstatefull{\gstate}{g}{\emptyarray}{\transeffects}}{\callstack}}}
\end{mathpar}

The $\SUICIDE$ instruction deletes the currently executing account. The $\SUICIDE$ command takes one argument from the stack specifying $a_{\textit{ben}}$ the address of the beneficiary that should get the balance of the suiciding account. 

We distinguish the cases where the beneficiary is an existing account and where it still needs to be created. In the latter an additional fee is charged. 

\begin{mathpar}
\infer{
\curropcode{\mstate}{\exenv} = \SUICIDE \\
\access{\mstate}{\stack} = \cons{a_\textit{ben}}{s} \\
a = a_\textit{ben} \mod  2^{160} \\
\gstate(a) \neq \bot \\
\simvalid{\access{\mstate}{\gas}}{5000}{\size{s}} \\
g = \access{\mstate}{\gas} - 5000 \\
\gstate' = \updategstate{\updategstate{\gstate}{\access{\exenv}{\activeaccount}}{\update{\gstate(\access{\exenv}{\activeaccount})}{\accountbalance}{0}}}{a}{\inc{\gstate(a)}{\accountbalance}{\access{\access{\gstate}{(\access{\exenv}{\activeaccount})}}{\accountbalance}}} \\ 
r = \cond{(\access{\exenv}{\activeaccount} \in \access{\transenv}{\suicideset})}{24000}{0} \\
\transeffects' = \inc{\update{\transeffects}{\suicidesetc}{\access{\transeffects}{\suicidesetc} \cup \{ \access{\exenv}{\activeaccount} \} }}{\refundbalance}{r}
}
{\sstep{\transenv}{\cons{\regstatefull{\mstate}{\exenv}{\gstate}{\transeffects}}{\callstack}}{\cons{\haltstatefull{\gstate'}{g}{\emptyarray}{\transeffects'}}{\callstack}}}
\end{mathpar}


\begin{mathpar}
\infer{
\curropcode{\mstate}{\exenv} = \SUICIDE \\
\access{\mstate}{\stack} = \cons{a_\textit{ben}}{s} \\
a = a_\textit{ben} \mod  2^{160} \\
\gstate(a) = \bot \\
\simvalid{\access{\mstate}{\gas}}{37000}{\size{s}} \\
g = \access{\mstate}{\gas} - 37000 \\
\gstate' = \updategstate{\updategstate{\gstate}{\access{\exenv}{\activeaccount}}{\update{\gstate(\access{\exenv}{\activeaccount})}{\accountbalance}{0}}}{a}{\account{0}{\access{\gstate(\activeaccount)}{\accountbalance}}{\lam{x}{0}}{\emptyarray}} \\ 
r = \cond{(\access{\exenv}{\activeaccount} \in \access{\transenv}{\suicideset})}{0}{24000} \\
\transeffects' = \inc{\update{\transeffects}{\suicidesetc}{\access{\transeffects}{\suicidesetc} \cup \{ \access{\exenv}{\activeaccount} \} }}{\refundbalance}{r}
}
{\sstep{\transenv}{\cons{\regstatefull{\mstate}{\exenv}{\gstate}{\transeffects}}{\callstack}}{\cons{\haltstatefull{\gstate'}{g}{\emptyarray}{\transeffects'}}{\callstack}}}
\end{mathpar}

\begin{mathpar}
\infer{
\curropcode{\mstate}{\exenv}= \SUICIDE \\
\access{\mstate}{\stack} = \cons{a_\textit{ben}}{s} \\
a = a_\textit{ben} \mod  2^{160} \\
\costs = \cond{(\gstate(a) = \none)}{37000}{5000} \\
\neg \simvalid{\access{\mstate}{\gas}}{\costs}{\size{s}}}
{\sstep{\transenv}{\cons{\regstatefull{\mstate}{\exenv}{\gstate}{\transeffects}}{\callstack}}{\cons{\excstate}{\callstack}}}
\end{mathpar}

\begin{mathpar}
\infer{
\curropcode{\mstate}{\exenv}= \SUICIDE \\
\size{\access{\mstate}{\stack}} < 1}
{\sstep{\transenv}{\cons{\regstatefull{\mstate}{\exenv}{\gstate}{\transeffects}}{\callstack}}{\cons{\excstate}{\callstack}}}
\end{mathpar}


There is a designated invalid instruction that always causes an exception 

\begin{mathpar}
\infer{
\curropcode{\mstate}{\exenv}= \INVALID}
{\sstep{\transenv}{\cons{\regstatefull{\mstate}{\exenv}{\gstate}{\transeffects}}{\callstack}}{\cons{\excstate}{\callstack}}}
\end{mathpar}

\paragraph{Calling}

The $\CALL$ command initiates a the execution of a (potentially different) account's code. To this end, it gets as parameters the gas $g$ to be spent on the execution, the address $\recipient$ of the destination account, the value $\valu$ to be transferred to the destination account. Additionally a fragment in the local memory containing input data for the called code is specified (by $\io$ and $\is$) and another fragment where the return values of the call are expected (specified by $\oo$ and $\os$). 
If the recipient $\recipient$ exists, the balance of the calling account $\access{\exenv}{\activeaccount}$ is sufficient to transfer $\valu$ and the call stack limit is not reached yet,  the recipient $\recipient$ gets the value $\valu$ transferred from the calling account $\access{\exenv}{\activeaccount}$. The input data $\inputdata$ to the call are read from the local memory and written to the execution environment. Additionally the execution environment is updated with the information on the originator $\sender$, the owner of the currently executed code $\activeaccount$ and the code to be executed (that is the code of the called account). 
The execution of the called code then starts in the updated execution environment and with an empty machine state. 

We introduce some functions for simplifying the cost calculations. 
First, we introduce a function that calculates the base costs for executing a $\CALL$ command (not including costs for memory consumption and the amount of gas given to the callee).

\begin{align*}
\basecosts{\valu}{\flag} &= 700 + (\cond{\valu = 0}{0}{6500}) +(\cond{\flag = 0}{25000}{0})
\end{align*}

The base costs include a fixed amount ($700$ wei) for calling and additional fees depending on whether ether is transferred or a new account needs to get created. 

Next, we introduce the function computing the amount of wei given to a call. This value depends on the amount of ether transferred during the call, on the amount of gas specified on the stack that should be given to the call as well as on the amount of local gas still available to the caller and the fact whether a new contract needs to be created or not. 
\begin{align*}
\gascapacity{\valu}{\flag}{g}{\lgas} &= \\
& \textit{let} \; c_\textit{ex} = 700 + (\cond{\valu = 0}{0}{9000}) + (\cond{\flag = 0}{25000}{0}) \\
&  \textit{in} \; (\cond{c_{\textit{ex}} > \lgas}{g}{\mini{g}{\funL{\lgas - c_{\textit{ex}}}}}) + (\cond{\valu=0}{0}{2300})
\end{align*}
The information on the transfer value and the existence of the called account influence the amount of fixed costs the caller needs to pay for the call independent of the execution of the callee contract. 
Actually the amount of gas specified on the stack should be given to the callee, but if the local gas runs too low (namely if the fixed amount to pay already uses too much of the callee's local gas) instead only a predefined fraction of the local gas is given to the call. 

We distinguish the cases where a new account needs to get created as the called address does not belong to an existing account and the one where the called account is existing. 

First we consider the case where the called account already exists: 
\begin{mathpar}
\infer{
\curropcode{\mstate}{\exenv}=\CALL  \\
\access{\mstate}{\stack}=\cons{g}{\cons{\recipient}{\cons{\valu}{\cons{\io}{\cons{\is}{\cons{\oo}{\cons{\os}{s}}}}}}}   \\
\recipient_a = \recipient \mod 2^{160} \\
\gstate (\recipient_a) \neq \none\\ 
\size{A} + 1 \leq 1024 \\
\access{\getaccount{\gstate}{\access{\exenv}{\activeaccount}}}{\balance} \geq \valu \\
\aw = \memext{\memext{\access{\mstate}{\actw}}{\io}{\is}}{\oo}{\os} \\ 
c_{\textit{call}} = \gascapacity{\valu}{1}{g}{\access{\mstate}{\gas}} \\
c = \basecosts{\valu}{1} + \costmem{\access{\mstate}{\actw}}{\aw} +c_{\textit{call}} \\
\simvalid{\access{\mstate}{\gas}}{c}{\size{s} + 1} \\
\gstate' = \updategstate{\updategstate{\gstate}{\recipient_a}{\inc{\getaccount{\gstate}{\recipient_a}}{\balance}{\valu}}}{\access{\exenv}{\activeaccount}}{\dec{\getaccount{\gstate}{\access{\exenv}{\activeaccount}}}{\balance}{\valu}} \\
d =\getinterval{\access{\mstate}{\memo}}{\io}{\io + \is -1}  \\ 
\mstate' =\smstate{c_{\textit{call}}}{0}{\emptymemory}{0}{\emptystack}  \\ 
\exenv' =\update{\update{\update{\update{\update{\exenv}{\sender}{\access{\exenv}{\activeaccount}}}{\activeaccount}{\recipient_a}}{\tvalue}{\valu}}{\inputdata}{d}}{\activecode}{\access{\getaccount{\gstate}{\recipient_a}}{\accountcode}}\\
} 
{\sstep{\transenv}{\cons{\regstatefull{\mstate}{\exenv}{\gstate}{\transeffects}}{\callstack}}{\cons{\regstatefull{\mstate'}{\exenv'}{\gstate'}{\transeffects}}{\cons{\regstatefull{\mstate}{\exenv}{\gstate}{\transeffects}}{\callstack}}}}
\end{mathpar}

Next, we consider the case where the called account does not exist. In this case an account with the called address (and the empty code) gets created in executed. 

\begin{mathpar}
\infer{
\curropcode{\mstate}{\exenv}=\CALL  \\
\access{\mstate}{\stack}=\cons{g}{\cons{\recipient}{\cons{\valu}{\cons{\io}{\cons{\is}{\cons{\oo}{\cons{\os}{s}}}}}}}   \\
\recipient_a = \recipient \mod 2^{160} \\
\getaccount{\gstate}{\recipient_a} = \none\\ 
\size{A} + 1 \leq 1024 \\
\access{\getaccount{\gstate}{\access{\exenv}{\activeaccount}}}{\balance} \geq \valu \\
\aw = \memext{\memext{\access{\mstate}{\actw}}{\io}{\is}}{\oo}{\os} \\ 
c_{\textit{call}} = \gascapacity{\valu}{0}{g}{\access{\mstate}{\gas}} \\
c = \basecosts{\valu}{0} + \costmem{\access{\mstate}{\actw}}{\aw} +c_{\textit{call}} \\
\simvalid{\access{\mstate}{\gas}}{c}{\size{s} + 1} \\
\gstate' = \updategstate{\updategstate{\gstate}{\recipient_a}{\account{0}{\valu}{\lam{x}{0}}{\emptyarray}}}{\access{\exenv}{\activeaccount}}{\dec{\getaccount{\gstate}{\access{\exenv}{\activeaccount}}}{\balance}{\valu}} \\
d =\getinterval{\access{\mstate}{\memo}}{\io}{\io + \is -1}  \\ 
\mstate' =\smstate{c_{\textit{call}}}{0}{\emptymemory}{0}{\emptystack}  \\ 
\exenv' =\update{\update{\update{\update{\update{\exenv}{\sender}{\access{\exenv}{\activeaccount}}}{\activeaccount}{\recipient_a}}{\tvalue}{\valu}}{\inputdata}{d}}{\activecode}{\emptyarray}\\
} 
{\sstep{\transenv}{\cons{\regstatefull{\mstate}{\exenv}{\gstate}{\transeffects}}{\callstack}}{\cons{\regstatefull{\mstate'}{\exenv'}{\gstate'}{\transeffects}}{\cons{\regstatefull{\mstate}{\exenv}{\gstate}{\transeffects}}{\callstack}}}}
\end{mathpar}

If the executing account $\access{\exenv}{\activeaccount}$ does not hold the amount of wei specified to be transferred by the $\CALL$ instruction ($\valu$) or if the call stack limit of $1024$ would be reached by performing the call, the call does not get executed. In the small step semantics this is modelled by throwing an exception on the callee level. 

\begin{mathpar}
\infer{
\curropcode{\mstate}{\exenv} =\CALL \\
\access{\mstate}{\stack}=\cons{g}{\cons{\recipient}{\cons{\valu}{\cons{\io}{\cons{\is}{\cons{\oo}{\cons{\os}{s}}}}}}}   \\
\recipient_a = \recipient \mod 2^{160} \\
\flag = \cond{(\getaccount{\gstate}{\recipient_a} = \none)}{0}{1} \\ 
\aw = \memext{\memext{\access{\mstate}{\actw}}{\io}{\is}}{\oo}{\os} \\
c_{\textit{call}} = \gascapacity{\valu}{\flag}{g}{\access{\mstate}{\gas}} \\
c = \basecosts{\valu}{\flag} + \costmem{\access{\mstate}{\actw}}{\aw} +c_{\textit{call}} \\
\simvalid{\access{\mstate}{\gas}}{c}{\size{s} + 1} \\
(\valu> \access{\getaccount{\gstate}{(\access{\exenv}{\activeaccount})}}{\accountbalance} \lor \size{A} +1 \geq 1024)}
{\sstep{\transenv}{\cons{\regstatefull{\mstate}{\exenv}{\gstate}{\transeffects}}{\callstack}}{\cons{\excstate}{\cons{\regstatefull{\mstate}{\exenv}{\gstate}{\transeffects}}{\callstack}}}}
\end{mathpar}

If the execution runs out of gas or the stack limit is exceeded, an exception is thrown: 
\begin{mathpar}
\infer{
\curropcode{\mstate}{\exenv} =\CALL \\
\access{\mstate}{\stack}=\cons{g}{\cons{\recipient}{\cons{\valu}{\cons{\io}{\cons{\is}{\cons{\oo}{\cons{\os}{s}}}}}}}   \\
\recipient_a = \recipient \mod 2^{160} \\
\flag = \cond{(\getaccount{\gstate}{\recipient_a} = \none)}{0}{1} \\ 
\aw = \memext{\memext{\access{\mstate}{\actw}}{\io}{\is}}{\oo}{\os} \\
c_{\textit{call}} = \gascapacity{\valu}{\flag}{g}{\access{\mstate}{\gas}} \\
c = \basecosts{\valu}{\flag} + \costmem{\access{\mstate}{\actw}}{\aw} +c_{\textit{call}} \\
\neg \simvalid{\access{\mstate}{\gas}}{c}{\size{\access{\mstate}{\stack}}  -6} \\
}
{\sstep{\transenv}{\cons{\regstatefull{\mstate}{\exenv}{\gstate}{\transeffects}}{\callstack}}{\cons{\excstate}{\callstack}}}

\infer{
(\curropcode{\mstate}{\exenv} =\CALL \lor   \curropcode{\mstate}{\exenv} =\CALLCODE)  \\
\size{\access{\mstate}{\stack}} < 7 \\ 
}
{\sstep{\transenv}{\cons{\regstatefull{\mstate}{\exenv}{\gstate}{\transeffects}}{\callstack}}{\cons{\excstate}{\callstack}}}
\end{mathpar}

For returning from a call, there are several options: 

\begin{enumerate}
\item The execution of the called code ends with $\RETURN$. In this case the call was successful. The current stack specifies the fragment of the local memory that contains the return value. The return value is copied to the caller's local memory as specified on the caller's stack and the execution proceeds in the global state left by the callee. The caller gets the remaining gas of the caller's execution refunded. To indicate success $1$ is written to the caller's stack. 
\item The execution of the called code ends with $\STOP$ or $\SUICIDE$. In this case the return value of the execution is the empty data $\emptyarray$ that is written to the local memory. This essentially means that nothing is written to the caller's local memory. 
\item The execution of the called code ends with an exception. In this case the remaining arguments are removed from the caller's stack and instead $0$ is written to the caller's stack. The caller does not get the remaining gas refunded 
\end{enumerate}

As the first two cases can be treated analogously, we just need two rules for returning from a call. 

\begin{mathpar}
\infer{
\curropcode{\mstate}{\exenv} =\CALL   \\
\access{\mstate}{\stack}=\cons{g}{\cons{\recipient}{\cons{\valu}{\cons{\io}{\cons{\is}{\cons{\oo}{\cons{\os}{s}}}}}}}   \\
\recipient_a = \recipient \mod 2^{160} \\
\flag = \cond{\access{\gstate}{\recipient_a} = \none}{0}{1} \\  
\aw = \memext{\memext{\access{\mstate}{\actw}}{\io}{\is}}{\oo}{\os} \\
c_{\textit{call}} = \gascapacity{\valu}{\flag}{g}{\access{\mstate}{\gas}} \\
c = \basecosts{\valu}{\flag} + \costmem{\access{\mstate}{\actw}}{\aw} +c_{\textit{call}} \\
\mstate' =\update{\inc{\inc{\update{\update{\mstate}{\actw}{\aw}}{\stack}{\cons{1}{s}}}{\pc}{1}}{\gas}{\lgas - c}}{\memo}{\updateinterval{\access{\mstate}{\memo}}{\oo}{\oo + s -1}{d}}
}
{\sstep{\transenv}{\cons{\haltstatefull{\gstate'}{\transeffects'}{\lgas}{d}}{\cons{\regstatefull{\mstate}{\exenv}{\gstate}{\transeffects}}{\callstack}}}{\cons{\regstatefull{\mstate'}{\exenv}{\gstate'}{\transeffects'}}{\callstack}}}
\end{mathpar}

\begin{mathpar}
\infer{
\curropcode{\mstate}{\exenv} =\CALL  \\
\access{\mstate}{\stack}=\cons{g}{\cons{\recipient}{\cons{\valu}{\cons{\io}{\cons{\is}{\cons{\oo}{\cons{\os}{s}}}}}}}   \\
\recipient_a = \recipient \mod 2^{160} \\
\flag = \cond{\getaccount{\gstate}{\recipient_a} = \none}{0}{1} \\  
\aw = \memext{\memext{\access{\mstate}{\actw}}{\io}{\is}}{\oo}{\os} \\
c_{\textit{call}} = \gascapacity{\valu}{\flag}{g}{\access{\mstate}{\gas}} \\
c = \basecosts{\valu}{\flag} + \costmem{\access{\mstate}{\actw}}{\aw} +c_{\textit{call}} \\
\mstate' =\dec{\inc{\update{\update{\mstate}{\actw}{\aw}}{\stack}{\cons{0}{s}}}{\pc}{1}}{\gas}{c}
}
{\sstep{\transenv}{\cons{\excstate}{\cons{\regstatefull{\mstate}{\exenv}{\gstate}{\transeffects}}{\callstack}}}{\cons{\regstatefull{\mstate'}{\exenv}{\gstate}{\transeffects}}{\callstack}}}
\end{mathpar}

The two other instructions for calling ($\CALLCODE$ and $\DELEGATECALL$) are similar to $\CALL$. 

The $\CALLCODE$ instruction only differs in the fact that the control flow is not handed over to the called contract, but only its code is executed in the environment of the calling account. This means in particular that the amount of money transferred is only relevant as a guard for the call, but does not need to be actually transferred. In addition, in case that the account whose code should be executed does not exists, this account is not created, but only the empty code is run. However, still the amount of Ether specified on the stack influences the execution cost. 

\begin{mathpar}
\infer{
\curropcode{\mstate}{\exenv}=\CALLCODE  \\
\access{\mstate}{\stack}=\cons{g}{\cons{\recipient}{\cons{\valu}{\cons{\io}{\cons{\is}{\cons{\oo}{\cons{\os}{s}}}}}}}   \\
\recipient_a = \recipient \mod 2^{160} \\
\gstate (\recipient_a) \neq \none\\ 
\size{A} + 1 \leq 1024 \\
\access{\getaccount{\gstate}{\access{\exenv}{\activeaccount}}}{\balance} \geq \valu \\
\aw = \memext{\memext{\access{\mstate}{\actw}}{\io}{\is}}{\oo}{\os} \\ 
c_{\textit{call}} = \gascapacity{\valu}{1}{g}{\access{\mstate}{\gas}} \\
c = \basecosts{\valu}{1} + \costmem{\access{\mstate}{\actw}}{\aw} +c_{\textit{call}} \\
\simvalid{\access{\mstate}{\gas}}{c}{\size{s} + 1} \\
d =\getinterval{\access{\mstate}{\memo}}{\io}{\io + \is -1}  \\ 
\mstate' =\smstate{c_{\textit{call}}}{0}{\emptymemory}{0}{\emptystack}  \\ 
\exenv' =\update{\update{\update{\update{\exenv}{\sender}{\access{\exenv}{\activeaccount}}}{\tvalue}{\valu}}{\inputdata}{d}}{\activecode}{\access{\getaccount{\gstate}{\recipient_a}}{\accountcode}}\\
} 
{\sstep{\transenv}{\cons{\regstatefull{\mstate}{\exenv}{\gstate}{\transeffects}}{\callstack}}{\cons{\regstatefull{\mstate'}{\exenv'}{\gstate}{\transeffects}}{\cons{\regstatefull{\mstate}{\exenv}{\gstate}{\transeffects}}{\callstack}}}}
\end{mathpar}

\begin{mathpar}
\infer{
\curropcode{\mstate}{\exenv}=\CALLCODE  \\
\access{\mstate}{\stack}=\cons{g}{\cons{\recipient}{\cons{\valu}{\cons{\io}{\cons{\is}{\cons{\oo}{\cons{\os}{s}}}}}}}   \\
\recipient_a = \recipient \mod 2^{160} \\
\getaccount{\gstate}{\recipient_a} = \none\\ 
\size{A} + 1 \leq 1024 \\
\access{\getaccount{\gstate}{\access{\exenv}{\activeaccount}}}{\balance} \geq \valu \\
\aw = \memext{\memext{\access{\mstate}{\actw}}{\io}{\is}}{\oo}{\os} \\ 
c_{\textit{call}} = \gascapacity{\valu}{1}{g}{\access{\mstate}{\gas}} \\
c = \basecosts{\valu}{1} + \costmem{\access{\mstate}{\actw}}{\aw} +c_{\textit{call}} \\
\simvalid{\access{\mstate}{\gas}}{c}{\size{s} + 1} \\
d =\getinterval{\access{\mstate}{\memo}}{\io}{\io + \is -1}  \\ 
\mstate' =\smstate{c_{\textit{call}}}{0}{\emptymemory}{0}{\emptystack}  \\ 
\exenv' =\update{\update{\update{\update{\exenv}{\sender}{\access{\exenv}{\activeaccount}}}{\tvalue}{\valu}}{\inputdata}{d}}{\activecode}{\emptyarray}\\
} 
{\sstep{\transenv}{\cons{\regstatefull{\mstate}{\exenv}{\gstate}{\transeffects}}{\callstack}}{\cons{\regstatefull{\mstate'}{\exenv'}{\gstate}{\transeffects}}{\cons{\regstatefull{\mstate}{\exenv}{\gstate}{\transeffects}}{\callstack}}}}
\end{mathpar}

\begin{mathpar}
\infer{
 \curropcode{\mstate}{\exenv} =\CALLCODE \\
\access{\mstate}{\stack}=\cons{g}{\cons{\recipient}{\cons{\valu}{\cons{\io}{\cons{\is}{\cons{\oo}{\cons{\os}{s}}}}}}}   \\
\recipient_a = \recipient \mod 2^{160} \\
\aw = \memext{\memext{\access{\mstate}{\actw}}{\io}{\is}}{\oo}{\os} \\
c_{\textit{call}} = \gascapacity{\valu}{1}{g}{\access{\mstate}{\gas}} \\
c = \basecosts{\valu}{1} + \costmem{\access{\mstate}{\actw}}{\aw} +c_{\textit{call}} \\
\simvalid{\access{\mstate}{\gas}}{c}{\size{s} + 1} \\
(\valu> \access{\getaccount{\gstate}{(\access{\exenv}{\activeaccount})}}{\accountbalance} \lor \size{A} +1 \geq 1024)}
{\sstep{\transenv}{\cons{\regstatefull{\mstate}{\exenv}{\gstate}{\transeffects}}{\callstack}}{\cons{\excstate}{\cons{\regstatefull{\mstate}{\exenv}{\gstate}{\transeffects}}{\callstack}}}}
\end{mathpar}

\begin{mathpar}
\infer{
\curropcode{\mstate}{\exenv} =\CALLCODE \\
\access{\mstate}{\stack}=\cons{g}{\cons{\recipient}{\cons{\valu}{\cons{\io}{\cons{\is}{\cons{\oo}{\cons{\os}{s}}}}}}}   \\
\recipient_a = \recipient \mod 2^{160} \\
\aw = \memext{\memext{\access{\mstate}{\actw}}{\io}{\is}}{\oo}{\os} \\
c_{\textit{call}} = \gascapacity{\valu}{1}{g}{\access{\mstate}{\gas}} \\
c = \basecosts{\valu}{1} + \costmem{\access{\mstate}{\actw}}{\aw} +c_{\textit{call}} \\
\neg \simvalid{\access{\mstate}{\gas}}{c}{\size{\access{\mstate}{\stack}}  -6} \\
}
{\sstep{\transenv}{\cons{\regstatefull{\mstate}{\exenv}{\gstate}{\transeffects}}{\callstack}}{\cons{\excstate}{\callstack}}}
\end{mathpar}

\begin{mathpar}
\infer{
 \curropcode{\mstate}{\exenv} =\CALLCODE  \\
\access{\mstate}{\stack}=\cons{g}{\cons{\recipient}{\cons{\valu}{\cons{\io}{\cons{\is}{\cons{\oo}{\cons{\os}{s}}}}}}}   \\
\recipient_a = \recipient \mod 2^{160} \\
\aw = \memext{\memext{\access{\mstate}{\actw}}{\io}{\is}}{\oo}{\os} \\
c_{\textit{call}} = \gascapacity{\valu}{1}{g}{\access{\mstate}{\gas}} \\
c = \basecosts{\valu}{1} + \costmem{\access{\mstate}{\actw}}{\aw} +c_{\textit{call}} \\
\mstate' =\update{\inc{\inc{\update{\update{\mstate}{\actw}{\aw}}{\stack}{\cons{1}{s}}}{\pc}{1}}{\gas}{\lgas - c}}{\memo}{\updateinterval{\access{\mstate}{\memo}}{\oo}{\oo + s -1}{d}}
}
{\sstep{\transenv}{\cons{\haltstatefull{\gstate'}{\transeffects'}{\lgas}{d}}{\cons{\regstatefull{\mstate}{\exenv}{\gstate}{\transeffects}}{\callstack}}}{\cons{\regstatefull{\mstate'}{\exenv}{\gstate'}{\transeffects'}}{\callstack}}}
\end{mathpar}

\begin{mathpar}
\infer{
\curropcode{\mstate}{\exenv} =\CALLCODE  \\
\access{\mstate}{\stack}=\cons{g}{\cons{\recipient}{\cons{\valu}{\cons{\io}{\cons{\is}{\cons{\oo}{\cons{\os}{s}}}}}}}   \\
\recipient_a = \recipient \mod 2^{160} \\
\aw = \memext{\memext{\access{\mstate}{\actw}}{\io}{\is}}{\oo}{\os} \\
c_{\textit{call}} = \gascapacity{\valu}{1}{g}{\access{\mstate}{\gas}} \\
c = \basecosts{\valu}{1} + \costmem{\access{\mstate}{\actw}}{\aw} +c_{\textit{call}} \\
\mstate' =\dec{\inc{\update{\update{\mstate}{\actw}{\aw}}{\stack}{\cons{0}{s}}}{\pc}{1}}{\gas}{c}
}
{\sstep{\transenv}{\cons{\excstate}{\cons{\regstatefull{\mstate}{\exenv}{\gstate}{\transeffects}}{\callstack}}}{\cons{\regstatefull{\mstate'}{\exenv}{\gstate}{\transeffects}}{\callstack}}}
\end{mathpar}

The $\DELEGATECALL$ instruction does not only keep the executing account of the current call, but also the transferred value and and the sender information. 
For this reason the value to be transferred does not need to be specified in the argument in this case. For this reason and because the cost calculation differs (not using the argument value, but the one from the environment) all rules from $\CALL$ needs to be replicated. Still, the general idea is very similar. 


\begin{mathpar}
\infer{
\curropcode{\mstate}{\exenv}=\DELEGATECALL \\
\access{\mstate}{\stack}=\cons{g}{\cons{\recipient}{\cons{\io}{\cons{\is}{\cons{\oo}{\cons{\os}{s}}}}}}   \\
\recipient_a = \recipient \mod 2^{160} \\
\gstate (\recipient_a) \neq \none\\ 
\size{A} + 1 \leq 1024 \\
\aw = \memext{\memext{\access{\mstate}{\actw}}{\io}{\is}}{\oo}{\os} \\ 
c_{\textit{call}} = \gascapacity{0}{1}{g}{\access{\mstate}{\gas}} \\
c = \basecosts{0}{1} + \costmem{\access{\mstate}{\actw}}{\aw} +c_{\textit{call}} \\
\simvalid{\access{\mstate}{\gas}}{c}{\size{s} + 1} \\
d =\getinterval{\access{\mstate}{\memo}}{\io}{\io + \is -1}  \\ 
\mstate' =\smstate{c_{\textit{call}}}{0}{\emptymemory}{0}{\emptystack}  \\ 
\exenv' =\update{\update{\exenv}{\inputdata}{d}}{\activecode}{\access{\getaccount{\gstate}{\recipient_a}}{\accountcode}}
} 
{\sstep{\transenv}{\cons{\regstatefull{\mstate}{\exenv}{\gstate}{\transeffects}}{\callstack}}{\cons{\regstatefull{\mstate'}{\exenv'}{\gstate}{\transeffects}}{\cons{\regstatefull{\mstate}{\exenv}{\gstate}{\transeffects}}{\callstack}}}}
\end{mathpar}

\begin{mathpar}
\infer{
\curropcode{\mstate}{\exenv}=\DELEGATECALL  \\
\access{\mstate}{\stack}=\cons{g}{\cons{\recipient}{\cons{\io}{\cons{\is}{\cons{\oo}{\cons{\os}{s}}}}}}  \\
\recipient_a = \recipient \mod 2^{160} \\
\getaccount{\gstate}{\recipient_a} = \none\\ 
\size{A} + 1 \leq 1024 \\
\aw = \memext{\memext{\access{\mstate}{\actw}}{\io}{\is}}{\oo}{\os} \\ 
c_{\textit{call}} = \gascapacity{0}{1}{g}{\access{\mstate}{\gas}} \\
c = \basecosts{0}{1} + \costmem{\access{\mstate}{\actw}}{\aw} +c_{\textit{call}} \\
\simvalid{\access{\mstate}{\gas}}{c}{\size{s} + 1} \\
d =\getinterval{\access{\mstate}{\memo}}{\io}{\io + \is -1}  \\ 
\mstate' =\smstate{c_{\textit{call}}}{0}{\emptymemory}{0}{\emptystack}  \\ 
\exenv' =\update{\update{\exenv}{\inputdata}{d}}{\activecode}{\emptyarray}\\
} 
{\sstep{\transenv}{\cons{\regstatefull{\mstate}{\exenv}{\gstate}{\transeffects}}{\callstack}}{\cons{\regstatefull{\mstate'}{\exenv'}{\gstate}{\transeffects}}{\cons{\regstatefull{\mstate}{\exenv}{\gstate}{\transeffects}}{\callstack}}}}
\end{mathpar}

\begin{mathpar}
\infer{
\curropcode{\mstate}{\exenv} =\DELEGATECALL \\
\access{\mstate}{\stack}=\cons{g}{\cons{\recipient}{\cons{\io}{\cons{\is}{\cons{\oo}{\cons{\os}{s}}}}}}  \\
\recipient_a = \recipient \mod 2^{160} \\
\aw = \memext{\memext{\access{\mstate}{\actw}}{\io}{\is}}{\oo}{\os} \\
c_{\textit{call}} = \gascapacity{0}{1}{g}{\access{\mstate}{\gas}} \\
c = \basecosts{0}{1} + \costmem{\access{\mstate}{\actw}}{\aw} +c_{\textit{call}} \\
\simvalid{\access{\mstate}{\gas}}{c}{\size{s} + 1} \\
\size{A} +1 \geq 1024}
{\sstep{\transenv}{\cons{\regstatefull{\mstate}{\exenv}{\gstate}{\transeffects}}{\callstack}}{\cons{\excstate}{\cons{\regstatefull{\mstate}{\exenv}{\gstate}{\transeffects}}{\callstack}}}}
\end{mathpar}

\begin{mathpar}
\infer{
\curropcode{\mstate}{\exenv} =\DELEGATECALL \\
\access{\mstate}{\stack}=\cons{g}{\cons{\recipient}{\cons{\io}{\cons{\is}{\cons{\oo}{\cons{\os}{s}}}}}}  \\
\recipient_a = \recipient \mod 2^{160} \\
\aw = \memext{\memext{\access{\mstate}{\actw}}{\io}{\is}}{\oo}{\os} \\
c_{\textit{call}} = \gascapacity{0}{1}{g}{\access{\mstate}{\gas}} \\
c = \basecosts{0}{1} + \costmem{\access{\mstate}{\actw}}{\aw} +c_{\textit{call}} \\
\neg \simvalid{\access{\mstate}{\gas}}{c}{\size{\access{\mstate}{\stack}}  -6} \\
}
{\sstep{\transenv}{\cons{\regstatefull{\mstate}{\exenv}{\gstate}{\transeffects}}{\callstack}}{\cons{\excstate}{\callstack}}}

\infer{
\curropcode{\mstate}{\exenv} =\DELEGATECALL \\
\size{\access{\mstate}{\stack}} < 6 \\ 
}
{\sstep{\transenv}{\cons{\regstatefull{\mstate}{\exenv}{\gstate}{\transeffects}}{\callstack}}{\cons{\excstate}{\callstack}}}
\end{mathpar}

\begin{mathpar}
\infer{
\curropcode{\mstate}{\exenv} =\DELEGATECALL \\
\access{\mstate}{\stack}=\cons{g}{\cons{\recipient}{\cons{\io}{\cons{\is}{\cons{\oo}{\cons{\os}{s}}}}}}  \\
\recipient_a = \recipient \mod 2^{160} \\
\aw = \memext{\memext{\access{\mstate}{\actw}}{\io}{\is}}{\oo}{\os} \\
c_{\textit{call}} = \gascapacity{0}{1}{g}{\access{\mstate}{\gas}} \\
c = \basecosts{0}{1} + \costmem{\access{\mstate}{\actw}}{\aw} +c_{\textit{call}} \\
\mstate' =\update{\inc{\inc{\update{\update{\mstate}{\actw}{\aw}}{\stack}{\cons{1}{s}}}{\pc}{1}}{\gas}{\lgas - c}}{\memo}{\updateinterval{\access{\mstate}{\memo}}{\oo}{\oo + s -1}{d}}
}
{\sstep{\transenv}{\cons{\haltstatefull{\gstate'}{\transeffects'}{\lgas}{d}}{\cons{\regstatefull{\mstate}{\exenv}{\gstate}{\transeffects}}{\callstack}}}{\cons{\regstatefull{\mstate'}{\exenv}{\gstate'}{\transeffects'}}{\callstack}}}
\end{mathpar}

\begin{mathpar}
\infer{
\curropcode{\mstate}{\exenv} =\DELEGATECALL \\
\access{\mstate}{\stack}=\cons{g}{\cons{\recipient}{\cons{\io}{\cons{\is}{\cons{\oo}{\cons{\os}{s}}}}}}  \\
\recipient_a = \recipient \mod 2^{160} \\
\aw = \memext{\memext{\access{\mstate}{\actw}}{\io}{\is}}{\oo}{\os} \\
c_{\textit{call}} = \gascapacity{0}{1}{g}{\access{\mstate}{\gas}} \\
c = \basecosts{0}{1} + \costmem{\access{\mstate}{\actw}}{\aw} +c_{\textit{call}} \\
\mstate' =\dec{\inc{\update{\update{\mstate}{\actw}{\aw}}{\stack}{\cons{0}{s}}}{\pc}{1}}{\gas}{c}
}
{\sstep{\transenv}{\cons{\excstate}{\cons{\regstatefull{\mstate}{\exenv}{\gstate}{\transeffects}}{\callstack}}}{\cons{\regstatefull{\mstate'}{\exenv}{\gstate}{\transeffects}}{\callstack}}}
\end{mathpar}

\paragraph{Contract creation}
The $\CREATE$ command initiates the creation of a new contract. 
The creation of a new contract is initiated if the call stack limit is not reached yet and if the initial balance $\valu$ that should be initially transferred to the new account does not exceed the balance of the sender (the account owning the currently executed code). 
In this case address $\freshaddress$ of the new account is created in dependence of the sender's address $\access{\exenv}{\activeaccount}$ and the sender's addresses current nonce incremented by one. 
If there already exists an account with the address, the balance of this account is transferred to the newly created one. 
Additionally, the new account gets the specified amount $\valu$ of ether transferred from the sender. 

Finally the execution of the contract starts by executing the initialization code $i$ ($i$ can be found in the local memory $\access{\mstate}{\memo}$, its location is specified by the arguments $\io$ and $\is$ on the stack). The owner of the initialization code is the newly created account $\freshaddress$. The owner $\access{\exenv}{\addr}$ of the calling code will be recorded as the initiator $\access{\exenv}{\sender}$ of the initialization code execution. 
The value $\valu$ transferred to the new account is given in the environment parameter $\access{\exenv}{\tvalue}$.
The execution starts in the empty machine state with the program counter and the number of active words set to $0$, in the empty memory $\lam{x}{0}$ (the function mapping each number to $0^{256}$) and the empty stack $\nil$. 
The original global state $\gstate$ is recorded in the caller state in order to be able to restore it in the case of an exception in the initiation code execution. 

\begin{mathpar}
\infer{
\curropcode{\mstate}{\exenv} =\CREATE \\
\access{\mstate}{\stack} = \cons{\valu}{\cons{\io}{\cons{\is}{s}}} \\
\aw = \memext{\access{\mstate}{\actw}}{\io}{\is}\\
c= \costmem{\access{\mstate}{\actw}}{\aw} + 32000 \\
\simvalid{\access{\mstate}{\gas}}{c}{\size{s} + 1} \\
\valu \leq \access{\getaccount{\gstate}{\access{\exenv}{\activeaccount}}}{\accountbalance} \\
\size{\callstack} + 1 \leq 1024 \\
\freshaddress=\getfreshaddress{\access{\exenv}{\activeaccount}}{\access{\getaccount{\gstate}{\access{\exenv}{\activeaccount}}}{\accountnonce}} \\ 
\getaccount{\gstate}{\freshaddress} = \none \\ 
\gstate' =\updategstate{
				\updategstate{
					\gstate}
					{\freshaddress}
					{\accountstate{0}{\valu}{\lam{x}{0}}{\emptyarray}}
				}
				{\access{\exenv}{\activeaccount}}
				{\inc{
					\dec{
						\getaccount{\gstate}{\access{\exenv}{\activeaccount}}}
						{\accountbalance}
						{\valu}}	
					{\accountnonce}
					{1}} \\
i=\getinterval{\access{\mstate}{\memo}}{\io}{\io + \is - 1} \\ 
\exenv' = \update{\update{\update{\update{\update{\exenv}{\sender}{\access{\exenv}{\activeaccount}}}{\activeaccount}{\freshaddress}}{\tvalue}{\valu}}{\activecode}{i}}{\inputdata}{\emptyarray} \\ 
\mstate' = \smstate{\funL{\access{\mstate}{\gas} - c}}{0}{\lam{x}{0}}{0}{\nil}}
{\sstep{\transenv}{\cons{\regstate{\mstate}{\exenv}{\gstate}{\transeffects}}{\callstack}}{\cons{\regstate{\mstate'}{\exenv'}{\gstate'}{\transeffects}}{\cons{\regstate{\mstate}{\exenv}{\gstate}{\transeffects}}{\callstack}}}}
\end{mathpar}

Actually it should not happen that the newly created address $\freshaddress$ already exists. By making $\freshaddress$ dependent on the active account's address and it's nonce (which can be seen as an internal counter on the number of new accounts already created by this account), it should be ensured that the resulting address is unique. 
However, in practice, the function $\getfreshaddress{\cdot}{\cdot}$ is realized by a hash function which requires to deal with collisions. 
For the cases where accidentally an existing address is created, the balance of the corresponding account is saved in the newly created one. 

\begin{mathpar}
\infer{
\curropcode{\mstate}{\exenv} =\CREATE \\
\access{\mstate}{\stack} = \cons{\valu}{\cons{\io}{\cons{\is}{s}}} \\
\aw = \memext{\access{\mstate}{\actw}}{\io}{\is}\\
c= \costmem{\access{\mstate}{\actw}}{\aw} + 32000 \\
\simvalid{\access{\mstate}{\gas}}{c}{\size{s} + 1} \\
\valu \leq \access{\getaccount{\gstate}{\access{\exenv}{\activeaccount}}}{\accountbalance} \\
\size{\callstack} + 1 \leq 1024 \\
\freshaddress=\getfreshaddress{\access{\exenv}{\activeaccount}}{\access{\getaccount{\gstate}{\access{\exenv}{\activeaccount}}}{\accountnonce}} \\ 
\getaccount{\gstate}{\freshaddress} \neq \none \\ 
b = \access{\getaccount{\gstate}{\freshaddress}}{\accountbalance} + \valu\\
\gstate' =\updategstate{
				\updategstate{
					\gstate}
					{\freshaddress}
					{\accountstate{0}{b}{\lam{x}{0}}{\emptyarray}}
				}
				{\access{\exenv}{\activeaccount}}
				{\inc{
					\dec{
						\getaccount{\gstate}{\access{\exenv}{\activeaccount}}}
						{\accountbalance}
						{\valu}}	
					{\accountnonce}
					{1}} \\
i=\getinterval{\access{\mstate}{\memo}}{\io}{\io + \is - 1} \\ 
\exenv' = \update{\update{\update{\update{\update{\exenv}{\sender}{\access{\exenv}{\activeaccount}}}{\activeaccount}{\freshaddress}}{\tvalue}{\valu}}{\activecode}{i}}{\inputdata}{\emptyarray} \\ 
\mstate' = \smstate{\funL{\access{\mstate}{\gas} - c}}{0}{\lam{x}{0}}{0}{\nil}}
{\sstep{\transenv}{\cons{\regstatefull{\mstate}{\exenv}{\gstate}{\transeffects}}{\callstack}}{\cons{\regstatefull{\mstate'}{\exenv'}{\gstate'}{\transeffects}}{\cons{\regstatefull{\mstate}{\exenv}{\gstate}{\transeffects}}{\callstack}}}}
\end{mathpar}

Similarly to the $\CALL$ case, the execution of the $\CREATE$ instruction can fail at call time in the case that either the value $\valu$ to be transferred to the newly created account exceeds the calling account's balance or if the call stack limit is reached. 

\begin{mathpar}
\infer{
\curropcode{\mstate}{\exenv} =\CREATE \\
\access{\mstate}{\stack} = \cons{\valu}{\cons{\io}{\cons{\is}{s}}} \\
\aw = \memext{\access{\mstate}{\actw}}{\io}{\is}\\
c= \costmem{\access{\mstate}{\actw}}{\aw} + 32000 \\
\simvalid{\access{\mstate}{\gas}}{c}{\size{s} + 1} \\
(\valu > \access{\getaccount{\gstate}{\access{\exenv}{\activeaccount}}}{\accountbalance} \lor
\size{\callstack} + 1 > 1024)}
{\sstep{\transenv}{\cons{\regstatefull{\mstate}{\exenv}{\gstate}{\transeffects}}{\callstack}}{\cons{\excstate}{\cons{\regstatefull{\mstate}{\exenv}{\gstate}{\transeffects}}{\callstack}}}}
\end{mathpar}

In addition the usual out-of-gas exception and violations of the stack limits need to be considered: 

\begin{mathpar}
\infer{
\curropcode{\mstate}{\exenv} =\CREATE \\
\access{\mstate}{\stack} = \cons{\valu}{\cons{\io}{\cons{\is}{s}}} \\
\aw = \memext{\access{\mstate}{\actw}}{\io}{\is}\\
c= \costmem{\access{\mstate}{\actw}}{\aw} + 32000 \\
\neg \simvalid{\access{\mstate}{\gas}}{c}{\size{s} + 1}}
{\sstep{\transenv}{\cons{\regstatefull{\mstate}{\exenv}{\gstate}{\transeffects}}{\callstack}}{\cons{\excstate}{\callstack}}}
\end{mathpar}

\begin{mathpar}
\infer{
\curropcode{\mstate}{\exenv} =\CREATE \\
\access{\mstate}{\stack} < 3}
{\sstep{\transenv}{\cons{\regstatefull{\mstate}{\exenv}{\gstate}{\transeffects}}{\callstack}}{\cons{\excstate}{\callstack}}}
\end{mathpar}

To return from contract creation we need to consider different cases: 
\begin{enumerate}
\item 
The initialization code ends with a $\RETURN$.
In this case contract creation was successful. The return value specifies the code of the new contract. This code will be executed when the contract is called later on.  
To indicate success and to make the newly created contract accessible to the caller, the address of the new contract account is written to the stack. 
The caller of the contract creation needs to proceed with the remaining gas from the contract creation and additionally needs to pay a final contract creation cost depending on the length of the contract body code. 
\item 
The initialization code ends with $\STOP$ or $\SUICIDE$.
In this case contract creation was theoretically successful, but no practical usable contract was created as calls to this contract do not cause code to be executed. 
Nevertheless the final contract creation cost needs to be paid. 
\item 
The initialization code causes an exception.  
In this case the contract creation was not successful. The former global state is restored and therefore all side effects of the contract creation are deleted. 
To indicate the failure of the contract creation the number $0$ is written to the stack of the caller. Additionally all gas of the caller state is deleted. 
\end{enumerate}

Cases one and two result in regular halting of the callee. The command specific changes affecting the global state, the remaining gas and the output data are recorded in the halting state. 
In the case of contract creation, a final fee is charged that depends on the size of the return data. 
If the gas remaining from the execution of the initialization code is not sufficient to pay the additional fee, an exception occurs. 

\begin{mathpar}
\infer{
\curropcode{\mstate}{\exenv} =\CREATE \\
\access{\mstate}{\stack} = \cons{\valu}{\cons{\io}{\cons{\is}{s}}} \\
\aw = \memext{\access{\mstate}{\actw}}{\io}{\is}\\
c= \costmem{\access{\mstate}{\actw}}{\aw} + 32000 \\
c_{\textit{final}} = 200 \cdot \size{d} \\ 
\lgas \geq c_{\textit{final}} \\ 
\freshaddress=\getfreshaddress{\access{\exenv}{\activeaccount}}{\access{\getaccount{\gstate}{\access{\exenv}{\activeaccount}}}{\accountnonce}} \\ 
\mstate' =\update{\inc{ \inc{\update{\mstate}{\stack}{\cons{\freshaddress}{s}}}{\pc}{1}}{\gas}{\lgas-c-c_{\textit{final}}}}{\actw}{\aw} \\
\gstate'' = \updategstate{\gstate'}{\freshaddress}{\update{\getaccount{\gstate'}{\freshaddress}}{\accountcode}{d}}\\
}
{\sstep{\transenv}{\cons{\haltstatefull{\gstate'}{\transeffects'}{\lgas}{d}}{\cons{\regstatefull{\mstate}{\exenv}{\gstate}{\transeffects}}{\callstack}}}{\cons{\regstatefull{\mstate'}{\exenv}{\gstate''}{\transeffects'}}{\callstack}}}
\end{mathpar}
 

\begin{mathpar}
\infer{
\curropcode{\mstate}{\exenv} =\CREATE \\
c_{\textit{final}} = 200 \cdot \size{d} \\ 
\lgas < c_{\textit{final}}}
{\sstep{\transenv}{\cons{\haltstatefull{\gstate'}{\transeffects'}{\lgas}{d}}{\cons{\regstatefull{\mstate}{\exenv}{\gstate}{\transeffects}}{\callstack}}}{\cons{\excstate}{\callstack}}}
\end{mathpar}

In the case of exceptional halting of the callee, as in the $\CALL$ case, the remaining gas is not refunded and the global state as well as the transaction effects are reverted. 
\begin{mathpar}
\infer{
\curropcode{\mstate}{\exenv} =\CREATE \\
\access{\mstate}{\stack} = \cons{\valu}{\cons{\io}{\cons{\is}{s}}} \\
\aw = \memext{\access{\mstate}{\actw}}{\io}{\is}\\
c= \costmem{\access{\mstate}{\actw}}{\aw} + 32000 \\
\mstate' =\update{\inc{\dec{\update{\mstate}{\stack}{\cons{0}{s}}}{\pc}{1}}{\gas}{c}}{\actw}{\aw} \\
}
{\sstep{\transenv}{\cons{\excstate}{\cons{\regstatefull{\mstate}{\exenv}{\gstate}{\transeffects}}{\callstack}}}{\cons{\regstatefull{\mstate'}{\exenv}{\gstate}{\transeffects}}{\callstack}}}
\end{mathpar}
\fi

\end{document}